\documentclass[
    reprint, twocolumn,
    aps,prl,10pt,amsmath,amssymb,
    longbibliography,superscriptaddress,citeautoscript,
    preprintnumbers]{revtex4-2}

\usepackage{xr}
\makeatletter
\newcommand*{\addFileDependency}[1]{
  \typeout{(#1)}
  \@addtofilelist{#1}
  \IfFileExists{#1}{}{\typeout{No file #1.}}
}
\makeatother

\newcommand*{\myexternaldocument}[1]{
    \externaldocument{#1}
    \addFileDependency{#1.tex}
    \addFileDependency{#1.aux}
}

\myexternaldocument{supp}

\usepackage[colorlinks=true, linkcolor=blue, citecolor=blue, urlcolor=blue]{hyperref}
\usepackage{oubraces}
\usepackage{bm,bbm,mathbbol}
\usepackage{physics}
\usepackage{mathtools}
\usepackage{amsmath, amsthm, amssymb, amsfonts}
\usepackage{bbold}
\usepackage{esvect}
\usepackage[dvipsnames]{xcolor}
\usepackage{epsfig}
\usepackage{array}
\usepackage{multirow}
\usepackage{graphicx}
\usepackage[normalem]{ulem}
\usepackage{float}
\usepackage[section]{placeins}
\usepackage{afterpage}
\usepackage{xifthen}
\usepackage{tikz}

\usepackage{siunitx}
\usepackage[caption=false]{subfig}
\usepackage{multirow}
\usepackage{placeins}
\usepackage{cancel}
\usepackage{spreadtab}

\newcommand{\mb}[1]{{\bm{\mathrm{#1}}}}  

\newcommand{\veps}[0]{\varepsilon}

\newcommand{\mcO}[0]{{\mathcal{O}}}

\newcommand{\stkout}[1]{\ifmmode\text{\sout{\ensuremath{#1}}}\else\sout{#1}\fi}

\newcommand{\Eq}[1]{Eq.~\eqref{#1}}

\newcommand{\nk}[0]{{n\mathbf{k}}}

\newcommand{\mkq}[0]{{m\mathbf{k+q}}}
\newcommand{\qnu}[0]{{\mathbf{q}\nu}}

\newcommand{\nint}{\!\int\!} 

\newcommand{\intdq}{{ \nint \frac{d\mb{q}}{\Omega^{\rm BZ}} }}

\renewcommand{\Im}{\operatorname{Im}}

\usepackage{orcidlink}

\begin{document}

\title{Self-consistent electron lifetimes for electron-phonon scattering}

\author{Jae-Mo Lihm\,\orcidlink{0000-0003-0900-0405}}
\email{jaemo.lihm@gmail.com}
\affiliation{Department of Physics and Astronomy, Seoul National University, Seoul 08826, Korea}
\affiliation{Center for Correlated Electron Systems, Institute for Basic Science, Seoul 08826, Korea}
\affiliation{Center for Theoretical Physics, Seoul National University, Seoul 08826, Korea}
\author{Samuel Ponc\'e\,\orcidlink{0000-0003-1159-8389}}
\email{samuel.ponce@uclouvain.be}
\affiliation{
European Theoretical Spectroscopy Facility, Institute of Condensed Matter and Nanosciences, Universit\'e catholique de Louvain, Chemin des \'Etoiles 8, B-1348 Louvain-la-Neuve, Belgium. 	
}
\affiliation{
WEL Research Institute, avenue Pasteur, 6, 1300 Wavre, Belgique	
}
\author{Cheol-Hwan Park\,\orcidlink{0000-0003-1584-6896}}
\email{cheolhwan@snu.ac.kr}
\affiliation{Department of Physics and Astronomy, Seoul National University, Seoul 08826, Korea}
\affiliation{Center for Correlated Electron Systems, Institute for Basic Science, Seoul 08826, Korea}
\affiliation{Center for Theoretical Physics, Seoul National University, Seoul 08826, Korea}

\date{\today}

\begin{abstract}
Acoustic phonons in piezoelectric materials strongly couple to electrons through a macroscopic electric field.
We show that this coupling leads to a momentum-dependent divergence of the Fan--Migdal electron linewidth.
We then develop a self-consistent theory for calculating electron linewidths, which not only removes this piezoelectric divergence but also considerably modifies the linewidth in nonpiezoelectric, polar materials.
Our predictions await immediate experimental confirmation, and this self-consistent method should be broadly used in interpreting various experiments on the electronic properties of real materials.
\end{abstract}

\maketitle

Piezoelectricity, the change in polarization in response to a mechanical strain, is a common phenomenon allowed in 20 out of 21 noncentrosymmetric point groups~\cite{Halasyamani1998}.
It induces a coupling between electrons and acoustic phonons mediated by a long-range electric field~\cite{1972MartinPiezo}.
In recent years, there has been a significant increase in the interest in the calculation of piezoelectric electron-phonon coupling (EPC) and its effect on electronic properties~\cite{Vanderbilt2000,Hamann2005,Stengel2013,2019RoyoPiezo,2017LiuPiezo,2020BruninPiezo1,2020BruninPiezo2,2020JhalaniPiezo,2020ParkPiezo,Ponce2021,2021GanoseAMSET,2023PoncePiezo1,2023PoncePiezo2}.
The combination of Wannier interpolation~\cite{1997Marzari,2001Souza,Giustino2007,2012Marzari} with analytically calculated long-range Fr\"ohlich and piezoelectric EPC in terms of dipole and quadrupole moments~\cite{2015VerdiFrohlich,2017LiuPiezo,2020BruninPiezo1,2020BruninPiezo2,2020JhalaniPiezo,2020ParkPiezo,Ponce2021,2021GanoseAMSET,2023PoncePiezo1,2023PoncePiezo2} enables a fully \textit{ab initio} study of the effects of EPC on the spectral and transport properties of piezoelectric materials.

Interestingly, in the presence of piezoelectric EPC, the electron linewidth diverges when calculated from the perturbative Fan--Migdal formula~\cite{Ridley1982}, the current standard for calculating carrier linewidths from first principles~\cite{2017Giustino, Ponce2020}.
This divergence should occur when using a vanishing broadening~\cite{Ponce2018}, an adaptive broadening~\cite{Li2014, Ponce2021, 2023PoncePiezo2}, or even the tetrahedron integration method~\cite{Bloechl1994, 2020BruninPiezo2}.
Even though this divergence implies that calculations of the linewidths cannot be converged, it has not been considered in first-principles calculations of piezoelectric materials~\cite{2017LiuPiezo,2020BruninPiezo1,2020BruninPiezo2,2020JhalaniPiezo,2020ParkPiezo,2023PoncePiezo1,2023PoncePiezo2}.
Calculations with a finite broadening would lead to results that depend sensitively on artificial computational parameters~\cite{Ponce2018}.

In this work, we study the limitation of the one-shot calculation of electron lifetimes in piezoelectric materials and prove that a self-consistent broadening of the electrons is required.
First, we find that the piezoelectric divergence depends on the electron wavevector and confirm this finding with first-principles calculations.
Then, we develop a formalism based on the quasiparticle approximation to self-consistently calculate the linewidths and apply it to study the EPC in cubic boron nitride (c-BN), Si, NaCl, and PbTe.
We find that self-consistency dominates the regularization of piezoelectric EPC for weak to intermediate doping and quantitatively affects the results even at higher doping.
Furthermore, we find that self-consistency also affects the scattering due to polar optical phonons in both piezoelectric and non-piezoelectric materials, leading to a complex dependence of the linewidths on the electron energy, temperature, and doping.
Our prediction can be directly confirmed by angle-resolved photoemission experiments~\cite{Bostwick2007, Park2007, Park2009}.
The self-consistent linewidths should be broadly applied to the computations of transport, optical, and spectroscopic properties.

The Fan--Migdal formula for the phonon-induced electron linewidth at band $n$ and wavevector $\mb{k}$ reads~\cite{MahanBook,Grimvall1981,Ponce2020}
\begin{multline} \label{eq:gamma}
    \gamma_\nk = \frac{2\pi}{\hbar}  \sum_{m\nu} \intdq | g_{mn\nu}(\mb{k,q})|^2 \sum_\pm \Big\{ \big[ n_{\qnu}  \\
    + f^\pm (\varepsilon_\mkq) \big]  \delta( \veps_{n\mb{k}} - \veps_{\mkq} \pm  \hbar \omega_{\qnu}) \Big\},
\end{multline}
where $\Omega^{\rm BZ}$ is the volume of the Brillouin zone, $n_\qnu$ the phonon occupation at wavevector $\mb{q}$ and mode $\nu$,
$f^+(\varepsilon_\mkq)$ is the electron occupation, $f^-(\varepsilon_\mkq) = 1 - f^+(\varepsilon_\mkq)$,
$g_{mn\nu}(\mb{k,q})$ the electron-phonon matrix element, and $\veps_\nk$ and $\hbar\omega_\qnu$ the electron and phonon energies, respectively.
The linewidth $\gamma_\nk$ is the inverse of the carrier lifetime $\tau_\nk$ and twice the imaginary part of the electron self-energy $\Sigma_{n\mathbf{k}}$,
$\gamma_\nk = 1 / \tau_\nk = 2 \abs{\Im \Sigma_{n\mathbf{k}}} / \hbar$~\cite{Ponce2016}.

To analyze the linewidths due to piezoelectric EPC, we start with a three-dimensional isotropic long-wavelength model:
$\veps_k = \frac{\hbar^2 k^2}{2m}$, $\omega_q = q v^{\rm ph}$,
where $m$ is the electron effective mass and $v^{\rm ph}$ the phonon velocity.
We set $\veps_k=0$ at the conduction-band minimum (CBM) throughout this paper.
We consider the piezoelectric EPC with longitudinal acoustic phonons~\cite{Ridley1982}, which originates from the long-range Coulomb interaction and is discontinuous at $\mathbf{q}=0$.

Let us regularize the delta function in \Eq{eq:gamma} with a Lorentzian
\begin{equation} \label{eq:lorentzian}
\delta_{\eta}(\varepsilon) = \frac{1}{\pi} \frac{\eta/2}{\varepsilon^2 + (\eta / 2)^2}\,.
\end{equation}
Using other regularization functions such as Gaussians does not alter the conclusions qualitatively.
Evaluating the integral in \Eq{eq:gamma} yields
\begin{equation} \label{eq:gamma_eta}
    \gamma_k \propto  \begin{cases}
        \dfrac{1}{k} \ln \dfrac{\veps_k}{\eta} & \text{ if $v^{\rm el}_k > v^{\rm ph}$ ($k > m v^{\rm ph}/\hbar$)}, \\
        \dfrac{1}{k}\ln \dfrac{v^{\rm ph} + v^{\rm el}_k}{v^{\rm ph} - v^{\rm el}_k} & \text{ if $v^{\rm el}_k < v^{\rm ph}$ ($k < m v^{\rm ph}/\hbar$)},
    \end{cases}
\end{equation}
where $v^{\rm el}_k = \hbar k / m$ is the band velocity of the electron at wavevector $\mb{k}$
(see Eqs.~(S15) and (S26) of the SM~\cite{Supp}).
\nocite{1968ArltPiezo, StefanucciBook, Abramovitch2023, Royo2020, 1980GenzAdaptive, JuliaHCubature, JuliaQuadGK, 1994DalCorsoPiezo, 1997BernardiniPiezo, 1998SaghiSzaboPiezo, Stengel2013, Giannozzi2017, Gonze2016, Gonze2020, 2013HamannONCVPSP, Setten2018, Perdew1996, 2020PizziWannier90, Giustino2007, 2017BezansonJulia, 1970Mermin, BornHuang1954, 2021ZhouPerturbo, 1992RestaPolarization, 1993KingSmithPolarization, 1993VanderbiltPolarization, 1994RestaPolarization, 1994GrimsditchElastic, Gonze1997a, Wu2005}

For the first case, $v^{\rm el}_k > v^{\rm ph}$, the linewidth diverges logarithmically in the zero-broadening limit $\eta \to 0^+$.
A similar logarithmic divergence has been found using an \textit{artificial} infrared cutoff of phonon wavevector $q$ instead of the finite broadening (see Sec.~3.6 of Ref.~\cite{Ridley1982}).
In the second case, $v^{\rm el}_k < v^{\rm ph}$, the linewidth converges to a finite value.
Due to the $1/k$ prefactor in \Eq{eq:gamma_eta}, for a sufficiently small $\eta$, the linewidth is peaked around $k = mv^{\rm ph} / \hbar$ where $v^{\rm el}_k = v^{\rm ph}$.
This dichotomy of the Brillouin zone into a divergent and convergent region is an important finding of this work.

\begin{figure}[b]
  \centering
  \includegraphics[width=1.0\columnwidth]{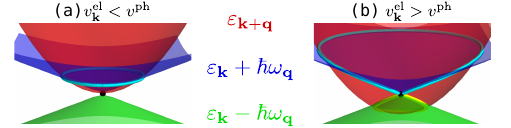}
  \includegraphics[width=1.0\columnwidth]{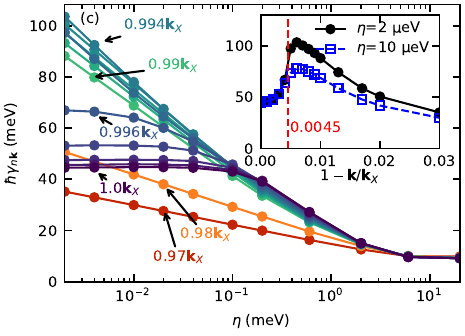}
  \caption{
  (a, b) Dispersion of electrons and phonons. We show two-dimensional dispersions for a simpler visualization. The $\mb{q}$ points where energy-conserving phonon absorption or emission is allowed is indicated by cyan and yellow, respectively.
  (c) Linewidths at the conduction band of intrinsic c-BN at $T=300$~K as a function of broadening for $\mb{k}$ points along the $\Gamma \mathrm{X}$ line. 
  Inset: linewidths as a function of wavevectors for two broadening values. 
  The vertical red dashed line indicates the onset of the divergence.
  }
  \label{fig:lifetime_fixed}
\end{figure}

This logarithmic divergence originates from the absorption and emission of acoustic phonons with an infinitesimal wavevector.
Figures~\ref{fig:lifetime_fixed}(a) and \ref{fig:lifetime_fixed}(b) illustrate the dispersion that disallows and allows the infinitesimal-wavevector scattering, respectively; only the latter leads to the divergence.
The criterion for divergence can be generalized to an arbitrary three-dimensional dispersion.
To have infinitesimal-wavevector scattering, one needs a direction $\hat{q}$ where the electron and acoustic phonon velocities are identical:
\begin{equation} \label{eq:onset_general}
    \mb{v}^{\rm el}_\nk \cdot \hat{q} = \mb{v}^{\rm ph}_{\mb{0}\nu} \cdot \hat{q}
    \text{ for some $\hat{q}$ and $\nu \in {1,2,3}$}.
\end{equation}

We confirmed this behavior by performing \textit{ab initio} calculations whose details are provided in the SM~\cite{Supp}.
Figure~\ref{fig:lifetime_fixed}(c) shows the \textit{ab initio} linewidths for the conduction-band states of c-BN.
We performed an ultradense nonuniform sampling near $\Gamma$, with a density reaching that of a $(5 \times 10^7)^3$ homogeneous $\mathbf{q}$-point grid.
The CBM is located at $\mb{k}_{\rm X} = \frac{2\pi}{a} \hat{z}$, where $a$ is the cubic lattice parameter.
For a $\mb{k}$ point on the $\Gamma \mathrm{X}$ line, the condition \Eq{eq:onset_general} for $\hat{q} \parallel \mb{k}_{\rm X}$ reads
\begin{equation}
    \abs{{(\mb{k} - \mb{k}_{\rm X}}) \cdot \hat{z}} = \frac{m_z v_z^{\rm ph}}{\hbar} = \frac{m_z}{\hbar} \sqrt{\frac{C_{44}}{\rho}}
    = 0.0080~\mathrm{\AA}^{-1},
\end{equation}
where $C_{44}$ is the shear modulus and $\rho$ the mass density.
By scanning all $\hat{q}$ directions, we find a lower bound of $\abs{k_z - \frac{2\pi}{a}} > 0.0078 \mathrm{\AA}^{-1} = 0.0045 \frac{2\pi}{a}$ for the presence of the logarithmic divergence
where \Eq{eq:onset_general} is satisfied at $\hat{q} = (\sin 0.1\pi, 0, \cos 0.1\pi)$ due to a small anisotropy in the dispersion.
Figure~\ref{fig:lifetime_fixed}(c) verifies this criterion, as the linewidths at $\abs{\mb{k} - \mb{k}_{\rm X}} \leq 0.004 \abs{\mb{k}_{\rm X}}$ converge to a finite value in the $\eta \to 0^+$ limit, while those at $\abs{\mb{k} - \mb{k}_{\rm X}} \geq 0.005 \abs{\mb{k}_{\rm X}}$ diverge logarithmically in $\eta$.

Since this divergence is logarithmic, it is easy to misinterpret it as a convergence, which explains why the divergence was not reported in any of the previous \textit{ab initio} calculations.
For a given broadening $\eta$, capturing the correct momentum dependence of the linewidths requires resolving phonon modes with energy as low as $\eta$ and wavevectors as small as $\eta / \hbar v^{\rm ph}$.
For c-BN, one would need a $\mb{q}$-point density of $1/6000^3$, albeit only at the zone center, to get the correct momentum dependence at $\eta = 0.2~$meV, where the divergence becomes barely noticeable [Fig.~\ref{fig:lifetime_fixed}(c)].
Such a grid is more than three orders of magnitude denser than the ones commonly used in \textit{ab initio} calculations~\cite{Ponce2020, Lee2023EPW}.
We emphasize that this divergence is an artifact of the zero-broadening formula and does not reflect experimental observations.

We remark that since the piezoelectric scattering is dominated by small-$q$ phonons, it conserves the electron momentum and therefore does not contribute to the electrical resistivity when calculated with the full Boltzmann transport equation.
Hence, the mobility of c-BN calculated using the Boltzmann transport equation converges rapidly with the density of the $\mb{q}$ points~\cite{Ponce2021}.

We now study the role of self-consistent broadening in regularizing this divergence.
Since the linewidths broaden the electronic spectral function, the energy-conserving delta function in \Eq{eq:gamma} should be broadened accordingly.
By evaluating the Fan--Migdal self-energy under the quasiparticle approximation and taking the imaginary part of the self-energy at the quasiparticle energy as the linewidth, we derive a self-consistent formula for the linewidth (see Sec.~\ref{sec:supp_sc}~\cite{Supp}):
\begin{multline} \label{eq:gamma_sc}
    \gamma_\nk = \frac{2\pi}{\hbar} \sum_{m\nu} \intdq \abs{g_{mn\nu}(\mb{k,q})}^2 \sum_{\pm} \Bigl\{ \bigl[ n_{\qnu} \\
	+ f^\pm(\veps_\nk \pm \hbar\omega_\qnu) \bigr] \delta_{\hbar\gamma_\mkq}( \veps_\nk - \veps_{\mkq} \pm \hbar \omega_{\qnu}) \Bigr\}.
\end{multline}
The self-consistent broadening $\gamma_\nk$ is calculated for each temperature.

There are two key differences compared to the non-self-consistent formula, \Eq{eq:gamma}.
First, the delta function is replaced with a Lorentzian [Eq.~\eqref{eq:lorentzian}] with a self-consistent width $\hbar\gamma_\mkq$.
Second, the occupation function $f^\pm(\veps_\nk \pm \hbar\omega_\qnu)$ is used instead of $f^\pm(\veps_\mkq)$.
The two occupations can be used interchangeably in the limit $\hbar\gamma_\mkq \ll k_{\rm B} T$ because of the energy-conserving delta function.
Otherwise, the two formulas give different results.
For example, at $T=0$, only the correct formula using $f^\pm(\veps_\nk \pm \hbar\omega_\qnu)$ gives a vanishing linewidth at the Fermi surface.
The approximation $f^\pm(\veps_\nk \pm \hbar\omega_\qnu) \approx f^\pm(\veps_\mkq)$ has been adopted in Ref.~\cite{Xu2020SelfConsistent} for a self-consistent calculation of electron lifetimes.

The renormalization of the phonon energies and spectral function may also be similarly included, by modifying the phonon frequency and occupation factors in \Eq{eq:gamma_sc}.
However, in the low- to intermediate-doping regime, which we focus on in this work, the adiabatic density-functional perturbation theory is expected to work well for the phonons~\cite{2017Giustino}, and the phonon renormalization can be viewed as a secondary effect.
In fact, in the intrinsic, undoped case, the phonon dispersion does not get renormalized, while the electron spectral functions are still strongly renormalized.


\begin{figure}[t]
  \centering
  \includegraphics[width=1.0\columnwidth]{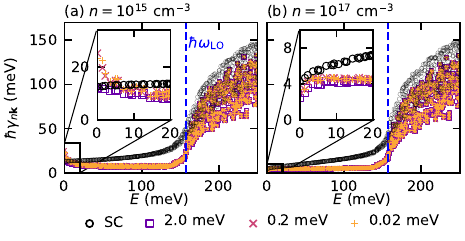}
  \caption{
  Electron linewidth of c-BN at $T=300$~K at different doping levels $n$.
  Black circles show the self-consistent (SC) linewidths while the others show those computed at a fixed broadening.
  Blue vertical dashed lines indicate the LO phonon frequency.
  }
  \label{fig:lifetime_screen}
\end{figure}

Another mechanism that regularizes the logarithmic divergence is the free-carrier screening~\cite{Ridley1982}.
In doped semiconductors, free carriers screen the long-range Coulomb interaction to make it finite ranged.
This screening is important for longitudinal optical (LO) phonons~\cite{2017Verdi,2022Kandolf,2022Macheda3D,2022Macheda2D} and ionized impurity potentials~\cite{2021LuImpurity,2023LeveilleeImpurity}.
For the piezoelectric EPC, the screening regularizes the $\mcO(q^0)$ discontinuity of the EPC at $q=0$ to a smoothly decaying function with a characteristic wavevector given by the Thomas--Fermi wavevector.
In metals, free carriers will completely screen the piezoelectric EPC.
However, when the doping is not too heavy, the piezoelectric EPC will still make a large contribution to the linewidth.
In this case, self-consistency plays an important role.

Figure~\ref{fig:lifetime_screen} compares the linewidths at two different doping levels.
The doping is included via the change in the electron occupation factor and the free-carrier screening of EPC~[\Eq{eq:supp_g_screen}], while the electron and phonon dispersions are taken from calculations on the undoped system.
We identify two distinct regimes.
For a low doping $n \leq 10^{16}~\mathrm{cm}^{-3}$, the self-consistent linewidths display a monotonic energy dependence in the shown energy range.
In contrast, the fixed-broadening linewidths display a sharp peak at an energy below 5~meV, which originates from the piezoelectric divergence regularized by the free-carrier screening.
In this regime, self-consistency plays a dominant role in regularizing the piezoelectric scattering.
For a higher doping $n \geq 10^{17}~\mathrm{cm}^{-3}$, the low-energy peak disappears but quantitative differences between the fixed-broadening and self-consistent linewidths remain, especially at energies $\veps_\nk \lesssim \hbar \omega_{\rm LO}$.
These differences are due to the LO phonon scattering, which is heavily broadened due to a large linewidth of the electrons above the LO phonon energy.
We note that free-carrier screening of the LO phonons ($\omega \approx $ 160~meV) is negligible at the considered carrier densities ($n \leq 10^{18}~\mathrm{cm}^{-3}$) since the LO phonon frequency is much higher than the plasma frequency ($\omega \lesssim $ 25~meV)~\cite{2017Verdi}.

\begin{figure}[b]
    \centering
    \includegraphics[width=1.0\columnwidth]{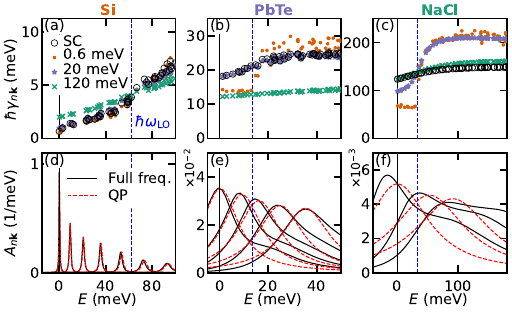}
    \caption{
    (a-c) Linewidths at the conduction band of Si, PbTe, and NaCl at $T=300~$K and $n=10^{13}~\mathrm{cm}^{-3}$ with self-consistent lifetime (empty circle) and fixed broadening (filled symbols).
    (d-f) Full-frequency and quasiparticle (QP) spectral functions along the $\Gamma$X line for $\abs{\mb{k}} / \abs{\mb{k}_\mathrm{X}}$ values
    (d) 0.82, 0.86, 0.88, 0.9, 0.92, 0.94, and 0.96 for Si;
    (e) 1, 0.97, 0.96, 0.95, and 0.94 for PbTe; and
    (f) 0, 0.06, and 0.08 for NaCl,
    from left to right.}
    \label{fig:lifetime_Si_PbTe}
\end{figure}

This analysis leads to our main finding that self-consistency may qualitatively change the energy dependence of electron linewidths.
To demonstrate that such a strong effect is a generic phenomenon that does not require piezoelectricity, we apply our theory of self-consistent linewidths to three non-piezoelectric materials, Si, PbTe, and NaCl.
We compare the self-consistent calculation with fixed-broadening calculations with three different broadenings, chosen from the smallest self-consistent linewidth of each material.
Figures~\ref{fig:lifetime_Si_PbTe}(a-c) show that while the linewidths of Si are affected only slightly by broadening parameters and self-consistency, those of PbTe and NaCl vary significantly.
Compared to the $\eta = 0.6~\mathrm{meV}$ results, self-consistency substantially smoothens the sharp rise of electron linewidths at the LO-phonon energy.
We also find that while fine-tuning can yield fixed-broadening results comparable to the self-consistent calculation, the value of the ``optimal'' broadening parameter is highly material dependent, spanning over two orders of magnitude, and cannot be determined \emph{a priori}.
The deviations of over several tens of meV's are well above the current resolution for angle-resolved photoemission spectroscopy and can be directly confirmed from experiments~\cite{Bostwick2007,Park2007,Park2009}.

Figures~\ref{fig:lifetime_Si_PbTe}(d-f) show the spectral functions obtained from a one-shot calculation of the frequency-dependent self-energy using the converged self-consistent broadening (see Sec.~\ref{sec:sc_spectral} for details~\cite{Supp}).
For Si and PbTe, the two spectral functions agree very well, especially in the low-energy part, providing a strong validation of the quasiparticle approximation.
We also find a similar agreement for c-BN (Fig.~\ref{fig:spectral_function_cBN}).
For NaCl, where the EPC is the strongest, the deviation is larger, but the quasiparticle approximation still captures the width of the spectral function.

\begin{figure}[tb]
    \centering
    \includegraphics[width=1.0\columnwidth]{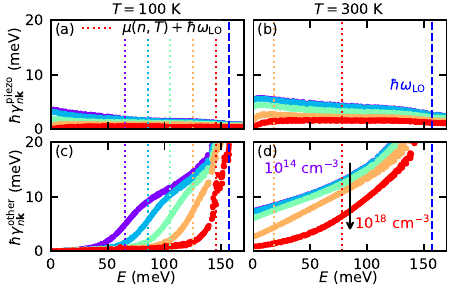}
    \caption{(a, b) Piezoelectric and (c, d) other contributions to the linewidth of c-BN at varying doping and using self-consistent broadening.
    Consecutive curves show results for systems with ten times different doping levels $n$, going from violet (10$^{14}$~cm$^{-3}$) to red (10$^{18}$~cm$^{-3}$).
    The values of $\mu(n,T)+\omega_{\text LO}$, where $\mu$ is the chemical potential, are indicated by vertical dotted lines in color. 
    }
    \label{fig:lifetime_temperature}
\end{figure}

The doping dependence of the linewidth has two major contributions, the piezoelectric scattering from acoustic phonons and the scattering from the LO phonons.
To understand their respective roles, we use a long-wavelength model that solely contains the acoustic phonons and their piezoelectric EPC with parameters computed from the piezoelectric and elastic tensors (see Sec.~\ref{sec:supp_hybrid}).
We write the total linewidth as the sum of the piezoelectric contribution and the remainder, i.e.,
\begin{equation}
    \gamma_\nk = \gamma_\nk^{\rm piezo} + \gamma_\nk^{\rm other}.
\end{equation}
Figures~\ref{fig:lifetime_temperature}(a,b) show that $\gamma^{\rm piezo}_\nk$ decreases with doping due to the free-carrier screening of the piezoelectric EPC.
In contrast, since the screening is ineffective for the LO phonon, and the doping dependence of $\gamma^{\rm other}_\nk$ in Figs.~\ref{fig:lifetime_temperature}(c,d) originates from the change in the chemical potential.
At $k_{\rm B} T \ll \hbar \omega_{\rm LO}$, where the LO-phonon occupation is negligible, the linewidth is almost zero for states with energy below $\mu + \omega_{\rm LO}$.
This behavior is a consequence of the general principle that at $T=0$ the emission of a phonon with energy $\hbar\omega_{\rm LO}$ is not allowed for states with energy inside the window $[\mu - \hbar\omega_{\rm LO},\ \mu + \hbar\omega_{\rm LO}]$, since for an electron with energy $\veps_\nk < \mu + \hbar\omega_{\rm LO}$ the final state has energy $\veps_\nk - \hbar\omega_{\rm LO} < \mu$ and is fully occupied, forbidding the transition.
At finite temperatures, this feature is broadened, resulting in a smooth increase centered at $\mu + \hbar\omega_{\rm LO}$, as shown in Figs.~\ref{fig:lifetime_temperature}(c,d).
If a fixed broadening is used (Fig.~\ref{fig:lifetime_temperature_fixed}), or if $f(\veps_\nk \pm \hbar\omega_\qnu)$ in \Eq{eq:gamma_sc} is approximated by $f(\veps_\mkq)$ (Fig.~\ref{fig:lifetime_temperature_occ_kq}), this feature disappears.


Self-consistency in the EPC can affect many electronic properties such as the broadening of the electron spectral function measured from angle-resolved photoemission spectroscopy~\cite{2003EigurenARPES,2008GiustinoARPES,2021LiARPES},
optical absorption spectra~\cite{2012NoffsingerIndabs,2014PatrickIndabs,2021LihmWFPT}, phonon-mediated superconductivity~\cite{Margine2013s,Eliashberg1960s,Eliashberg1961s},
and phonon-limited transport~\cite{2016ZhouMobility,Ponce2018,Ponce2020}.
Incorporating self-consistency in the study of these quantities and reexamining \textit{ab initio} calculations would be desirable.
The effect of the self-consistent linewidth on the ionized impurity linewidths~\cite{2021LuImpurity,2023LeveilleeImpurity} could also be studied.
The effect of piezoelectric EPC on low-dimensional materials, for which one of the acoustic phonon branches displays a quadratic dispersion~\cite{Carrete2016Phonon2d,Lin2022Phonon}, is another interesting venue for future research.
Our work also forms the foundation for going beyond the simple perturbative treatment of EPC in first-principles studies.
A straightforward generalization to full spectral self-consistency~\cite{2022Panday} could capture the reduction in the quasiparticle weight and the appearance of satellite peaks~\cite{2017Verdi,2018Nery,2022Kandolf}.

In conclusion, we show that the calculation of the electron lifetime for piezoelectric materials in the zero-broadening limit breaks down and needs to be replaced with a self-consistent method.
We implement a self-consistent equation for the electron linewidth using the quasiparticle approximation and apply it to c-BN, Si, PbTe, and NaCl.
We find that self-consistency plays a central role in regularizing the piezoelectric EPC for a wide range of experimentally relevant doping and strongly affects the doping dependence of the LO-phonon scattering.
Therefore, our theory should be broadly applied to various experiments on electronic properties.
Finally, our predictions on the energy-dependent electron linewidths, which are qualitatively different from conventional calculations, await immediate experimental confirmation.

\begin{acknowledgments}
J.-M.L. and C.-H.P. were supported by the Creative-Pioneering Research Program through Seoul National University, Korean NRF No. 2023R1A2C1007297, and the Institute for Basic Science (No. IBSR009-D1).
S.P. acknowledges support from the Fonds de la Recherche Scientifique de Belgique (FRS-FNRS) and by the Walloon Region in the strategic axe FRFS-WEL-T. 
Computational resources have been provided by KISTI (KSC-2022-CRE-0407), the PRACE award granting access to MareNostrum4 at Barcelona Supercomputing Center (BSC), Spain and Discoverer in SofiaTech, Bulgaria (OptoSpin Project ID 2020225411), the Consortium des \'Equipements de Calcul Intensif (C\'ECI), funded by the FRS-FNRS under Grant No. 2.5020.11, and by the Walloon Region, as well as computational resources awarded on the Belgian share of the EuroHPC LUMI supercomputer.
\end{acknowledgments}

\appendix

\FloatBarrier  

\makeatletter\@input{xy.tex}\makeatother
\bibliography{Bibliography}

\begin{thebibliography}{83}%
\makeatletter
\providecommand \@ifxundefined [1]{%
 \@ifx{#1\undefined}
}%
\providecommand \@ifnum [1]{%
 \ifnum #1\expandafter \@firstoftwo
 \else \expandafter \@secondoftwo
 \fi
}%
\providecommand \@ifx [1]{%
 \ifx #1\expandafter \@firstoftwo
 \else \expandafter \@secondoftwo
 \fi
}%
\providecommand \natexlab [1]{#1}%
\providecommand \enquote  [1]{``#1''}%
\providecommand \bibnamefont  [1]{#1}%
\providecommand \bibfnamefont [1]{#1}%
\providecommand \citenamefont [1]{#1}%
\providecommand \href@noop [0]{\@secondoftwo}%
\providecommand \href [0]{\begingroup \@sanitize@url \@href}%
\providecommand \@href[1]{\@@startlink{#1}\@@href}%
\providecommand \@@href[1]{\endgroup#1\@@endlink}%
\providecommand \@sanitize@url [0]{\catcode `\\12\catcode `\$12\catcode
  `\&12\catcode `\#12\catcode `\^12\catcode `\_12\catcode `\%12\relax}%
\providecommand \@@startlink[1]{}%
\providecommand \@@endlink[0]{}%
\providecommand \url  [0]{\begingroup\@sanitize@url \@url }%
\providecommand \@url [1]{\endgroup\@href {#1}{\urlprefix }}%
\providecommand \urlprefix  [0]{URL }%
\providecommand \Eprint [0]{\href }%
\providecommand \doibase [0]{https://doi.org/}%
\providecommand \selectlanguage [0]{\@gobble}%
\providecommand \bibinfo  [0]{\@secondoftwo}%
\providecommand \bibfield  [0]{\@secondoftwo}%
\providecommand \translation [1]{[#1]}%
\providecommand \BibitemOpen [0]{}%
\providecommand \bibitemStop [0]{}%
\providecommand \bibitemNoStop [0]{.\EOS\space}%
\providecommand \EOS [0]{\spacefactor3000\relax}%
\providecommand \BibitemShut  [1]{\csname bibitem#1\endcsname}%
\let\auto@bib@innerbib\@empty
\bibitem [{\citenamefont {Halasyamani}\ and\ \citenamefont
  {Poeppelmeier}(1998)}]{Halasyamani1998}%
  \BibitemOpen
  \bibfield  {author} {\bibinfo {author} {\bibfnamefont {P.~S.}\ \bibnamefont
  {Halasyamani}}\ and\ \bibinfo {author} {\bibfnamefont {K.~R.}\ \bibnamefont
  {Poeppelmeier}},\ }\bibfield  {title} {\bibinfo {title} {Noncentrosymmetric
  oxides},\ }\href {https://doi.org/10.1021/cm980140w} {\bibfield  {journal}
  {\bibinfo  {journal} {Chem. Mater.}\ }\textbf {\bibinfo {volume} {10}},\
  \bibinfo {pages} {2753} (\bibinfo {year} {1998})}\BibitemShut {NoStop}%
\bibitem [{\citenamefont {Martin}(1972)}]{1972MartinPiezo}%
  \BibitemOpen
  \bibfield  {author} {\bibinfo {author} {\bibfnamefont {R.~M.}\ \bibnamefont
  {Martin}},\ }\bibfield  {title} {\bibinfo {title} {Piezoelectricity},\ }\href
  {https://doi.org/10.1103/PhysRevB.5.1607} {\bibfield  {journal} {\bibinfo
  {journal} {Phys. Rev. B}\ }\textbf {\bibinfo {volume} {5}},\ \bibinfo {pages}
  {1607} (\bibinfo {year} {1972})}\BibitemShut {NoStop}%
\bibitem [{\citenamefont {Vanderbilt}(2000)}]{Vanderbilt2000}%
  \BibitemOpen
  \bibfield  {author} {\bibinfo {author} {\bibfnamefont {D.}~\bibnamefont
  {Vanderbilt}},\ }\bibfield  {title} {\bibinfo {title} {Berry-phase theory of
  proper piezoelectric response},\ }\href
  {https://doi.org/https://doi.org/10.1016/S0022-3697(99)00273-5} {\bibfield
  {journal} {\bibinfo  {journal} {J. Phys. Chem. Solids}\ }\textbf {\bibinfo
  {volume} {61}},\ \bibinfo {pages} {147} (\bibinfo {year} {2000})}\BibitemShut
  {NoStop}%
\bibitem [{\citenamefont {Hamann}\ \emph {et~al.}(2005)\citenamefont {Hamann},
  \citenamefont {Wu}, \citenamefont {Rabe},\ and\ \citenamefont
  {Vanderbilt}}]{Hamann2005}%
  \BibitemOpen
  \bibfield  {author} {\bibinfo {author} {\bibfnamefont {D.~R.}\ \bibnamefont
  {Hamann}}, \bibinfo {author} {\bibfnamefont {X.}~\bibnamefont {Wu}}, \bibinfo
  {author} {\bibfnamefont {K.~M.}\ \bibnamefont {Rabe}},\ and\ \bibinfo
  {author} {\bibfnamefont {D.}~\bibnamefont {Vanderbilt}},\ }\bibfield  {title}
  {\bibinfo {title} {Metric tensor formulation of strain in density-functional
  perturbation theory},\ }\href {https://doi.org/10.1103/physrevb.71.035117}
  {\bibfield  {journal} {\bibinfo  {journal} {Phys. Rev. B}\ }\textbf {\bibinfo
  {volume} {71}},\ \bibinfo {pages} {035117} (\bibinfo {year}
  {2005})}\BibitemShut {NoStop}%
\bibitem [{\citenamefont {Stengel}(2013)}]{Stengel2013}%
  \BibitemOpen
  \bibfield  {author} {\bibinfo {author} {\bibfnamefont {M.}~\bibnamefont
  {Stengel}},\ }\bibfield  {title} {\bibinfo {title} {Flexoelectricity from
  density-functional perturbation theory},\ }\href
  {https://doi.org/10.1103/PhysRevB.88.174106} {\bibfield  {journal} {\bibinfo
  {journal} {Phys. Rev. B}\ }\textbf {\bibinfo {volume} {88}},\ \bibinfo
  {pages} {174106} (\bibinfo {year} {2013})}\BibitemShut {NoStop}%
\bibitem [{\citenamefont {Royo}\ and\ \citenamefont
  {Stengel}(2019)}]{2019RoyoPiezo}%
  \BibitemOpen
  \bibfield  {author} {\bibinfo {author} {\bibfnamefont {M.}~\bibnamefont
  {Royo}}\ and\ \bibinfo {author} {\bibfnamefont {M.}~\bibnamefont {Stengel}},\
  }\bibfield  {title} {\bibinfo {title} {First-principles theory of spatial
  dispersion: Dynamical quadrupoles and flexoelectricity},\ }\href
  {https://doi.org/10.1103/PhysRevX.9.021050} {\bibfield  {journal} {\bibinfo
  {journal} {Phys. Rev. X}\ }\textbf {\bibinfo {volume} {9}},\ \bibinfo {pages}
  {021050} (\bibinfo {year} {2019})}\BibitemShut {NoStop}%
\bibitem [{\citenamefont {Liu}\ \emph {et~al.}(2017)\citenamefont {Liu},
  \citenamefont {Zhou}, \citenamefont {Liao}, \citenamefont {Singh},\ and\
  \citenamefont {Chen}}]{2017LiuPiezo}%
  \BibitemOpen
  \bibfield  {author} {\bibinfo {author} {\bibfnamefont {T.-H.}\ \bibnamefont
  {Liu}}, \bibinfo {author} {\bibfnamefont {J.}~\bibnamefont {Zhou}}, \bibinfo
  {author} {\bibfnamefont {B.}~\bibnamefont {Liao}}, \bibinfo {author}
  {\bibfnamefont {D.~J.}\ \bibnamefont {Singh}},\ and\ \bibinfo {author}
  {\bibfnamefont {G.}~\bibnamefont {Chen}},\ }\bibfield  {title} {\bibinfo
  {title} {First-principles mode-by-mode analysis for electron-phonon
  scattering channels and mean free path spectra in {{GaAs}}},\ }\href
  {https://doi.org/10.1103/PhysRevB.95.075206} {\bibfield  {journal} {\bibinfo
  {journal} {Phys. Rev. B}\ }\textbf {\bibinfo {volume} {95}},\ \bibinfo
  {pages} {075206} (\bibinfo {year} {2017})}\BibitemShut {NoStop}%
\bibitem [{\citenamefont {Brunin}\ \emph
  {et~al.}(2020{\natexlab{a}})\citenamefont {Brunin}, \citenamefont {Miranda},
  \citenamefont {Giantomassi}, \citenamefont {Royo}, \citenamefont {Stengel},
  \citenamefont {Verstraete}, \citenamefont {Gonze}, \citenamefont
  {Rignanese},\ and\ \citenamefont {Hautier}}]{2020BruninPiezo1}%
  \BibitemOpen
  \bibfield  {author} {\bibinfo {author} {\bibfnamefont {G.}~\bibnamefont
  {Brunin}}, \bibinfo {author} {\bibfnamefont {H.~P.~C.}\ \bibnamefont
  {Miranda}}, \bibinfo {author} {\bibfnamefont {M.}~\bibnamefont
  {Giantomassi}}, \bibinfo {author} {\bibfnamefont {M.}~\bibnamefont {Royo}},
  \bibinfo {author} {\bibfnamefont {M.}~\bibnamefont {Stengel}}, \bibinfo
  {author} {\bibfnamefont {M.~J.}\ \bibnamefont {Verstraete}}, \bibinfo
  {author} {\bibfnamefont {X.}~\bibnamefont {Gonze}}, \bibinfo {author}
  {\bibfnamefont {G.-M.}\ \bibnamefont {Rignanese}},\ and\ \bibinfo {author}
  {\bibfnamefont {G.}~\bibnamefont {Hautier}},\ }\bibfield  {title} {\bibinfo
  {title} {Electron-phonon beyond {Fr\"ohlich}: Dynamical quadrupoles in polar
  and covalent solids},\ }\href
  {https://doi.org/10.1103/PhysRevLett.125.136601} {\bibfield  {journal}
  {\bibinfo  {journal} {Phys. Rev. Lett.}\ }\textbf {\bibinfo {volume} {125}},\
  \bibinfo {pages} {136601} (\bibinfo {year} {2020}{\natexlab{a}})}\BibitemShut
  {NoStop}%
\bibitem [{\citenamefont {Brunin}\ \emph
  {et~al.}(2020{\natexlab{b}})\citenamefont {Brunin}, \citenamefont {Miranda},
  \citenamefont {Giantomassi}, \citenamefont {Royo}, \citenamefont {Stengel},
  \citenamefont {Verstraete}, \citenamefont {Gonze}, \citenamefont
  {Rignanese},\ and\ \citenamefont {Hautier}}]{2020BruninPiezo2}%
  \BibitemOpen
  \bibfield  {author} {\bibinfo {author} {\bibfnamefont {G.}~\bibnamefont
  {Brunin}}, \bibinfo {author} {\bibfnamefont {H.~P.~C.}\ \bibnamefont
  {Miranda}}, \bibinfo {author} {\bibfnamefont {M.}~\bibnamefont
  {Giantomassi}}, \bibinfo {author} {\bibfnamefont {M.}~\bibnamefont {Royo}},
  \bibinfo {author} {\bibfnamefont {M.}~\bibnamefont {Stengel}}, \bibinfo
  {author} {\bibfnamefont {M.~J.}\ \bibnamefont {Verstraete}}, \bibinfo
  {author} {\bibfnamefont {X.}~\bibnamefont {Gonze}}, \bibinfo {author}
  {\bibfnamefont {G.-M.}\ \bibnamefont {Rignanese}},\ and\ \bibinfo {author}
  {\bibfnamefont {G.}~\bibnamefont {Hautier}},\ }\bibfield  {title} {\bibinfo
  {title} {Phonon-limited electron mobility in {{Si}}, {{GaAs}}, and {{GaP}}
  with exact treatment of dynamical quadrupoles},\ }\href
  {https://doi.org/10.1103/PhysRevB.102.094308} {\bibfield  {journal} {\bibinfo
   {journal} {Phys. Rev. B}\ }\textbf {\bibinfo {volume} {102}},\ \bibinfo
  {pages} {094308} (\bibinfo {year} {2020}{\natexlab{b}})}\BibitemShut
  {NoStop}%
\bibitem [{\citenamefont {Jhalani}\ \emph {et~al.}(2020)\citenamefont
  {Jhalani}, \citenamefont {Zhou}, \citenamefont {Park}, \citenamefont
  {Dreyer},\ and\ \citenamefont {Bernardi}}]{2020JhalaniPiezo}%
  \BibitemOpen
  \bibfield  {author} {\bibinfo {author} {\bibfnamefont {V.~A.}\ \bibnamefont
  {Jhalani}}, \bibinfo {author} {\bibfnamefont {J.-J.}\ \bibnamefont {Zhou}},
  \bibinfo {author} {\bibfnamefont {J.}~\bibnamefont {Park}}, \bibinfo {author}
  {\bibfnamefont {C.~E.}\ \bibnamefont {Dreyer}},\ and\ \bibinfo {author}
  {\bibfnamefont {M.}~\bibnamefont {Bernardi}},\ }\bibfield  {title} {\bibinfo
  {title} {Piezoelectric electron-phonon interaction from \emph{ab initio}
  dynamical quadrupoles: Impact on charge transport in wurtzite {{GaN}}},\
  }\href {https://doi.org/10.1103/PhysRevLett.125.136602} {\bibfield  {journal}
  {\bibinfo  {journal} {Phys. Rev. Lett.}\ }\textbf {\bibinfo {volume} {125}},\
  \bibinfo {pages} {136602} (\bibinfo {year} {2020})}\BibitemShut {NoStop}%
\bibitem [{\citenamefont {Park}\ \emph {et~al.}(2020)\citenamefont {Park},
  \citenamefont {Zhou}, \citenamefont {Jhalani}, \citenamefont {Dreyer},\ and\
  \citenamefont {Bernardi}}]{2020ParkPiezo}%
  \BibitemOpen
  \bibfield  {author} {\bibinfo {author} {\bibfnamefont {J.}~\bibnamefont
  {Park}}, \bibinfo {author} {\bibfnamefont {J.-J.}\ \bibnamefont {Zhou}},
  \bibinfo {author} {\bibfnamefont {V.~A.}\ \bibnamefont {Jhalani}}, \bibinfo
  {author} {\bibfnamefont {C.~E.}\ \bibnamefont {Dreyer}},\ and\ \bibinfo
  {author} {\bibfnamefont {M.}~\bibnamefont {Bernardi}},\ }\bibfield  {title}
  {\bibinfo {title} {Long-range quadrupole electron-phonon interaction from
  first principles},\ }\href {https://doi.org/10.1103/PhysRevB.102.125203}
  {\bibfield  {journal} {\bibinfo  {journal} {Phys. Rev. B}\ }\textbf {\bibinfo
  {volume} {102}},\ \bibinfo {pages} {125203} (\bibinfo {year}
  {2020})}\BibitemShut {NoStop}%
\bibitem [{\citenamefont {Ponc\'e}\ \emph {et~al.}(2021)\citenamefont
  {Ponc\'e}, \citenamefont {Macheda}, \citenamefont {Margine}, \citenamefont
  {Marzari}, \citenamefont {Bonini},\ and\ \citenamefont
  {Giustino}}]{Ponce2021}%
  \BibitemOpen
  \bibfield  {author} {\bibinfo {author} {\bibfnamefont {S.}~\bibnamefont
  {Ponc\'e}}, \bibinfo {author} {\bibfnamefont {F.}~\bibnamefont {Macheda}},
  \bibinfo {author} {\bibfnamefont {E.~R.}\ \bibnamefont {Margine}}, \bibinfo
  {author} {\bibfnamefont {N.}~\bibnamefont {Marzari}}, \bibinfo {author}
  {\bibfnamefont {N.}~\bibnamefont {Bonini}},\ and\ \bibinfo {author}
  {\bibfnamefont {F.}~\bibnamefont {Giustino}},\ }\bibfield  {title} {\bibinfo
  {title} {First-principles predictions of {{Hall}} and drift mobilities in
  semiconductors},\ }\href {https://doi.org/10.1103/PhysRevResearch.3.043022}
  {\bibfield  {journal} {\bibinfo  {journal} {Phys. Rev. Research}\ }\textbf
  {\bibinfo {volume} {3}},\ \bibinfo {pages} {043022} (\bibinfo {year}
  {2021})}\BibitemShut {NoStop}%
\bibitem [{\citenamefont {Ganose}\ \emph {et~al.}(2021)\citenamefont {Ganose},
  \citenamefont {Park}, \citenamefont {Faghaninia}, \citenamefont
  {{Woods-Robinson}}, \citenamefont {Persson},\ and\ \citenamefont
  {Jain}}]{2021GanoseAMSET}%
  \BibitemOpen
  \bibfield  {author} {\bibinfo {author} {\bibfnamefont {A.~M.}\ \bibnamefont
  {Ganose}}, \bibinfo {author} {\bibfnamefont {J.}~\bibnamefont {Park}},
  \bibinfo {author} {\bibfnamefont {A.}~\bibnamefont {Faghaninia}}, \bibinfo
  {author} {\bibfnamefont {R.}~\bibnamefont {{Woods-Robinson}}}, \bibinfo
  {author} {\bibfnamefont {K.~A.}\ \bibnamefont {Persson}},\ and\ \bibinfo
  {author} {\bibfnamefont {A.}~\bibnamefont {Jain}},\ }\bibfield  {title}
  {\bibinfo {title} {Efficient calculation of carrier scattering rates from
  first principles},\ }\href {https://doi.org/10.1038/s41467-021-22440-5}
  {\bibfield  {journal} {\bibinfo  {journal} {Nat. Commun.}\ }\textbf {\bibinfo
  {volume} {12}},\ \bibinfo {pages} {2222} (\bibinfo {year}
  {2021})}\BibitemShut {NoStop}%
\bibitem [{\citenamefont {Ponc{\'e}}\ \emph
  {et~al.}(2023{\natexlab{a}})\citenamefont {Ponc{\'e}}, \citenamefont {Royo},
  \citenamefont {Gibertini}, \citenamefont {Marzari},\ and\ \citenamefont
  {Stengel}}]{2023PoncePiezo1}%
  \BibitemOpen
  \bibfield  {author} {\bibinfo {author} {\bibfnamefont {S.}~\bibnamefont
  {Ponc{\'e}}}, \bibinfo {author} {\bibfnamefont {M.}~\bibnamefont {Royo}},
  \bibinfo {author} {\bibfnamefont {M.}~\bibnamefont {Gibertini}}, \bibinfo
  {author} {\bibfnamefont {N.}~\bibnamefont {Marzari}},\ and\ \bibinfo {author}
  {\bibfnamefont {M.}~\bibnamefont {Stengel}},\ }\bibfield  {title} {\bibinfo
  {title} {Accurate prediction of {{Hall}} mobilities in two-dimensional
  materials through gauge-covariant quadrupolar contributions},\ }\href
  {https://doi.org/10.1103/PhysRevLett.130.166301} {\bibfield  {journal}
  {\bibinfo  {journal} {Phys. Rev. Lett.}\ }\textbf {\bibinfo {volume} {130}},\
  \bibinfo {pages} {166301} (\bibinfo {year} {2023}{\natexlab{a}})}\BibitemShut
  {NoStop}%
\bibitem [{\citenamefont {Ponc{\'e}}\ \emph
  {et~al.}(2023{\natexlab{b}})\citenamefont {Ponc{\'e}}, \citenamefont {Royo},
  \citenamefont {Stengel}, \citenamefont {Marzari},\ and\ \citenamefont
  {Gibertini}}]{2023PoncePiezo2}%
  \BibitemOpen
  \bibfield  {author} {\bibinfo {author} {\bibfnamefont {S.}~\bibnamefont
  {Ponc{\'e}}}, \bibinfo {author} {\bibfnamefont {M.}~\bibnamefont {Royo}},
  \bibinfo {author} {\bibfnamefont {M.}~\bibnamefont {Stengel}}, \bibinfo
  {author} {\bibfnamefont {N.}~\bibnamefont {Marzari}},\ and\ \bibinfo {author}
  {\bibfnamefont {M.}~\bibnamefont {Gibertini}},\ }\bibfield  {title} {\bibinfo
  {title} {Long-range electrostatic contribution to electron-phonon couplings
  and mobilities of two-dimensional and bulk materials},\ }\href
  {https://doi.org/10.1103/PhysRevB.107.155424} {\bibfield  {journal} {\bibinfo
   {journal} {Phys. Rev. B}\ }\textbf {\bibinfo {volume} {107}},\ \bibinfo
  {pages} {155424} (\bibinfo {year} {2023}{\natexlab{b}})}\BibitemShut
  {NoStop}%
\bibitem [{\citenamefont {Marzari}\ and\ \citenamefont
  {Vanderbilt}(1997)}]{1997Marzari}%
  \BibitemOpen
  \bibfield  {author} {\bibinfo {author} {\bibfnamefont {N.}~\bibnamefont
  {Marzari}}\ and\ \bibinfo {author} {\bibfnamefont {D.}~\bibnamefont
  {Vanderbilt}},\ }\bibfield  {title} {\bibinfo {title} {Maximally-localized
  generalized {{Wannier}} functions for composite energy bands},\ }\href
  {https://doi.org/10.1103/PhysRevB.56.12847} {\bibfield  {journal} {\bibinfo
  {journal} {Phys. Rev. B}\ }\textbf {\bibinfo {volume} {56}},\ \bibinfo
  {pages} {12847} (\bibinfo {year} {1997})}\BibitemShut {NoStop}%
\bibitem [{\citenamefont {Souza}\ \emph {et~al.}(2001)\citenamefont {Souza},
  \citenamefont {Marzari},\ and\ \citenamefont {Vanderbilt}}]{2001Souza}%
  \BibitemOpen
  \bibfield  {author} {\bibinfo {author} {\bibfnamefont {I.}~\bibnamefont
  {Souza}}, \bibinfo {author} {\bibfnamefont {N.}~\bibnamefont {Marzari}},\
  and\ \bibinfo {author} {\bibfnamefont {D.}~\bibnamefont {Vanderbilt}},\
  }\bibfield  {title} {\bibinfo {title} {Maximally localized {{Wannier}}
  functions for entangled energy bands},\ }\href
  {https://doi.org/10.1103/PhysRevB.65.035109} {\bibfield  {journal} {\bibinfo
  {journal} {Phys. Rev. B}\ }\textbf {\bibinfo {volume} {65}},\ \bibinfo
  {pages} {035109} (\bibinfo {year} {2001})}\BibitemShut {NoStop}%
\bibitem [{\citenamefont {Giustino}\ \emph {et~al.}(2007)\citenamefont
  {Giustino}, \citenamefont {Cohen},\ and\ \citenamefont
  {Louie}}]{Giustino2007}%
  \BibitemOpen
  \bibfield  {author} {\bibinfo {author} {\bibfnamefont {F.}~\bibnamefont
  {Giustino}}, \bibinfo {author} {\bibfnamefont {M.~L.}\ \bibnamefont
  {Cohen}},\ and\ \bibinfo {author} {\bibfnamefont {S.~G.}\ \bibnamefont
  {Louie}},\ }\bibfield  {title} {\bibinfo {title} {Electron-phonon interaction
  using {Wannier} functions},\ }\href
  {https://doi.org/10.1103/PhysRevB.76.165108} {\bibfield  {journal} {\bibinfo
  {journal} {Phys. Rev. B}\ }\textbf {\bibinfo {volume} {76}},\ \bibinfo
  {pages} {165108} (\bibinfo {year} {2007})}\BibitemShut {NoStop}%
\bibitem [{\citenamefont {Marzari}\ \emph {et~al.}(2012)\citenamefont
  {Marzari}, \citenamefont {Mostofi}, \citenamefont {Yates}, \citenamefont
  {Souza},\ and\ \citenamefont {Vanderbilt}}]{2012Marzari}%
  \BibitemOpen
  \bibfield  {author} {\bibinfo {author} {\bibfnamefont {N.}~\bibnamefont
  {Marzari}}, \bibinfo {author} {\bibfnamefont {A.~A.}\ \bibnamefont
  {Mostofi}}, \bibinfo {author} {\bibfnamefont {J.~R.}\ \bibnamefont {Yates}},
  \bibinfo {author} {\bibfnamefont {I.}~\bibnamefont {Souza}},\ and\ \bibinfo
  {author} {\bibfnamefont {D.}~\bibnamefont {Vanderbilt}},\ }\bibfield  {title}
  {\bibinfo {title} {Maximally localized {{Wannier}} functions: {{Theory}} and
  applications},\ }\href {https://doi.org/10.1103/RevModPhys.84.1419}
  {\bibfield  {journal} {\bibinfo  {journal} {Rev. Mod. Phys.}\ }\textbf
  {\bibinfo {volume} {84}},\ \bibinfo {pages} {1419} (\bibinfo {year}
  {2012})}\BibitemShut {NoStop}%
\bibitem [{\citenamefont {Verdi}\ and\ \citenamefont
  {Giustino}(2015)}]{2015VerdiFrohlich}%
  \BibitemOpen
  \bibfield  {author} {\bibinfo {author} {\bibfnamefont {C.}~\bibnamefont
  {Verdi}}\ and\ \bibinfo {author} {\bibfnamefont {F.}~\bibnamefont
  {Giustino}},\ }\bibfield  {title} {\bibinfo {title} {{Fr\"ohlich}
  electron-phonon vertex from first principles},\ }\href
  {https://doi.org/10.1103/PhysRevLett.115.176401} {\bibfield  {journal}
  {\bibinfo  {journal} {Phys. Rev. Lett.}\ }\textbf {\bibinfo {volume} {115}},\
  \bibinfo {pages} {176401} (\bibinfo {year} {2015})}\BibitemShut {NoStop}%
\bibitem [{\citenamefont {Ridley}(1982)}]{Ridley1982}%
  \BibitemOpen
  \bibfield  {author} {\bibinfo {author} {\bibfnamefont {B.~K.}\ \bibnamefont
  {Ridley}},\ }\href@noop {} {\emph {\bibinfo {title} {Quantum processes in
  semiconductors}}}\ (\bibinfo  {publisher} {Oxford University Press},\
  \bibinfo {address} {Oxford},\ \bibinfo {year} {1982})\BibitemShut {NoStop}%
\bibitem [{\citenamefont {Giustino}(2017)}]{2017Giustino}%
  \BibitemOpen
  \bibfield  {author} {\bibinfo {author} {\bibfnamefont {F.}~\bibnamefont
  {Giustino}},\ }\bibfield  {title} {\bibinfo {title} {Electron-phonon
  interactions from first principles},\ }\href
  {https://doi.org/10.1103/RevModPhys.89.015003} {\bibfield  {journal}
  {\bibinfo  {journal} {Rev. Mod. Phys.}\ }\textbf {\bibinfo {volume} {89}},\
  \bibinfo {pages} {015003} (\bibinfo {year} {2017})}\BibitemShut {NoStop}%
\bibitem [{\citenamefont {Ponc{\'{e}}}\ \emph {et~al.}(2020)\citenamefont
  {Ponc{\'{e}}}, \citenamefont {Li}, \citenamefont {Reichardt},\ and\
  \citenamefont {Giustino}}]{Ponce2020}%
  \BibitemOpen
  \bibfield  {author} {\bibinfo {author} {\bibfnamefont {S.}~\bibnamefont
  {Ponc{\'{e}}}}, \bibinfo {author} {\bibfnamefont {W.}~\bibnamefont {Li}},
  \bibinfo {author} {\bibfnamefont {S.}~\bibnamefont {Reichardt}},\ and\
  \bibinfo {author} {\bibfnamefont {F.}~\bibnamefont {Giustino}},\ }\bibfield
  {title} {\bibinfo {title} {First-principles calculations of charge carrier
  mobility and conductivity in bulk semiconductors and two-dimensional
  materials},\ }\href {https://doi.org/10.1088/1361-6633/ab6a43} {\bibfield
  {journal} {\bibinfo  {journal} {Rep. Prog. Phys.}\ }\textbf {\bibinfo
  {volume} {83}},\ \bibinfo {pages} {036501} (\bibinfo {year}
  {2020})}\BibitemShut {NoStop}%
\bibitem [{\citenamefont {Ponc\'e}\ \emph {et~al.}(2018)\citenamefont
  {Ponc\'e}, \citenamefont {Margine},\ and\ \citenamefont
  {Giustino}}]{Ponce2018}%
  \BibitemOpen
  \bibfield  {author} {\bibinfo {author} {\bibfnamefont {S.}~\bibnamefont
  {Ponc\'e}}, \bibinfo {author} {\bibfnamefont {E.~R.}\ \bibnamefont
  {Margine}},\ and\ \bibinfo {author} {\bibfnamefont {F.}~\bibnamefont
  {Giustino}},\ }\bibfield  {title} {\bibinfo {title} {Towards predictive
  many-body calculations of phonon-limited carrier mobilities in
  semiconductors},\ }\href {https://doi.org/10.1103/PhysRevB.97.121201}
  {\bibfield  {journal} {\bibinfo  {journal} {Phys. Rev. B}\ }\textbf {\bibinfo
  {volume} {97}},\ \bibinfo {pages} {121201} (\bibinfo {year}
  {2018})}\BibitemShut {NoStop}%
\bibitem [{\citenamefont {Li}\ \emph {et~al.}(2014)\citenamefont {Li},
  \citenamefont {Carrete}, \citenamefont {Katcho},\ and\ \citenamefont
  {Mingo}}]{Li2014}%
  \BibitemOpen
  \bibfield  {author} {\bibinfo {author} {\bibfnamefont {W.}~\bibnamefont
  {Li}}, \bibinfo {author} {\bibfnamefont {J.}~\bibnamefont {Carrete}},
  \bibinfo {author} {\bibfnamefont {N.~A.}\ \bibnamefont {Katcho}},\ and\
  \bibinfo {author} {\bibfnamefont {N.}~\bibnamefont {Mingo}},\ }\bibfield
  {title} {\bibinfo {title} {{{ShengBTE}}: A solver of the {{Boltzmann}}
  transport equation for phonons},\ }\href
  {https://doi.org/http://dx.doi.org/10.1016/j.cpc.2014.02.015} {\bibfield
  {journal} {\bibinfo  {journal} {Comput. Phys. Commun.}\ }\textbf {\bibinfo
  {volume} {185}},\ \bibinfo {pages} {1747 } (\bibinfo {year}
  {2014})}\BibitemShut {NoStop}%
\bibitem [{\citenamefont {Bl\"ochl}\ \emph {et~al.}(1994)\citenamefont
  {Bl\"ochl}, \citenamefont {Jepsen},\ and\ \citenamefont
  {Andersen}}]{Bloechl1994}%
  \BibitemOpen
  \bibfield  {author} {\bibinfo {author} {\bibfnamefont {P.~E.}\ \bibnamefont
  {Bl\"ochl}}, \bibinfo {author} {\bibfnamefont {O.}~\bibnamefont {Jepsen}},\
  and\ \bibinfo {author} {\bibfnamefont {O.~K.}\ \bibnamefont {Andersen}},\
  }\bibfield  {title} {\bibinfo {title} {Improved tetrahedron method for
  {Brillouin}-zone integrations},\ }\href
  {https://doi.org/10.1103/PhysRevB.49.16223} {\bibfield  {journal} {\bibinfo
  {journal} {Phys. Rev. B}\ }\textbf {\bibinfo {volume} {49}},\ \bibinfo
  {pages} {16223} (\bibinfo {year} {1994})}\BibitemShut {NoStop}%
\bibitem [{\citenamefont {Bostwick}\ \emph {et~al.}(2007)\citenamefont
  {Bostwick}, \citenamefont {Ohta}, \citenamefont {Seyller}, \citenamefont
  {Horn},\ and\ \citenamefont {Rotenberg}}]{Bostwick2007}%
  \BibitemOpen
  \bibfield  {author} {\bibinfo {author} {\bibfnamefont {A.}~\bibnamefont
  {Bostwick}}, \bibinfo {author} {\bibfnamefont {T.}~\bibnamefont {Ohta}},
  \bibinfo {author} {\bibfnamefont {T.}~\bibnamefont {Seyller}}, \bibinfo
  {author} {\bibfnamefont {K.}~\bibnamefont {Horn}},\ and\ \bibinfo {author}
  {\bibfnamefont {E.}~\bibnamefont {Rotenberg}},\ }\bibfield  {title} {\bibinfo
  {title} {Quasiparticle dynamics in graphene},\ }\href
  {https://doi.org/10.1038/nphys477} {\bibfield  {journal} {\bibinfo  {journal}
  {Nat. Phys.}\ }\textbf {\bibinfo {volume} {3}},\ \bibinfo {pages} {36}
  (\bibinfo {year} {2007})}\BibitemShut {NoStop}%
\bibitem [{\citenamefont {Park}\ \emph {et~al.}(2007)\citenamefont {Park},
  \citenamefont {Giustino}, \citenamefont {Cohen},\ and\ \citenamefont
  {Louie}}]{Park2007}%
  \BibitemOpen
  \bibfield  {author} {\bibinfo {author} {\bibfnamefont {C.-H.}\ \bibnamefont
  {Park}}, \bibinfo {author} {\bibfnamefont {F.}~\bibnamefont {Giustino}},
  \bibinfo {author} {\bibfnamefont {M.~L.}\ \bibnamefont {Cohen}},\ and\
  \bibinfo {author} {\bibfnamefont {S.~G.}\ \bibnamefont {Louie}},\ }\bibfield
  {title} {\bibinfo {title} {Velocity renormalization and carrier lifetime in
  graphene from the electron-phonon interaction},\ }\href
  {https://doi.org/10.1103/PhysRevLett.99.086804} {\bibfield  {journal}
  {\bibinfo  {journal} {Phys. Rev. Lett.}\ }\textbf {\bibinfo {volume} {99}},\
  \bibinfo {pages} {086804} (\bibinfo {year} {2007})}\BibitemShut {NoStop}%
\bibitem [{\citenamefont {Park}\ \emph {et~al.}(2009)\citenamefont {Park},
  \citenamefont {Giustino}, \citenamefont {Spataru}, \citenamefont {Cohen},\
  and\ \citenamefont {Louie}}]{Park2009}%
  \BibitemOpen
  \bibfield  {author} {\bibinfo {author} {\bibfnamefont {C.-H.}\ \bibnamefont
  {Park}}, \bibinfo {author} {\bibfnamefont {F.}~\bibnamefont {Giustino}},
  \bibinfo {author} {\bibfnamefont {C.~D.}\ \bibnamefont {Spataru}}, \bibinfo
  {author} {\bibfnamefont {M.~L.}\ \bibnamefont {Cohen}},\ and\ \bibinfo
  {author} {\bibfnamefont {S.~G.}\ \bibnamefont {Louie}},\ }\bibfield  {title}
  {\bibinfo {title} {First-principles study of electron linewidths in
  graphene},\ }\href {https://doi.org/10.1103/PhysRevLett.102.076803}
  {\bibfield  {journal} {\bibinfo  {journal} {Phys. Rev. Lett.}\ }\textbf
  {\bibinfo {volume} {102}},\ \bibinfo {pages} {076803} (\bibinfo {year}
  {2009})}\BibitemShut {NoStop}%
\bibitem [{\citenamefont {Mahan}(2000)}]{MahanBook}%
  \BibitemOpen
  \bibfield  {author} {\bibinfo {author} {\bibfnamefont {G.~D.}\ \bibnamefont
  {Mahan}},\ }\href@noop {} {\emph {\bibinfo {title} {Many-particle
  physics}}},\ \bibinfo {edition} {3rd}\ ed.\ (\bibinfo  {publisher} {Kluwer
  Academic / Plenum Publishers},\ \bibinfo {address} {New York},\ \bibinfo
  {year} {2000})\BibitemShut {NoStop}%
\bibitem [{\citenamefont {Grimvall}(1981)}]{Grimvall1981}%
  \BibitemOpen
  \bibfield  {author} {\bibinfo {author} {\bibfnamefont {G.}~\bibnamefont
  {Grimvall}},\ }\href@noop {} {\emph {\bibinfo {title} {The electron-phonon
  interaction in metals}}}\ (\bibinfo  {publisher} {North-Holland},\ \bibinfo
  {address} {Amsterdam},\ \bibinfo {year} {1981})\BibitemShut {NoStop}%
\bibitem [{\citenamefont {Poncé}\ \emph {et~al.}(2016)\citenamefont {Poncé},
  \citenamefont {Margine}, \citenamefont {Verdi},\ and\ \citenamefont
  {Giustino}}]{Ponce2016}%
  \BibitemOpen
  \bibfield  {author} {\bibinfo {author} {\bibfnamefont {S.}~\bibnamefont
  {Poncé}}, \bibinfo {author} {\bibfnamefont {E.}~\bibnamefont {Margine}},
  \bibinfo {author} {\bibfnamefont {C.}~\bibnamefont {Verdi}},\ and\ \bibinfo
  {author} {\bibfnamefont {F.}~\bibnamefont {Giustino}},\ }\bibfield  {title}
  {\bibinfo {title} {{EPW}: Electron–phonon coupling, transport and
  superconducting properties using maximally localized {Wannier} functions},\
  }\href {https://doi.org/http://dx.doi.org/10.1016/j.cpc.2016.07.028}
  {\bibfield  {journal} {\bibinfo  {journal} {Comput. Phys. Commun.}\ }\textbf
  {\bibinfo {volume} {209}},\ \bibinfo {pages} {116 } (\bibinfo {year}
  {2016})}\BibitemShut {NoStop}%
\bibitem [{Sup()}]{Supp}%
  \BibitemOpen
  \href@noop {} {}\bibinfo {note} {See the Supplemental Material at [URL will
  be inserted by publisher] for the derivation of the self-consistent theory
  and computational details.}\BibitemShut {Stop}%
\bibitem [{\citenamefont {Arlt}\ and\ \citenamefont
  {Quadflieg}(1968)}]{1968ArltPiezo}%
  \BibitemOpen
  \bibfield  {author} {\bibinfo {author} {\bibfnamefont {G.}~\bibnamefont
  {Arlt}}\ and\ \bibinfo {author} {\bibfnamefont {P.}~\bibnamefont
  {Quadflieg}},\ }\bibfield  {title} {\bibinfo {title} {Piezoelectricity in
  {III}-{V} compounds with a phenomenological analysis of the piezoelectric
  effect},\ }\href {https://doi.org/10.1002/pssb.19680250131} {\bibfield
  {journal} {\bibinfo  {journal} {Phys. Status Solidi (b)}\ }\textbf {\bibinfo
  {volume} {25}},\ \bibinfo {pages} {323} (\bibinfo {year} {1968})}\BibitemShut
  {NoStop}%
\bibitem [{\citenamefont {Stefanucci}\ and\ \citenamefont
  {Van~Leeuwen}(2013)}]{StefanucciBook}%
  \BibitemOpen
  \bibfield  {author} {\bibinfo {author} {\bibfnamefont {G.}~\bibnamefont
  {Stefanucci}}\ and\ \bibinfo {author} {\bibfnamefont {R.}~\bibnamefont
  {Van~Leeuwen}},\ }\href@noop {} {\emph {\bibinfo {title} {Nonequilibrium
  many-body theory of quantum systems: a modern introduction}}}\ (\bibinfo
  {publisher} {Cambridge University Press},\ \bibinfo {address} {Cambridge},\
  \bibinfo {year} {2013})\BibitemShut {NoStop}%
\bibitem [{\citenamefont {Abramovitch}\ \emph {et~al.}(2023)\citenamefont
  {Abramovitch}, \citenamefont {Zhou}, \citenamefont {Mravlje}, \citenamefont
  {Georges},\ and\ \citenamefont {Bernardi}}]{Abramovitch2023}%
  \BibitemOpen
  \bibfield  {author} {\bibinfo {author} {\bibfnamefont {D.~J.}\ \bibnamefont
  {Abramovitch}}, \bibinfo {author} {\bibfnamefont {J.-J.}\ \bibnamefont
  {Zhou}}, \bibinfo {author} {\bibfnamefont {J.}~\bibnamefont {Mravlje}},
  \bibinfo {author} {\bibfnamefont {A.}~\bibnamefont {Georges}},\ and\ \bibinfo
  {author} {\bibfnamefont {M.}~\bibnamefont {Bernardi}},\ }\bibfield  {title}
  {\bibinfo {title} {Combining electron-phonon and dynamical mean-field theory
  calculations of correlated materials: Transport in the correlated metal
  {${\mathrm{Sr}}_{2}{\mathrm{RuO}}_{4}$}},\ }\href
  {https://doi.org/10.1103/PhysRevMaterials.7.093801} {\bibfield  {journal}
  {\bibinfo  {journal} {Phys. Rev. Mater.}\ }\textbf {\bibinfo {volume} {7}},\
  \bibinfo {pages} {093801} (\bibinfo {year} {2023})}\BibitemShut {NoStop}%
\bibitem [{\citenamefont {Royo}\ \emph {et~al.}(2020)\citenamefont {Royo},
  \citenamefont {Hahn},\ and\ \citenamefont {Stengel}}]{Royo2020}%
  \BibitemOpen
  \bibfield  {author} {\bibinfo {author} {\bibfnamefont {M.}~\bibnamefont
  {Royo}}, \bibinfo {author} {\bibfnamefont {K.~R.}\ \bibnamefont {Hahn}},\
  and\ \bibinfo {author} {\bibfnamefont {M.}~\bibnamefont {Stengel}},\
  }\bibfield  {title} {\bibinfo {title} {Using high multipolar orders to
  reconstruct the sound velocity in piezoelectrics from lattice dynamics},\
  }\href {https://doi.org/10.1103/PhysRevLett.125.217602} {\bibfield  {journal}
  {\bibinfo  {journal} {Phys. Rev. Lett.}\ }\textbf {\bibinfo {volume} {125}},\
  \bibinfo {pages} {217602} (\bibinfo {year} {2020})}\BibitemShut {NoStop}%
\bibitem [{\citenamefont {Genz}\ and\ \citenamefont
  {Malik}(1980)}]{1980GenzAdaptive}%
  \BibitemOpen
  \bibfield  {author} {\bibinfo {author} {\bibfnamefont {A.~C.}\ \bibnamefont
  {Genz}}\ and\ \bibinfo {author} {\bibfnamefont {A.~A.}\ \bibnamefont
  {Malik}},\ }\bibfield  {title} {\bibinfo {title} {Remarks on algorithm 006:
  An adaptive algorithm for numerical integration over an n-dimensional
  rectangular region},\ }\href {https://doi.org/10.1016/0771-050X(80)90039-X}
  {\bibfield  {journal} {\bibinfo  {journal} {J. Comput. Appl. Math.}\ }\textbf
  {\bibinfo {volume} {6}},\ \bibinfo {pages} {295} (\bibinfo {year}
  {1980})}\BibitemShut {NoStop}%
\bibitem [{Jul(2022)}]{JuliaHCubature}%
  \BibitemOpen
  \href@noop {} {\bibinfo {title} {{HCubature.jl} (v1.5.1)}},\ \bibinfo
  {howpublished} {\url{https://github.com/JuliaMath/HCubature.jl}} (\bibinfo
  {year} {2022})\BibitemShut {NoStop}%
\bibitem [{Jul(2023)}]{JuliaQuadGK}%
  \BibitemOpen
  \href@noop {} {\bibinfo {title} {{QuadGK.jl} (v2.7.0)}},\ \bibinfo
  {howpublished} {\url{https://github.com/JuliaMath/QuadGK.jl}} (\bibinfo
  {year} {2023})\BibitemShut {NoStop}%
\bibitem [{\citenamefont {Dal~Corso}\ \emph {et~al.}(1994)\citenamefont
  {Dal~Corso}, \citenamefont {Posternak}, \citenamefont {Resta},\ and\
  \citenamefont {Baldereschi}}]{1994DalCorsoPiezo}%
  \BibitemOpen
  \bibfield  {author} {\bibinfo {author} {\bibfnamefont {A.}~\bibnamefont
  {Dal~Corso}}, \bibinfo {author} {\bibfnamefont {M.}~\bibnamefont
  {Posternak}}, \bibinfo {author} {\bibfnamefont {R.}~\bibnamefont {Resta}},\
  and\ \bibinfo {author} {\bibfnamefont {A.}~\bibnamefont {Baldereschi}},\
  }\bibfield  {title} {\bibinfo {title} {Ab initio study of piezoelectricity
  and spontaneous polarization in {{ZnO}}},\ }\href
  {https://doi.org/10.1103/PhysRevB.50.10715} {\bibfield  {journal} {\bibinfo
  {journal} {Phys. Rev. B}\ }\textbf {\bibinfo {volume} {50}},\ \bibinfo
  {pages} {10715} (\bibinfo {year} {1994})}\BibitemShut {NoStop}%
\bibitem [{\citenamefont {Bernardini}\ \emph {et~al.}(1997)\citenamefont
  {Bernardini}, \citenamefont {Fiorentini},\ and\ \citenamefont
  {Vanderbilt}}]{1997BernardiniPiezo}%
  \BibitemOpen
  \bibfield  {author} {\bibinfo {author} {\bibfnamefont {F.}~\bibnamefont
  {Bernardini}}, \bibinfo {author} {\bibfnamefont {V.}~\bibnamefont
  {Fiorentini}},\ and\ \bibinfo {author} {\bibfnamefont {D.}~\bibnamefont
  {Vanderbilt}},\ }\bibfield  {title} {\bibinfo {title} {Spontaneous
  polarization and piezoelectric constants of {III}-{V} nitrides},\ }\href
  {https://doi.org/10.1103/PhysRevB.56.R10024} {\bibfield  {journal} {\bibinfo
  {journal} {Phys. Rev. B}\ }\textbf {\bibinfo {volume} {56}},\ \bibinfo
  {pages} {R10024} (\bibinfo {year} {1997})}\BibitemShut {NoStop}%
\bibitem [{\citenamefont {S\'aghi-Szab\'o}\ \emph {et~al.}(1998)\citenamefont
  {S\'aghi-Szab\'o}, \citenamefont {Cohen},\ and\ \citenamefont
  {Krakauer}}]{1998SaghiSzaboPiezo}%
  \BibitemOpen
  \bibfield  {author} {\bibinfo {author} {\bibfnamefont {G.}~\bibnamefont
  {S\'aghi-Szab\'o}}, \bibinfo {author} {\bibfnamefont {R.~E.}\ \bibnamefont
  {Cohen}},\ and\ \bibinfo {author} {\bibfnamefont {H.}~\bibnamefont
  {Krakauer}},\ }\bibfield  {title} {\bibinfo {title} {First-principles study
  of piezoelectricity in {{PbTiO$_3$}}},\ }\href
  {https://doi.org/10.1103/PhysRevLett.80.4321} {\bibfield  {journal} {\bibinfo
   {journal} {Phys. Rev. Lett.}\ }\textbf {\bibinfo {volume} {80}},\ \bibinfo
  {pages} {4321} (\bibinfo {year} {1998})}\BibitemShut {NoStop}%
\bibitem [{\citenamefont {Giannozzi}\ \emph {et~al.}(2017)\citenamefont
  {Giannozzi} \emph {et~al.}}]{Giannozzi2017}%
  \BibitemOpen
  \bibfield  {author} {\bibinfo {author} {\bibfnamefont {P.}~\bibnamefont
  {Giannozzi}} \emph {et~al.},\ }\bibfield  {title} {\bibinfo {title} {Advanced
  capabilities for materials modelling with {Quantum} {ESPRESSO}},\ }\href
  {https://doi.org/10.1088/1361-648x/aa8f79} {\bibfield  {journal} {\bibinfo
  {journal} {J. Condens. Matter Phys.}\ }\textbf {\bibinfo {volume} {29}},\
  \bibinfo {pages} {465901} (\bibinfo {year} {2017})}\BibitemShut {NoStop}%
\bibitem [{\citenamefont {Gonze}\ \emph {et~al.}(2016)\citenamefont {Gonze}
  \emph {et~al.}}]{Gonze2016}%
  \BibitemOpen
  \bibfield  {author} {\bibinfo {author} {\bibfnamefont {X.}~\bibnamefont
  {Gonze}} \emph {et~al.},\ }\bibfield  {title} {\bibinfo {title} {Recent
  developments in the {ABINIT} software package},\ }\href
  {https://doi.org/http://dx.doi.org/10.1016/j.cpc.2016.04.003} {\bibfield
  {journal} {\bibinfo  {journal} {Comput. Phys. Commun.}\ }\textbf {\bibinfo
  {volume} {205}},\ \bibinfo {pages} {106 } (\bibinfo {year}
  {2016})}\BibitemShut {NoStop}%
\bibitem [{\citenamefont {Gonze}\ \emph {et~al.}(2020)\citenamefont {Gonze}
  \emph {et~al.}}]{Gonze2020}%
  \BibitemOpen
  \bibfield  {author} {\bibinfo {author} {\bibfnamefont {X.}~\bibnamefont
  {Gonze}} \emph {et~al.},\ }\bibfield  {title} {\bibinfo {title} {The
  {{Abinit}} project: Impact, environment and recent developments},\ }\href
  {https://doi.org/https://doi.org/10.1016/j.cpc.2019.107042} {\bibfield
  {journal} {\bibinfo  {journal} {Comput. Phys. Commun.}\ }\textbf {\bibinfo
  {volume} {248}},\ \bibinfo {pages} {107042} (\bibinfo {year}
  {2020})}\BibitemShut {NoStop}%
\bibitem [{\citenamefont {Hamann}(2013)}]{2013HamannONCVPSP}%
  \BibitemOpen
  \bibfield  {author} {\bibinfo {author} {\bibfnamefont {D.~R.}\ \bibnamefont
  {Hamann}},\ }\bibfield  {title} {\bibinfo {title} {Optimized norm-conserving
  {Vanderbilt} pseudopotentials},\ }\href
  {https://doi.org/10.1103/PhysRevB.88.085117} {\bibfield  {journal} {\bibinfo
  {journal} {Phys. Rev. B}\ }\textbf {\bibinfo {volume} {88}},\ \bibinfo
  {pages} {085117} (\bibinfo {year} {2013})}\BibitemShut {NoStop}%
\bibitem [{\citenamefont {van Setten}\ \emph {et~al.}(2018)\citenamefont {van
  Setten}, \citenamefont {Giantomassi}, \citenamefont {Bousquet}, \citenamefont
  {Verstraete}, \citenamefont {Hamann}, \citenamefont {Gonze},\ and\
  \citenamefont {Rignanese}}]{Setten2018}%
  \BibitemOpen
  \bibfield  {author} {\bibinfo {author} {\bibfnamefont {M.}~\bibnamefont {van
  Setten}}, \bibinfo {author} {\bibfnamefont {M.}~\bibnamefont {Giantomassi}},
  \bibinfo {author} {\bibfnamefont {E.}~\bibnamefont {Bousquet}}, \bibinfo
  {author} {\bibfnamefont {M.}~\bibnamefont {Verstraete}}, \bibinfo {author}
  {\bibfnamefont {D.}~\bibnamefont {Hamann}}, \bibinfo {author} {\bibfnamefont
  {X.}~\bibnamefont {Gonze}},\ and\ \bibinfo {author} {\bibfnamefont {G.-M.}\
  \bibnamefont {Rignanese}},\ }\bibfield  {title} {\bibinfo {title} {The
  {PseudoDojo}: Training and grading a 85 element optimized norm-conserving
  pseudopotential table},\ }\href
  {https://doi.org/https://doi.org/10.1016/j.cpc.2018.01.012} {\bibfield
  {journal} {\bibinfo  {journal} {Comput. Phys. Commun.}\ }\textbf {\bibinfo
  {volume} {226}},\ \bibinfo {pages} {39 } (\bibinfo {year}
  {2018})}\BibitemShut {NoStop}%
\bibitem [{\citenamefont {Perdew}\ \emph {et~al.}(1996)\citenamefont {Perdew},
  \citenamefont {Burke},\ and\ \citenamefont {Ernzerhof}}]{Perdew1996}%
  \BibitemOpen
  \bibfield  {author} {\bibinfo {author} {\bibfnamefont {J.~P.}\ \bibnamefont
  {Perdew}}, \bibinfo {author} {\bibfnamefont {K.}~\bibnamefont {Burke}},\ and\
  \bibinfo {author} {\bibfnamefont {M.}~\bibnamefont {Ernzerhof}},\ }\bibfield
  {title} {\bibinfo {title} {Generalized gradient approximation made simple},\
  }\href {https://doi.org/10.1103/PhysRevLett.77.3865} {\bibfield  {journal}
  {\bibinfo  {journal} {Phys. Rev. Lett.}\ }\textbf {\bibinfo {volume} {77}},\
  \bibinfo {pages} {3865} (\bibinfo {year} {1996})}\BibitemShut {NoStop}%
\bibitem [{\citenamefont {Pizzi}\ \emph {et~al.}(2020)\citenamefont {Pizzi}
  \emph {et~al.}}]{2020PizziWannier90}%
  \BibitemOpen
  \bibfield  {author} {\bibinfo {author} {\bibfnamefont {G.}~\bibnamefont
  {Pizzi}} \emph {et~al.},\ }\bibfield  {title} {\bibinfo {title} {Wannier90 as
  a community code: New features and applications},\ }\href
  {https://doi.org/10.1088/1361-648X/ab51ff} {\bibfield  {journal} {\bibinfo
  {journal} {J. Condens. Matter Phys.}\ }\textbf {\bibinfo {volume} {32}},\
  \bibinfo {pages} {165902} (\bibinfo {year} {2020})}\BibitemShut {NoStop}%
\bibitem [{\citenamefont {Bezanson}\ \emph {et~al.}(2017)\citenamefont
  {Bezanson}, \citenamefont {Edelman}, \citenamefont {Karpinski},\ and\
  \citenamefont {Shah}}]{2017BezansonJulia}%
  \BibitemOpen
  \bibfield  {author} {\bibinfo {author} {\bibfnamefont {J.}~\bibnamefont
  {Bezanson}}, \bibinfo {author} {\bibfnamefont {A.}~\bibnamefont {Edelman}},
  \bibinfo {author} {\bibfnamefont {S.}~\bibnamefont {Karpinski}},\ and\
  \bibinfo {author} {\bibfnamefont {V.~B.}\ \bibnamefont {Shah}},\ }\bibfield
  {title} {\bibinfo {title} {Julia: A fresh approach to numerical computing},\
  }\href {https://doi.org/10.1137/141000671} {\bibfield  {journal} {\bibinfo
  {journal} {SIAM Review}\ }\textbf {\bibinfo {volume} {59}},\ \bibinfo {pages}
  {65} (\bibinfo {year} {2017})}\BibitemShut {NoStop}%
\bibitem [{\citenamefont {Mermin}(1970)}]{1970Mermin}%
  \BibitemOpen
  \bibfield  {author} {\bibinfo {author} {\bibfnamefont {N.~D.}\ \bibnamefont
  {Mermin}},\ }\bibfield  {title} {\bibinfo {title} {Lindhard dielectric
  function in the relaxation-time approximation},\ }\href
  {https://doi.org/10.1103/PhysRevB.1.2362} {\bibfield  {journal} {\bibinfo
  {journal} {Phys. Rev. B}\ }\textbf {\bibinfo {volume} {1}},\ \bibinfo {pages}
  {2362} (\bibinfo {year} {1970})}\BibitemShut {NoStop}%
\bibitem [{\citenamefont {Born}\ and\ \citenamefont
  {Huang}(1996)}]{BornHuang1954}%
  \BibitemOpen
  \bibfield  {author} {\bibinfo {author} {\bibfnamefont {M.}~\bibnamefont
  {Born}}\ and\ \bibinfo {author} {\bibfnamefont {K.}~\bibnamefont {Huang}},\
  }\href@noop {} {\emph {\bibinfo {title} {Dynamical theory of crystal
  lattices}}}\ (\bibinfo  {publisher} {Oxford university press},\ \bibinfo
  {address} {Oxford},\ \bibinfo {year} {1996})\BibitemShut {NoStop}%
\bibitem [{\citenamefont {Zhou}\ \emph {et~al.}(2021)\citenamefont {Zhou},
  \citenamefont {Park}, \citenamefont {Lu}, \citenamefont {Maliyov},
  \citenamefont {Tong},\ and\ \citenamefont {Bernardi}}]{2021ZhouPerturbo}%
  \BibitemOpen
  \bibfield  {author} {\bibinfo {author} {\bibfnamefont {J.-J.}\ \bibnamefont
  {Zhou}}, \bibinfo {author} {\bibfnamefont {J.}~\bibnamefont {Park}}, \bibinfo
  {author} {\bibfnamefont {I.-T.}\ \bibnamefont {Lu}}, \bibinfo {author}
  {\bibfnamefont {I.}~\bibnamefont {Maliyov}}, \bibinfo {author} {\bibfnamefont
  {X.}~\bibnamefont {Tong}},\ and\ \bibinfo {author} {\bibfnamefont
  {M.}~\bibnamefont {Bernardi}},\ }\bibfield  {title} {\bibinfo {title}
  {Perturbo: A software package for ab initio electron–phonon interactions,
  charge transport and ultrafast dynamics},\ }\href
  {https://doi.org/https://doi.org/10.1016/j.cpc.2021.107970} {\bibfield
  {journal} {\bibinfo  {journal} {Comput. Phys. Commun.}\ }\textbf {\bibinfo
  {volume} {264}},\ \bibinfo {pages} {107970} (\bibinfo {year}
  {2021})}\BibitemShut {NoStop}%
\bibitem [{\citenamefont {Resta}(1992)}]{1992RestaPolarization}%
  \BibitemOpen
  \bibfield  {author} {\bibinfo {author} {\bibfnamefont {R.}~\bibnamefont
  {Resta}},\ }\bibfield  {title} {\bibinfo {title} {Theory of the electric
  polarization in crystals},\ }\href
  {https://doi.org/10.1080/00150199208016065} {\bibfield  {journal} {\bibinfo
  {journal} {Ferroelectrics}\ }\textbf {\bibinfo {volume} {136}},\ \bibinfo
  {pages} {51} (\bibinfo {year} {1992})}\BibitemShut {NoStop}%
\bibitem [{\citenamefont {{King-Smith}}\ and\ \citenamefont
  {Vanderbilt}(1993)}]{1993KingSmithPolarization}%
  \BibitemOpen
  \bibfield  {author} {\bibinfo {author} {\bibfnamefont {R.~D.}\ \bibnamefont
  {{King-Smith}}}\ and\ \bibinfo {author} {\bibfnamefont {D.}~\bibnamefont
  {Vanderbilt}},\ }\bibfield  {title} {\bibinfo {title} {Theory of polarization
  of crystalline solids},\ }\href {https://doi.org/10.1103/PhysRevB.47.1651}
  {\bibfield  {journal} {\bibinfo  {journal} {Phys. Rev. B}\ }\textbf {\bibinfo
  {volume} {47}},\ \bibinfo {pages} {1651} (\bibinfo {year}
  {1993})}\BibitemShut {NoStop}%
\bibitem [{\citenamefont {Vanderbilt}\ and\ \citenamefont
  {{King-Smith}}(1993)}]{1993VanderbiltPolarization}%
  \BibitemOpen
  \bibfield  {author} {\bibinfo {author} {\bibfnamefont {D.}~\bibnamefont
  {Vanderbilt}}\ and\ \bibinfo {author} {\bibfnamefont {R.~D.}\ \bibnamefont
  {{King-Smith}}},\ }\bibfield  {title} {\bibinfo {title} {Electric
  polarization as a bulk quantity and its relation to surface charge},\ }\href
  {https://doi.org/10.1103/PhysRevB.48.4442} {\bibfield  {journal} {\bibinfo
  {journal} {Phys. Rev. B}\ }\textbf {\bibinfo {volume} {48}},\ \bibinfo
  {pages} {4442} (\bibinfo {year} {1993})}\BibitemShut {NoStop}%
\bibitem [{\citenamefont {Resta}(1994)}]{1994RestaPolarization}%
  \BibitemOpen
  \bibfield  {author} {\bibinfo {author} {\bibfnamefont {R.}~\bibnamefont
  {Resta}},\ }\bibfield  {title} {\bibinfo {title} {Macroscopic polarization in
  crystalline dielectrics: The geometric phase approach},\ }\href
  {https://doi.org/10.1103/RevModPhys.66.899} {\bibfield  {journal} {\bibinfo
  {journal} {Rev. Mod. Phys.}\ }\textbf {\bibinfo {volume} {66}},\ \bibinfo
  {pages} {899} (\bibinfo {year} {1994})}\BibitemShut {NoStop}%
\bibitem [{\citenamefont {Grimsditch}\ \emph {et~al.}(1994)\citenamefont
  {Grimsditch}, \citenamefont {Zouboulis},\ and\ \citenamefont
  {Polian}}]{1994GrimsditchElastic}%
  \BibitemOpen
  \bibfield  {author} {\bibinfo {author} {\bibfnamefont {M.}~\bibnamefont
  {Grimsditch}}, \bibinfo {author} {\bibfnamefont {E.~S.}\ \bibnamefont
  {Zouboulis}},\ and\ \bibinfo {author} {\bibfnamefont {A.}~\bibnamefont
  {Polian}},\ }\bibfield  {title} {\bibinfo {title} {{Elastic constants of
  boron nitride}},\ }\href {https://doi.org/10.1063/1.357757} {\bibfield
  {journal} {\bibinfo  {journal} {Journal of Applied Physics}\ }\textbf
  {\bibinfo {volume} {76}},\ \bibinfo {pages} {832} (\bibinfo {year}
  {1994})}\BibitemShut {NoStop}%
\bibitem [{\citenamefont {Gonze}\ and\ \citenamefont {Lee}(1997)}]{Gonze1997a}%
  \BibitemOpen
  \bibfield  {author} {\bibinfo {author} {\bibfnamefont {X.}~\bibnamefont
  {Gonze}}\ and\ \bibinfo {author} {\bibfnamefont {C.}~\bibnamefont {Lee}},\
  }\bibfield  {title} {\bibinfo {title} {Dynamical matrices, born effective
  charges, dielectric permittivity tensors, and interatomic force constants
  from density-functional perturbation theory},\ }\href
  {https://doi.org/10.1103/PhysRevB.55.10355} {\bibfield  {journal} {\bibinfo
  {journal} {Phys. Rev. B}\ }\textbf {\bibinfo {volume} {55}},\ \bibinfo
  {pages} {10355} (\bibinfo {year} {1997})}\BibitemShut {NoStop}%
\bibitem [{\citenamefont {Wu}\ \emph {et~al.}(2005)\citenamefont {Wu},
  \citenamefont {Vanderbilt},\ and\ \citenamefont {Hamann}}]{Wu2005}%
  \BibitemOpen
  \bibfield  {author} {\bibinfo {author} {\bibfnamefont {X.}~\bibnamefont
  {Wu}}, \bibinfo {author} {\bibfnamefont {D.}~\bibnamefont {Vanderbilt}},\
  and\ \bibinfo {author} {\bibfnamefont {D.~R.}\ \bibnamefont {Hamann}},\
  }\bibfield  {title} {\bibinfo {title} {Systematic treatment of displacements,
  strains, and electric fields in density-functional perturbation theory},\
  }\href {https://doi.org/10.1103/PhysRevB.72.035105} {\bibfield  {journal}
  {\bibinfo  {journal} {Phys. Rev. B}\ }\textbf {\bibinfo {volume} {72}},\
  \bibinfo {pages} {035105} (\bibinfo {year} {2005})}\BibitemShut {NoStop}%
\bibitem [{\citenamefont {Lee}\ \emph {et~al.}(2023)\citenamefont {Lee},
  \citenamefont {Ponc{\'e}}, \citenamefont {Bushick}, \citenamefont
  {Hajinazar}, \citenamefont {{Lafuente-Bartolome}}, \citenamefont {Leveillee},
  \citenamefont {Lian}, \citenamefont {Lihm}, \citenamefont {Macheda},
  \citenamefont {Mori}, \citenamefont {Paudyal}, \citenamefont {Sio},
  \citenamefont {Tiwari}, \citenamefont {Zacharias}, \citenamefont {Zhang},
  \citenamefont {Bonini}, \citenamefont {Kioupakis}, \citenamefont {Margine},\
  and\ \citenamefont {Giustino}}]{Lee2023EPW}%
  \BibitemOpen
  \bibfield  {author} {\bibinfo {author} {\bibfnamefont {H.}~\bibnamefont
  {Lee}}, \bibinfo {author} {\bibfnamefont {S.}~\bibnamefont {Ponc{\'e}}},
  \bibinfo {author} {\bibfnamefont {K.}~\bibnamefont {Bushick}}, \bibinfo
  {author} {\bibfnamefont {S.}~\bibnamefont {Hajinazar}}, \bibinfo {author}
  {\bibfnamefont {J.}~\bibnamefont {{Lafuente-Bartolome}}}, \bibinfo {author}
  {\bibfnamefont {J.}~\bibnamefont {Leveillee}}, \bibinfo {author}
  {\bibfnamefont {C.}~\bibnamefont {Lian}}, \bibinfo {author} {\bibfnamefont
  {J.-M.}\ \bibnamefont {Lihm}}, \bibinfo {author} {\bibfnamefont
  {F.}~\bibnamefont {Macheda}}, \bibinfo {author} {\bibfnamefont
  {H.}~\bibnamefont {Mori}}, \bibinfo {author} {\bibfnamefont {H.}~\bibnamefont
  {Paudyal}}, \bibinfo {author} {\bibfnamefont {W.~H.}\ \bibnamefont {Sio}},
  \bibinfo {author} {\bibfnamefont {S.}~\bibnamefont {Tiwari}}, \bibinfo
  {author} {\bibfnamefont {M.}~\bibnamefont {Zacharias}}, \bibinfo {author}
  {\bibfnamefont {X.}~\bibnamefont {Zhang}}, \bibinfo {author} {\bibfnamefont
  {N.}~\bibnamefont {Bonini}}, \bibinfo {author} {\bibfnamefont
  {E.}~\bibnamefont {Kioupakis}}, \bibinfo {author} {\bibfnamefont {E.~R.}\
  \bibnamefont {Margine}},\ and\ \bibinfo {author} {\bibfnamefont
  {F.}~\bibnamefont {Giustino}},\ }\bibfield  {title} {\bibinfo {title}
  {Electron{\textendash}phonon physics from first principles using the {{EPW}}
  code},\ }\href {https://doi.org/10.1038/s41524-023-01107-3} {\bibfield
  {journal} {\bibinfo  {journal} {npj Comput. Mater.}\ }\textbf {\bibinfo
  {volume} {9}},\ \bibinfo {pages} {156} (\bibinfo {year} {2023})}\BibitemShut
  {NoStop}%
\bibitem [{\citenamefont {Xu}\ \emph {et~al.}(2020)\citenamefont {Xu},
  \citenamefont {Habib}, \citenamefont {Kumar}, \citenamefont {Wu},
  \citenamefont {Sundararaman},\ and\ \citenamefont
  {Ping}}]{Xu2020SelfConsistent}%
  \BibitemOpen
  \bibfield  {author} {\bibinfo {author} {\bibfnamefont {J.}~\bibnamefont
  {Xu}}, \bibinfo {author} {\bibfnamefont {A.}~\bibnamefont {Habib}}, \bibinfo
  {author} {\bibfnamefont {S.}~\bibnamefont {Kumar}}, \bibinfo {author}
  {\bibfnamefont {F.}~\bibnamefont {Wu}}, \bibinfo {author} {\bibfnamefont
  {R.}~\bibnamefont {Sundararaman}},\ and\ \bibinfo {author} {\bibfnamefont
  {Y.}~\bibnamefont {Ping}},\ }\bibfield  {title} {\bibinfo {title}
  {Spin-phonon relaxation from a universal ab initio density-matrix approach},\
  }\href {https://doi.org/10.1038/s41467-020-16063-5} {\bibfield  {journal}
  {\bibinfo  {journal} {Nat. Commun.}\ }\textbf {\bibinfo {volume} {11}},\
  \bibinfo {pages} {2780} (\bibinfo {year} {2020})}\BibitemShut {NoStop}%
\bibitem [{\citenamefont {Verdi}\ \emph {et~al.}(2017)\citenamefont {Verdi},
  \citenamefont {Caruso},\ and\ \citenamefont {Giustino}}]{2017Verdi}%
  \BibitemOpen
  \bibfield  {author} {\bibinfo {author} {\bibfnamefont {C.}~\bibnamefont
  {Verdi}}, \bibinfo {author} {\bibfnamefont {F.}~\bibnamefont {Caruso}},\ and\
  \bibinfo {author} {\bibfnamefont {F.}~\bibnamefont {Giustino}},\ }\bibfield
  {title} {\bibinfo {title} {Origin of the crossover from polarons to {{Fermi}}
  liquids in transition metal oxides},\ }\href
  {https://doi.org/10.1038/ncomms15769} {\bibfield  {journal} {\bibinfo
  {journal} {Nat. Commun.}\ }\textbf {\bibinfo {volume} {8}},\ \bibinfo {pages}
  {15769} (\bibinfo {year} {2017})}\BibitemShut {NoStop}%
\bibitem [{\citenamefont {Kandolf}\ \emph {et~al.}(2022)\citenamefont
  {Kandolf}, \citenamefont {Verdi},\ and\ \citenamefont
  {Giustino}}]{2022Kandolf}%
  \BibitemOpen
  \bibfield  {author} {\bibinfo {author} {\bibfnamefont {N.}~\bibnamefont
  {Kandolf}}, \bibinfo {author} {\bibfnamefont {C.}~\bibnamefont {Verdi}},\
  and\ \bibinfo {author} {\bibfnamefont {F.}~\bibnamefont {Giustino}},\
  }\bibfield  {title} {\bibinfo {title} {Many-body {Green's} function
  approaches to the doped {Fr\"ohlich} solid: Exact solutions and anomalous
  mass enhancement},\ }\href {https://doi.org/10.1103/PhysRevB.105.085148}
  {\bibfield  {journal} {\bibinfo  {journal} {Phys. Rev. B}\ }\textbf {\bibinfo
  {volume} {105}},\ \bibinfo {pages} {085148} (\bibinfo {year}
  {2022})}\BibitemShut {NoStop}%
\bibitem [{\citenamefont {Macheda}\ \emph {et~al.}(2022)\citenamefont
  {Macheda}, \citenamefont {Barone},\ and\ \citenamefont
  {Mauri}}]{2022Macheda3D}%
  \BibitemOpen
  \bibfield  {author} {\bibinfo {author} {\bibfnamefont {F.}~\bibnamefont
  {Macheda}}, \bibinfo {author} {\bibfnamefont {P.}~\bibnamefont {Barone}},\
  and\ \bibinfo {author} {\bibfnamefont {F.}~\bibnamefont {Mauri}},\ }\bibfield
   {title} {\bibinfo {title} {Electron-phonon interaction and
  longitudinal-transverse phonon splitting in doped semiconductors},\ }\href
  {https://doi.org/10.1103/PhysRevLett.129.185902} {\bibfield  {journal}
  {\bibinfo  {journal} {Phys. Rev. Lett.}\ }\textbf {\bibinfo {volume} {129}},\
  \bibinfo {pages} {185902} (\bibinfo {year} {2022})}\BibitemShut {NoStop}%
\bibitem [{\citenamefont {Macheda}\ \emph {et~al.}(2023)\citenamefont
  {Macheda}, \citenamefont {Sohier}, \citenamefont {Barone},\ and\
  \citenamefont {Mauri}}]{2022Macheda2D}%
  \BibitemOpen
  \bibfield  {author} {\bibinfo {author} {\bibfnamefont {F.}~\bibnamefont
  {Macheda}}, \bibinfo {author} {\bibfnamefont {T.}~\bibnamefont {Sohier}},
  \bibinfo {author} {\bibfnamefont {P.}~\bibnamefont {Barone}},\ and\ \bibinfo
  {author} {\bibfnamefont {F.}~\bibnamefont {Mauri}},\ }\bibfield  {title}
  {\bibinfo {title} {Electron-phonon interaction and phonon frequencies in
  two-dimensional doped semiconductors},\ }\href
  {https://doi.org/10.1103/PhysRevB.107.094308} {\bibfield  {journal} {\bibinfo
   {journal} {Phys. Rev. B}\ }\textbf {\bibinfo {volume} {107}},\ \bibinfo
  {pages} {094308} (\bibinfo {year} {2023})}\BibitemShut {NoStop}%
\bibitem [{\citenamefont {Lu}\ \emph {et~al.}(2022)\citenamefont {Lu},
  \citenamefont {Zhou}, \citenamefont {Park},\ and\ \citenamefont
  {Bernardi}}]{2021LuImpurity}%
  \BibitemOpen
  \bibfield  {author} {\bibinfo {author} {\bibfnamefont {I.-T.}\ \bibnamefont
  {Lu}}, \bibinfo {author} {\bibfnamefont {J.-J.}\ \bibnamefont {Zhou}},
  \bibinfo {author} {\bibfnamefont {J.}~\bibnamefont {Park}},\ and\ \bibinfo
  {author} {\bibfnamefont {M.}~\bibnamefont {Bernardi}},\ }\bibfield  {title}
  {\bibinfo {title} {First-principles ionized-impurity scattering and charge
  transport in doped materials},\ }\href
  {https://doi.org/10.1103/PhysRevMaterials.6.L010801} {\bibfield  {journal}
  {\bibinfo  {journal} {Phys. Rev. Mater.}\ }\textbf {\bibinfo {volume} {6}},\
  \bibinfo {pages} {L010801} (\bibinfo {year} {2022})}\BibitemShut {NoStop}%
\bibitem [{\citenamefont {Leveillee}\ \emph {et~al.}(2023)\citenamefont
  {Leveillee}, \citenamefont {Zhang}, \citenamefont {Kioupakis},\ and\
  \citenamefont {Giustino}}]{2023LeveilleeImpurity}%
  \BibitemOpen
  \bibfield  {author} {\bibinfo {author} {\bibfnamefont {J.}~\bibnamefont
  {Leveillee}}, \bibinfo {author} {\bibfnamefont {X.}~\bibnamefont {Zhang}},
  \bibinfo {author} {\bibfnamefont {E.}~\bibnamefont {Kioupakis}},\ and\
  \bibinfo {author} {\bibfnamefont {F.}~\bibnamefont {Giustino}},\ }\bibfield
  {title} {\bibinfo {title} {Ab initio calculation of carrier mobility in
  semiconductors including ionized-impurity scattering},\ }\href
  {https://doi.org/10.1103/PhysRevB.107.125207} {\bibfield  {journal} {\bibinfo
   {journal} {Phys. Rev. B}\ }\textbf {\bibinfo {volume} {107}},\ \bibinfo
  {pages} {125207} (\bibinfo {year} {2023})}\BibitemShut {NoStop}%
\bibitem [{\citenamefont {Eiguren}\ \emph {et~al.}(2003)\citenamefont
  {Eiguren}, \citenamefont {de~Gironcoli}, \citenamefont {Chulkov},
  \citenamefont {Echenique},\ and\ \citenamefont {Tosatti}}]{2003EigurenARPES}%
  \BibitemOpen
  \bibfield  {author} {\bibinfo {author} {\bibfnamefont {A.}~\bibnamefont
  {Eiguren}}, \bibinfo {author} {\bibfnamefont {S.}~\bibnamefont
  {de~Gironcoli}}, \bibinfo {author} {\bibfnamefont {E.~V.}\ \bibnamefont
  {Chulkov}}, \bibinfo {author} {\bibfnamefont {P.~M.}\ \bibnamefont
  {Echenique}},\ and\ \bibinfo {author} {\bibfnamefont {E.}~\bibnamefont
  {Tosatti}},\ }\bibfield  {title} {\bibinfo {title} {Electron-phonon
  interaction at the {Be(0001)} surface},\ }\href
  {https://doi.org/10.1103/PhysRevLett.91.166803} {\bibfield  {journal}
  {\bibinfo  {journal} {Phys. Rev. Lett.}\ }\textbf {\bibinfo {volume} {91}},\
  \bibinfo {pages} {166803} (\bibinfo {year} {2003})}\BibitemShut {NoStop}%
\bibitem [{\citenamefont {Giustino}\ \emph {et~al.}(2008)\citenamefont
  {Giustino}, \citenamefont {Cohen},\ and\ \citenamefont
  {Louie}}]{2008GiustinoARPES}%
  \BibitemOpen
  \bibfield  {author} {\bibinfo {author} {\bibfnamefont {F.}~\bibnamefont
  {Giustino}}, \bibinfo {author} {\bibfnamefont {M.~L.}\ \bibnamefont
  {Cohen}},\ and\ \bibinfo {author} {\bibfnamefont {S.~G.}\ \bibnamefont
  {Louie}},\ }\bibfield  {title} {\bibinfo {title} {Small phonon contribution
  to the photoemission kink in the copper oxide superconductors},\ }\href
  {https://doi.org/10.1038/nature06874} {\bibfield  {journal} {\bibinfo
  {journal} {Nature}\ }\textbf {\bibinfo {volume} {452}},\ \bibinfo {pages}
  {975} (\bibinfo {year} {2008})}\BibitemShut {NoStop}%
\bibitem [{\citenamefont {Li}\ \emph {et~al.}(2021)\citenamefont {Li},
  \citenamefont {Wu}, \citenamefont {Chan},\ and\ \citenamefont
  {Louie}}]{2021LiARPES}%
  \BibitemOpen
  \bibfield  {author} {\bibinfo {author} {\bibfnamefont {Z.}~\bibnamefont
  {Li}}, \bibinfo {author} {\bibfnamefont {M.}~\bibnamefont {Wu}}, \bibinfo
  {author} {\bibfnamefont {Y.-H.}\ \bibnamefont {Chan}},\ and\ \bibinfo
  {author} {\bibfnamefont {S.~G.}\ \bibnamefont {Louie}},\ }\bibfield  {title}
  {\bibinfo {title} {Unmasking the origin of kinks in the photoemission spectra
  of cuprate superconductors},\ }\href
  {https://doi.org/10.1103/PhysRevLett.126.146401} {\bibfield  {journal}
  {\bibinfo  {journal} {Phys. Rev. Lett.}\ }\textbf {\bibinfo {volume} {126}},\
  \bibinfo {pages} {146401} (\bibinfo {year} {2021})}\BibitemShut {NoStop}%
\bibitem [{\citenamefont {Noffsinger}\ \emph {et~al.}(2012)\citenamefont
  {Noffsinger}, \citenamefont {Kioupakis}, \citenamefont {Van~de Walle},
  \citenamefont {Louie},\ and\ \citenamefont {Cohen}}]{2012NoffsingerIndabs}%
  \BibitemOpen
  \bibfield  {author} {\bibinfo {author} {\bibfnamefont {J.}~\bibnamefont
  {Noffsinger}}, \bibinfo {author} {\bibfnamefont {E.}~\bibnamefont
  {Kioupakis}}, \bibinfo {author} {\bibfnamefont {C.~G.}\ \bibnamefont {Van~de
  Walle}}, \bibinfo {author} {\bibfnamefont {S.~G.}\ \bibnamefont {Louie}},\
  and\ \bibinfo {author} {\bibfnamefont {M.~L.}\ \bibnamefont {Cohen}},\
  }\bibfield  {title} {\bibinfo {title} {Phonon-assisted optical absorption in
  silicon from first principles},\ }\href
  {https://doi.org/10.1103/PhysRevLett.108.167402} {\bibfield  {journal}
  {\bibinfo  {journal} {Phys. Rev. Lett.}\ }\textbf {\bibinfo {volume} {108}},\
  \bibinfo {pages} {167402} (\bibinfo {year} {2012})}\BibitemShut {NoStop}%
\bibitem [{\citenamefont {Patrick}\ and\ \citenamefont
  {Giustino}(2014)}]{2014PatrickIndabs}%
  \BibitemOpen
  \bibfield  {author} {\bibinfo {author} {\bibfnamefont {C.~E.}\ \bibnamefont
  {Patrick}}\ and\ \bibinfo {author} {\bibfnamefont {F.}~\bibnamefont
  {Giustino}},\ }\bibfield  {title} {\bibinfo {title} {Unified theory of
  electron-phonon renormalization and phonon-assisted optical absorption},\
  }\href {https://doi.org/10.1088/0953-8984/26/36/365503} {\bibfield  {journal}
  {\bibinfo  {journal} {J. Condens. Matter Phys.}\ }\textbf {\bibinfo {volume}
  {26}},\ \bibinfo {pages} {365503} (\bibinfo {year} {2014})}\BibitemShut
  {NoStop}%
\bibitem [{\citenamefont {Lihm}\ and\ \citenamefont
  {Park}(2021)}]{2021LihmWFPT}%
  \BibitemOpen
  \bibfield  {author} {\bibinfo {author} {\bibfnamefont {J.-M.}\ \bibnamefont
  {Lihm}}\ and\ \bibinfo {author} {\bibfnamefont {C.-H.}\ \bibnamefont
  {Park}},\ }\bibfield  {title} {\bibinfo {title} {Wannier function
  perturbation theory: Localized representation and interpolation of wave
  function perturbation},\ }\href {https://doi.org/10.1103/PhysRevX.11.041053}
  {\bibfield  {journal} {\bibinfo  {journal} {Phys. Rev. X}\ }\textbf {\bibinfo
  {volume} {11}},\ \bibinfo {pages} {041053} (\bibinfo {year}
  {2021})}\BibitemShut {NoStop}%
\bibitem [{\citenamefont {Margine}\ and\ \citenamefont
  {Giustino}(2013)}]{Margine2013s}%
  \BibitemOpen
  \bibfield  {author} {\bibinfo {author} {\bibfnamefont {E.~R.}\ \bibnamefont
  {Margine}}\ and\ \bibinfo {author} {\bibfnamefont {F.}~\bibnamefont
  {Giustino}},\ }\bibfield  {title} {\bibinfo {title} {Anisotropic
  {Migdal}--{Eliashberg} theory using {Wannier} functions},\ }\href
  {https://doi.org/10.1103/PhysRevB.87.024505} {\bibfield  {journal} {\bibinfo
  {journal} {Phys. Rev. B}\ }\textbf {\bibinfo {volume} {87}},\ \bibinfo
  {pages} {024505} (\bibinfo {year} {2013})}\BibitemShut {NoStop}%
\bibitem [{\citenamefont {Eliashberg}(1960)}]{Eliashberg1960s}%
  \BibitemOpen
  \bibfield  {author} {\bibinfo {author} {\bibfnamefont {G.}~\bibnamefont
  {Eliashberg}},\ }\bibfield  {title} {\bibinfo {title} {Interactions between
  electrons and lattice vibrations in a superconductor},\ }\href
  {https://www.osti.gov/biblio/7354388} {\bibfield  {journal} {\bibinfo
  {journal} {Sov. Phys. JETP}\ }\textbf {\bibinfo {volume} {11}},\ \bibinfo
  {pages} {696} (\bibinfo {year} {1960})}\BibitemShut {NoStop}%
\bibitem [{\citenamefont {Eliashberg}(1961)}]{Eliashberg1961s}%
  \BibitemOpen
  \bibfield  {author} {\bibinfo {author} {\bibfnamefont {G.}~\bibnamefont
  {Eliashberg}},\ }\bibfield  {title} {\bibinfo {title} {Temperature {Green's}
  function for electrons in a superconductor},\ }\href@noop {} {\bibfield
  {journal} {\bibinfo  {journal} {Sov. Phys. JETP}\ }\textbf {\bibinfo {volume}
  {12}},\ \bibinfo {pages} {1000} (\bibinfo {year} {1961})}\BibitemShut
  {NoStop}%
\bibitem [{\citenamefont {Zhou}\ and\ \citenamefont
  {Bernardi}(2016)}]{2016ZhouMobility}%
  \BibitemOpen
  \bibfield  {author} {\bibinfo {author} {\bibfnamefont {J.-J.}\ \bibnamefont
  {Zhou}}\ and\ \bibinfo {author} {\bibfnamefont {M.}~\bibnamefont
  {Bernardi}},\ }\bibfield  {title} {\bibinfo {title} {Ab initio electron
  mobility and polar phonon scattering in {GaAs}},\ }\href
  {https://doi.org/10.1103/PhysRevB.94.201201} {\bibfield  {journal} {\bibinfo
  {journal} {Phys. Rev. B}\ }\textbf {\bibinfo {volume} {94}},\ \bibinfo
  {pages} {201201} (\bibinfo {year} {2016})}\BibitemShut {NoStop}%
\bibitem [{\citenamefont {Carrete}\ \emph {et~al.}(2016)\citenamefont
  {Carrete}, \citenamefont {Li}, \citenamefont {Lindsay}, \citenamefont
  {Broido}, \citenamefont {Gallego},\ and\ \citenamefont
  {Mingo}}]{Carrete2016Phonon2d}%
  \BibitemOpen
  \bibfield  {author} {\bibinfo {author} {\bibfnamefont {J.}~\bibnamefont
  {Carrete}}, \bibinfo {author} {\bibfnamefont {W.}~\bibnamefont {Li}},
  \bibinfo {author} {\bibfnamefont {L.}~\bibnamefont {Lindsay}}, \bibinfo
  {author} {\bibfnamefont {D.~A.}\ \bibnamefont {Broido}}, \bibinfo {author}
  {\bibfnamefont {L.~J.}\ \bibnamefont {Gallego}},\ and\ \bibinfo {author}
  {\bibfnamefont {N.}~\bibnamefont {Mingo}},\ }\bibfield  {title} {\bibinfo
  {title} {Physically founded phonon dispersions of few-layer materials and the
  case of borophene},\ }\href {https://doi.org/10.1080/21663831.2016.1174163}
  {\bibfield  {journal} {\bibinfo  {journal} {Mater. Res. Lett.}\ }\textbf
  {\bibinfo {volume} {4}},\ \bibinfo {pages} {204} (\bibinfo {year}
  {2016})}\BibitemShut {NoStop}%
\bibitem [{\citenamefont {Lin}\ \emph {et~al.}(2022)\citenamefont {Lin},
  \citenamefont {Ponc{\'e}},\ and\ \citenamefont {Marzari}}]{Lin2022Phonon}%
  \BibitemOpen
  \bibfield  {author} {\bibinfo {author} {\bibfnamefont {C.}~\bibnamefont
  {Lin}}, \bibinfo {author} {\bibfnamefont {S.}~\bibnamefont {Ponc{\'e}}},\
  and\ \bibinfo {author} {\bibfnamefont {N.}~\bibnamefont {Marzari}},\
  }\bibfield  {title} {\bibinfo {title} {General invariance and equilibrium
  conditions for lattice dynamics in {1D}, {2D}, and {3D} materials},\ }\href
  {https://doi.org/10.1038/s41524-022-00920-6} {\bibfield  {journal} {\bibinfo
  {journal} {npj Comput. Mater.}\ }\textbf {\bibinfo {volume} {8}},\ \bibinfo
  {pages} {236} (\bibinfo {year} {2022})}\BibitemShut {NoStop}%
\bibitem [{\citenamefont {Pandey}\ and\ \citenamefont
  {Littlewood}(2022)}]{2022Panday}%
  \BibitemOpen
  \bibfield  {author} {\bibinfo {author} {\bibfnamefont {B.}~\bibnamefont
  {Pandey}}\ and\ \bibinfo {author} {\bibfnamefont {P.~B.}\ \bibnamefont
  {Littlewood}},\ }\bibfield  {title} {\bibinfo {title} {Going beyond the
  cumulant approximation: Power series correction to the single-particle
  {Green's} function in the {{Holstein}} system},\ }\href
  {https://doi.org/10.1103/PhysRevLett.129.136401} {\bibfield  {journal}
  {\bibinfo  {journal} {Phys. Rev. Lett.}\ }\textbf {\bibinfo {volume} {129}},\
  \bibinfo {pages} {136401} (\bibinfo {year} {2022})}\BibitemShut {NoStop}%
\bibitem [{\citenamefont {Nery}\ \emph {et~al.}(2018)\citenamefont {Nery},
  \citenamefont {Allen}, \citenamefont {Antonius}, \citenamefont {Reining},
  \citenamefont {Miglio},\ and\ \citenamefont {Gonze}}]{2018Nery}%
  \BibitemOpen
  \bibfield  {author} {\bibinfo {author} {\bibfnamefont {J.~P.}\ \bibnamefont
  {Nery}}, \bibinfo {author} {\bibfnamefont {P.~B.}\ \bibnamefont {Allen}},
  \bibinfo {author} {\bibfnamefont {G.}~\bibnamefont {Antonius}}, \bibinfo
  {author} {\bibfnamefont {L.}~\bibnamefont {Reining}}, \bibinfo {author}
  {\bibfnamefont {A.}~\bibnamefont {Miglio}},\ and\ \bibinfo {author}
  {\bibfnamefont {X.}~\bibnamefont {Gonze}},\ }\bibfield  {title} {\bibinfo
  {title} {Quasiparticles and phonon satellites in spectral functions of
  semiconductors and insulators: {Cumulants} applied to the full
  first-principles theory and the {Fr\"ohlich} polaron},\ }\href
  {https://doi.org/10.1103/PhysRevB.97.115145} {\bibfield  {journal} {\bibinfo
  {journal} {Phys. Rev. B}\ }\textbf {\bibinfo {volume} {97}},\ \bibinfo
  {pages} {115145} (\bibinfo {year} {2018})}\BibitemShut {NoStop}%
\end{thebibliography}%


\begin{thebibliography}{51}%
\makeatletter
\providecommand \@ifxundefined [1]{%
 \@ifx{#1\undefined}
}%
\providecommand \@ifnum [1]{%
 \ifnum #1\expandafter \@firstoftwo
 \else \expandafter \@secondoftwo
 \fi
}%
\providecommand \@ifx [1]{%
 \ifx #1\expandafter \@firstoftwo
 \else \expandafter \@secondoftwo
 \fi
}%
\providecommand \natexlab [1]{#1}%
\providecommand \enquote  [1]{``#1''}%
\providecommand \bibnamefont  [1]{#1}%
\providecommand \bibfnamefont [1]{#1}%
\providecommand \citenamefont [1]{#1}%
\providecommand \href@noop [0]{\@secondoftwo}%
\providecommand \href [0]{\begingroup \@sanitize@url \@href}%
\providecommand \@href[1]{\@@startlink{#1}\@@href}%
\providecommand \@@href[1]{\endgroup#1\@@endlink}%
\providecommand \@sanitize@url [0]{\catcode `\\12\catcode `\$12\catcode
  `\&12\catcode `\#12\catcode `\^12\catcode `\_12\catcode `\%12\relax}%
\providecommand \@@startlink[1]{}%
\providecommand \@@endlink[0]{}%
\providecommand \url  [0]{\begingroup\@sanitize@url \@url }%
\providecommand \@url [1]{\endgroup\@href {#1}{\urlprefix }}%
\providecommand \urlprefix  [0]{URL }%
\providecommand \Eprint [0]{\href }%
\providecommand \doibase [0]{https://doi.org/}%
\providecommand \selectlanguage [0]{\@gobble}%
\providecommand \bibinfo  [0]{\@secondoftwo}%
\providecommand \bibfield  [0]{\@secondoftwo}%
\providecommand \translation [1]{[#1]}%
\providecommand \BibitemOpen [0]{}%
\providecommand \bibitemStop [0]{}%
\providecommand \bibitemNoStop [0]{.\EOS\space}%
\providecommand \EOS [0]{\spacefactor3000\relax}%
\providecommand \BibitemShut  [1]{\csname bibitem#1\endcsname}%
\let\auto@bib@innerbib\@empty
\bibitem [{\citenamefont {Arlt}\ and\ \citenamefont
  {Quadflieg}(1968)}]{1968ArltPiezo}%
  \BibitemOpen
  \bibfield  {author} {\bibinfo {author} {\bibfnamefont {G.}~\bibnamefont
  {Arlt}}\ and\ \bibinfo {author} {\bibfnamefont {P.}~\bibnamefont
  {Quadflieg}},\ }\bibfield  {title} {\bibinfo {title} {Piezoelectricity in
  {III}-{V} compounds with a phenomenological analysis of the piezoelectric
  effect},\ }\href {https://doi.org/10.1002/pssb.19680250131} {\bibfield
  {journal} {\bibinfo  {journal} {Phys. Status Solidi (b)}\ }\textbf {\bibinfo
  {volume} {25}},\ \bibinfo {pages} {323} (\bibinfo {year} {1968})}\BibitemShut
  {NoStop}%
\bibitem [{\citenamefont {Ridley}(1982)}]{Ridley1982}%
  \BibitemOpen
  \bibfield  {author} {\bibinfo {author} {\bibfnamefont {B.~K.}\ \bibnamefont
  {Ridley}},\ }\href@noop {} {\emph {\bibinfo {title} {Quantum processes in
  semiconductors}}}\ (\bibinfo  {publisher} {Oxford University Press},\
  \bibinfo {address} {Oxford},\ \bibinfo {year} {1982})\BibitemShut {NoStop}%
\bibitem [{\citenamefont {Giustino}(2017)}]{2017Giustino}%
  \BibitemOpen
  \bibfield  {author} {\bibinfo {author} {\bibfnamefont {F.}~\bibnamefont
  {Giustino}},\ }\bibfield  {title} {\bibinfo {title} {Electron-phonon
  interactions from first principles},\ }\href
  {https://doi.org/10.1103/RevModPhys.89.015003} {\bibfield  {journal}
  {\bibinfo  {journal} {Rev. Mod. Phys.}\ }\textbf {\bibinfo {volume} {89}},\
  \bibinfo {pages} {015003} (\bibinfo {year} {2017})}\BibitemShut {NoStop}%
\bibitem [{\citenamefont {Ponc{\'{e}}}\ \emph {et~al.}(2020)\citenamefont
  {Ponc{\'{e}}}, \citenamefont {Li}, \citenamefont {Reichardt},\ and\
  \citenamefont {Giustino}}]{Ponce2020}%
  \BibitemOpen
  \bibfield  {author} {\bibinfo {author} {\bibfnamefont {S.}~\bibnamefont
  {Ponc{\'{e}}}}, \bibinfo {author} {\bibfnamefont {W.}~\bibnamefont {Li}},
  \bibinfo {author} {\bibfnamefont {S.}~\bibnamefont {Reichardt}},\ and\
  \bibinfo {author} {\bibfnamefont {F.}~\bibnamefont {Giustino}},\ }\bibfield
  {title} {\bibinfo {title} {First-principles calculations of charge carrier
  mobility and conductivity in bulk semiconductors and two-dimensional
  materials},\ }\href {https://doi.org/10.1088/1361-6633/ab6a43} {\bibfield
  {journal} {\bibinfo  {journal} {Rep. Prog. Phys.}\ }\textbf {\bibinfo
  {volume} {83}},\ \bibinfo {pages} {036501} (\bibinfo {year}
  {2020})}\BibitemShut {NoStop}%
\bibitem [{\citenamefont {Stefanucci}\ and\ \citenamefont
  {Van~Leeuwen}(2013)}]{StefanucciBook}%
  \BibitemOpen
  \bibfield  {author} {\bibinfo {author} {\bibfnamefont {G.}~\bibnamefont
  {Stefanucci}}\ and\ \bibinfo {author} {\bibfnamefont {R.}~\bibnamefont
  {Van~Leeuwen}},\ }\href@noop {} {\emph {\bibinfo {title} {Nonequilibrium
  many-body theory of quantum systems: a modern introduction}}}\ (\bibinfo
  {publisher} {Cambridge University Press},\ \bibinfo {address} {Cambridge},\
  \bibinfo {year} {2013})\BibitemShut {NoStop}%
\bibitem [{\citenamefont {Abramovitch}\ \emph {et~al.}(2023)\citenamefont
  {Abramovitch}, \citenamefont {Zhou}, \citenamefont {Mravlje}, \citenamefont
  {Georges},\ and\ \citenamefont {Bernardi}}]{Abramovitch2023}%
  \BibitemOpen
  \bibfield  {author} {\bibinfo {author} {\bibfnamefont {D.~J.}\ \bibnamefont
  {Abramovitch}}, \bibinfo {author} {\bibfnamefont {J.-J.}\ \bibnamefont
  {Zhou}}, \bibinfo {author} {\bibfnamefont {J.}~\bibnamefont {Mravlje}},
  \bibinfo {author} {\bibfnamefont {A.}~\bibnamefont {Georges}},\ and\ \bibinfo
  {author} {\bibfnamefont {M.}~\bibnamefont {Bernardi}},\ }\bibfield  {title}
  {\bibinfo {title} {Combining electron-phonon and dynamical mean-field theory
  calculations of correlated materials: Transport in the correlated metal
  {${\mathrm{Sr}}_{2}{\mathrm{RuO}}_{4}$}},\ }\href
  {https://doi.org/10.1103/PhysRevMaterials.7.093801} {\bibfield  {journal}
  {\bibinfo  {journal} {Phys. Rev. Mater.}\ }\textbf {\bibinfo {volume} {7}},\
  \bibinfo {pages} {093801} (\bibinfo {year} {2023})}\BibitemShut {NoStop}%
\bibitem [{\citenamefont {Kandolf}\ \emph {et~al.}(2022)\citenamefont
  {Kandolf}, \citenamefont {Verdi},\ and\ \citenamefont
  {Giustino}}]{2022Kandolf}%
  \BibitemOpen
  \bibfield  {author} {\bibinfo {author} {\bibfnamefont {N.}~\bibnamefont
  {Kandolf}}, \bibinfo {author} {\bibfnamefont {C.}~\bibnamefont {Verdi}},\
  and\ \bibinfo {author} {\bibfnamefont {F.}~\bibnamefont {Giustino}},\
  }\bibfield  {title} {\bibinfo {title} {Many-body {Green's} function
  approaches to the doped {Fr\"ohlich} solid: Exact solutions and anomalous
  mass enhancement},\ }\href {https://doi.org/10.1103/PhysRevB.105.085148}
  {\bibfield  {journal} {\bibinfo  {journal} {Phys. Rev. B}\ }\textbf {\bibinfo
  {volume} {105}},\ \bibinfo {pages} {085148} (\bibinfo {year}
  {2022})}\BibitemShut {NoStop}%
\bibitem [{\citenamefont {Royo}\ \emph {et~al.}(2020)\citenamefont {Royo},
  \citenamefont {Hahn},\ and\ \citenamefont {Stengel}}]{Royo2020}%
  \BibitemOpen
  \bibfield  {author} {\bibinfo {author} {\bibfnamefont {M.}~\bibnamefont
  {Royo}}, \bibinfo {author} {\bibfnamefont {K.~R.}\ \bibnamefont {Hahn}},\
  and\ \bibinfo {author} {\bibfnamefont {M.}~\bibnamefont {Stengel}},\
  }\bibfield  {title} {\bibinfo {title} {Using high multipolar orders to
  reconstruct the sound velocity in piezoelectrics from lattice dynamics},\
  }\href {https://doi.org/10.1103/PhysRevLett.125.217602} {\bibfield  {journal}
  {\bibinfo  {journal} {Phys. Rev. Lett.}\ }\textbf {\bibinfo {volume} {125}},\
  \bibinfo {pages} {217602} (\bibinfo {year} {2020})}\BibitemShut {NoStop}%
\bibitem [{\citenamefont {Genz}\ and\ \citenamefont
  {Malik}(1980)}]{1980GenzAdaptive}%
  \BibitemOpen
  \bibfield  {author} {\bibinfo {author} {\bibfnamefont {A.~C.}\ \bibnamefont
  {Genz}}\ and\ \bibinfo {author} {\bibfnamefont {A.~A.}\ \bibnamefont
  {Malik}},\ }\bibfield  {title} {\bibinfo {title} {Remarks on algorithm 006:
  An adaptive algorithm for numerical integration over an n-dimensional
  rectangular region},\ }\href {https://doi.org/10.1016/0771-050X(80)90039-X}
  {\bibfield  {journal} {\bibinfo  {journal} {J. Comput. Appl. Math.}\ }\textbf
  {\bibinfo {volume} {6}},\ \bibinfo {pages} {295} (\bibinfo {year}
  {1980})}\BibitemShut {NoStop}%
\bibitem [{Jul(2022)}]{JuliaHCubature}%
  \BibitemOpen
  \href@noop {} {\bibinfo {title} {{HCubature.jl} (v1.5.1)}},\ \bibinfo
  {howpublished} {\url{https://github.com/JuliaMath/HCubature.jl}} (\bibinfo
  {year} {2022})\BibitemShut {NoStop}%
\bibitem [{Jul(2023)}]{JuliaQuadGK}%
  \BibitemOpen
  \href@noop {} {\bibinfo {title} {{QuadGK.jl} (v2.7.0)}},\ \bibinfo
  {howpublished} {\url{https://github.com/JuliaMath/QuadGK.jl}} (\bibinfo
  {year} {2023})\BibitemShut {NoStop}%
\bibitem [{\citenamefont {Dal~Corso}\ \emph {et~al.}(1994)\citenamefont
  {Dal~Corso}, \citenamefont {Posternak}, \citenamefont {Resta},\ and\
  \citenamefont {Baldereschi}}]{1994DalCorsoPiezo}%
  \BibitemOpen
  \bibfield  {author} {\bibinfo {author} {\bibfnamefont {A.}~\bibnamefont
  {Dal~Corso}}, \bibinfo {author} {\bibfnamefont {M.}~\bibnamefont
  {Posternak}}, \bibinfo {author} {\bibfnamefont {R.}~\bibnamefont {Resta}},\
  and\ \bibinfo {author} {\bibfnamefont {A.}~\bibnamefont {Baldereschi}},\
  }\bibfield  {title} {\bibinfo {title} {Ab initio study of piezoelectricity
  and spontaneous polarization in {{ZnO}}},\ }\href
  {https://doi.org/10.1103/PhysRevB.50.10715} {\bibfield  {journal} {\bibinfo
  {journal} {Phys. Rev. B}\ }\textbf {\bibinfo {volume} {50}},\ \bibinfo
  {pages} {10715} (\bibinfo {year} {1994})}\BibitemShut {NoStop}%
\bibitem [{\citenamefont {Bernardini}\ \emph {et~al.}(1997)\citenamefont
  {Bernardini}, \citenamefont {Fiorentini},\ and\ \citenamefont
  {Vanderbilt}}]{1997BernardiniPiezo}%
  \BibitemOpen
  \bibfield  {author} {\bibinfo {author} {\bibfnamefont {F.}~\bibnamefont
  {Bernardini}}, \bibinfo {author} {\bibfnamefont {V.}~\bibnamefont
  {Fiorentini}},\ and\ \bibinfo {author} {\bibfnamefont {D.}~\bibnamefont
  {Vanderbilt}},\ }\bibfield  {title} {\bibinfo {title} {Spontaneous
  polarization and piezoelectric constants of {III}-{V} nitrides},\ }\href
  {https://doi.org/10.1103/PhysRevB.56.R10024} {\bibfield  {journal} {\bibinfo
  {journal} {Phys. Rev. B}\ }\textbf {\bibinfo {volume} {56}},\ \bibinfo
  {pages} {R10024} (\bibinfo {year} {1997})}\BibitemShut {NoStop}%
\bibitem [{\citenamefont {S\'aghi-Szab\'o}\ \emph {et~al.}(1998)\citenamefont
  {S\'aghi-Szab\'o}, \citenamefont {Cohen},\ and\ \citenamefont
  {Krakauer}}]{1998SaghiSzaboPiezo}%
  \BibitemOpen
  \bibfield  {author} {\bibinfo {author} {\bibfnamefont {G.}~\bibnamefont
  {S\'aghi-Szab\'o}}, \bibinfo {author} {\bibfnamefont {R.~E.}\ \bibnamefont
  {Cohen}},\ and\ \bibinfo {author} {\bibfnamefont {H.}~\bibnamefont
  {Krakauer}},\ }\bibfield  {title} {\bibinfo {title} {First-principles study
  of piezoelectricity in {{PbTiO$_3$}}},\ }\href
  {https://doi.org/10.1103/PhysRevLett.80.4321} {\bibfield  {journal} {\bibinfo
   {journal} {Phys. Rev. Lett.}\ }\textbf {\bibinfo {volume} {80}},\ \bibinfo
  {pages} {4321} (\bibinfo {year} {1998})}\BibitemShut {NoStop}%
\bibitem [{\citenamefont {Stengel}(2013)}]{Stengel2013}%
  \BibitemOpen
  \bibfield  {author} {\bibinfo {author} {\bibfnamefont {M.}~\bibnamefont
  {Stengel}},\ }\bibfield  {title} {\bibinfo {title} {Flexoelectricity from
  density-functional perturbation theory},\ }\href
  {https://doi.org/10.1103/PhysRevB.88.174106} {\bibfield  {journal} {\bibinfo
  {journal} {Phys. Rev. B}\ }\textbf {\bibinfo {volume} {88}},\ \bibinfo
  {pages} {174106} (\bibinfo {year} {2013})}\BibitemShut {NoStop}%
\bibitem [{\citenamefont {Verdi}\ and\ \citenamefont
  {Giustino}(2015)}]{2015VerdiFrohlich}%
  \BibitemOpen
  \bibfield  {author} {\bibinfo {author} {\bibfnamefont {C.}~\bibnamefont
  {Verdi}}\ and\ \bibinfo {author} {\bibfnamefont {F.}~\bibnamefont
  {Giustino}},\ }\bibfield  {title} {\bibinfo {title} {{Fr\"ohlich}
  electron-phonon vertex from first principles},\ }\href
  {https://doi.org/10.1103/PhysRevLett.115.176401} {\bibfield  {journal}
  {\bibinfo  {journal} {Phys. Rev. Lett.}\ }\textbf {\bibinfo {volume} {115}},\
  \bibinfo {pages} {176401} (\bibinfo {year} {2015})}\BibitemShut {NoStop}%
\bibitem [{\citenamefont {Brunin}\ \emph
  {et~al.}(2020{\natexlab{a}})\citenamefont {Brunin}, \citenamefont {Miranda},
  \citenamefont {Giantomassi}, \citenamefont {Royo}, \citenamefont {Stengel},
  \citenamefont {Verstraete}, \citenamefont {Gonze}, \citenamefont
  {Rignanese},\ and\ \citenamefont {Hautier}}]{2020BruninPiezo1}%
  \BibitemOpen
  \bibfield  {author} {\bibinfo {author} {\bibfnamefont {G.}~\bibnamefont
  {Brunin}}, \bibinfo {author} {\bibfnamefont {H.~P.~C.}\ \bibnamefont
  {Miranda}}, \bibinfo {author} {\bibfnamefont {M.}~\bibnamefont
  {Giantomassi}}, \bibinfo {author} {\bibfnamefont {M.}~\bibnamefont {Royo}},
  \bibinfo {author} {\bibfnamefont {M.}~\bibnamefont {Stengel}}, \bibinfo
  {author} {\bibfnamefont {M.~J.}\ \bibnamefont {Verstraete}}, \bibinfo
  {author} {\bibfnamefont {X.}~\bibnamefont {Gonze}}, \bibinfo {author}
  {\bibfnamefont {G.-M.}\ \bibnamefont {Rignanese}},\ and\ \bibinfo {author}
  {\bibfnamefont {G.}~\bibnamefont {Hautier}},\ }\bibfield  {title} {\bibinfo
  {title} {Electron-phonon beyond {Fr\"ohlich}: Dynamical quadrupoles in polar
  and covalent solids},\ }\href
  {https://doi.org/10.1103/PhysRevLett.125.136601} {\bibfield  {journal}
  {\bibinfo  {journal} {Phys. Rev. Lett.}\ }\textbf {\bibinfo {volume} {125}},\
  \bibinfo {pages} {136601} (\bibinfo {year} {2020}{\natexlab{a}})}\BibitemShut
  {NoStop}%
\bibitem [{\citenamefont {Brunin}\ \emph
  {et~al.}(2020{\natexlab{b}})\citenamefont {Brunin}, \citenamefont {Miranda},
  \citenamefont {Giantomassi}, \citenamefont {Royo}, \citenamefont {Stengel},
  \citenamefont {Verstraete}, \citenamefont {Gonze}, \citenamefont
  {Rignanese},\ and\ \citenamefont {Hautier}}]{2020BruninPiezo2}%
  \BibitemOpen
  \bibfield  {author} {\bibinfo {author} {\bibfnamefont {G.}~\bibnamefont
  {Brunin}}, \bibinfo {author} {\bibfnamefont {H.~P.~C.}\ \bibnamefont
  {Miranda}}, \bibinfo {author} {\bibfnamefont {M.}~\bibnamefont
  {Giantomassi}}, \bibinfo {author} {\bibfnamefont {M.}~\bibnamefont {Royo}},
  \bibinfo {author} {\bibfnamefont {M.}~\bibnamefont {Stengel}}, \bibinfo
  {author} {\bibfnamefont {M.~J.}\ \bibnamefont {Verstraete}}, \bibinfo
  {author} {\bibfnamefont {X.}~\bibnamefont {Gonze}}, \bibinfo {author}
  {\bibfnamefont {G.-M.}\ \bibnamefont {Rignanese}},\ and\ \bibinfo {author}
  {\bibfnamefont {G.}~\bibnamefont {Hautier}},\ }\bibfield  {title} {\bibinfo
  {title} {Phonon-limited electron mobility in {{Si}}, {{GaAs}}, and {{GaP}}
  with exact treatment of dynamical quadrupoles},\ }\href
  {https://doi.org/10.1103/PhysRevB.102.094308} {\bibfield  {journal} {\bibinfo
   {journal} {Phys. Rev. B}\ }\textbf {\bibinfo {volume} {102}},\ \bibinfo
  {pages} {094308} (\bibinfo {year} {2020}{\natexlab{b}})}\BibitemShut
  {NoStop}%
\bibitem [{\citenamefont {Jhalani}\ \emph {et~al.}(2020)\citenamefont
  {Jhalani}, \citenamefont {Zhou}, \citenamefont {Park}, \citenamefont
  {Dreyer},\ and\ \citenamefont {Bernardi}}]{2020JhalaniPiezo}%
  \BibitemOpen
  \bibfield  {author} {\bibinfo {author} {\bibfnamefont {V.~A.}\ \bibnamefont
  {Jhalani}}, \bibinfo {author} {\bibfnamefont {J.-J.}\ \bibnamefont {Zhou}},
  \bibinfo {author} {\bibfnamefont {J.}~\bibnamefont {Park}}, \bibinfo {author}
  {\bibfnamefont {C.~E.}\ \bibnamefont {Dreyer}},\ and\ \bibinfo {author}
  {\bibfnamefont {M.}~\bibnamefont {Bernardi}},\ }\bibfield  {title} {\bibinfo
  {title} {Piezoelectric electron-phonon interaction from \emph{ab initio}
  dynamical quadrupoles: Impact on charge transport in wurtzite {{GaN}}},\
  }\href {https://doi.org/10.1103/PhysRevLett.125.136602} {\bibfield  {journal}
  {\bibinfo  {journal} {Phys. Rev. Lett.}\ }\textbf {\bibinfo {volume} {125}},\
  \bibinfo {pages} {136602} (\bibinfo {year} {2020})}\BibitemShut {NoStop}%
\bibitem [{\citenamefont {Park}\ \emph {et~al.}(2020)\citenamefont {Park},
  \citenamefont {Zhou}, \citenamefont {Jhalani}, \citenamefont {Dreyer},\ and\
  \citenamefont {Bernardi}}]{2020ParkPiezo}%
  \BibitemOpen
  \bibfield  {author} {\bibinfo {author} {\bibfnamefont {J.}~\bibnamefont
  {Park}}, \bibinfo {author} {\bibfnamefont {J.-J.}\ \bibnamefont {Zhou}},
  \bibinfo {author} {\bibfnamefont {V.~A.}\ \bibnamefont {Jhalani}}, \bibinfo
  {author} {\bibfnamefont {C.~E.}\ \bibnamefont {Dreyer}},\ and\ \bibinfo
  {author} {\bibfnamefont {M.}~\bibnamefont {Bernardi}},\ }\bibfield  {title}
  {\bibinfo {title} {Long-range quadrupole electron-phonon interaction from
  first principles},\ }\href {https://doi.org/10.1103/PhysRevB.102.125203}
  {\bibfield  {journal} {\bibinfo  {journal} {Phys. Rev. B}\ }\textbf {\bibinfo
  {volume} {102}},\ \bibinfo {pages} {125203} (\bibinfo {year}
  {2020})}\BibitemShut {NoStop}%
\bibitem [{\citenamefont {Ganose}\ \emph {et~al.}(2021)\citenamefont {Ganose},
  \citenamefont {Park}, \citenamefont {Faghaninia}, \citenamefont
  {{Woods-Robinson}}, \citenamefont {Persson},\ and\ \citenamefont
  {Jain}}]{2021GanoseAMSET}%
  \BibitemOpen
  \bibfield  {author} {\bibinfo {author} {\bibfnamefont {A.~M.}\ \bibnamefont
  {Ganose}}, \bibinfo {author} {\bibfnamefont {J.}~\bibnamefont {Park}},
  \bibinfo {author} {\bibfnamefont {A.}~\bibnamefont {Faghaninia}}, \bibinfo
  {author} {\bibfnamefont {R.}~\bibnamefont {{Woods-Robinson}}}, \bibinfo
  {author} {\bibfnamefont {K.~A.}\ \bibnamefont {Persson}},\ and\ \bibinfo
  {author} {\bibfnamefont {A.}~\bibnamefont {Jain}},\ }\bibfield  {title}
  {\bibinfo {title} {Efficient calculation of carrier scattering rates from
  first principles},\ }\href {https://doi.org/10.1038/s41467-021-22440-5}
  {\bibfield  {journal} {\bibinfo  {journal} {Nat. Commun.}\ }\textbf {\bibinfo
  {volume} {12}},\ \bibinfo {pages} {2222} (\bibinfo {year}
  {2021})}\BibitemShut {NoStop}%
\bibitem [{\citenamefont {Ponc{\'e}}\ \emph
  {et~al.}(2023{\natexlab{a}})\citenamefont {Ponc{\'e}}, \citenamefont {Royo},
  \citenamefont {Gibertini}, \citenamefont {Marzari},\ and\ \citenamefont
  {Stengel}}]{2023PoncePiezo1}%
  \BibitemOpen
  \bibfield  {author} {\bibinfo {author} {\bibfnamefont {S.}~\bibnamefont
  {Ponc{\'e}}}, \bibinfo {author} {\bibfnamefont {M.}~\bibnamefont {Royo}},
  \bibinfo {author} {\bibfnamefont {M.}~\bibnamefont {Gibertini}}, \bibinfo
  {author} {\bibfnamefont {N.}~\bibnamefont {Marzari}},\ and\ \bibinfo {author}
  {\bibfnamefont {M.}~\bibnamefont {Stengel}},\ }\bibfield  {title} {\bibinfo
  {title} {Accurate prediction of {{Hall}} mobilities in two-dimensional
  materials through gauge-covariant quadrupolar contributions},\ }\href
  {https://doi.org/10.1103/PhysRevLett.130.166301} {\bibfield  {journal}
  {\bibinfo  {journal} {Phys. Rev. Lett.}\ }\textbf {\bibinfo {volume} {130}},\
  \bibinfo {pages} {166301} (\bibinfo {year} {2023}{\natexlab{a}})}\BibitemShut
  {NoStop}%
\bibitem [{\citenamefont {Ponc{\'e}}\ \emph
  {et~al.}(2023{\natexlab{b}})\citenamefont {Ponc{\'e}}, \citenamefont {Royo},
  \citenamefont {Stengel}, \citenamefont {Marzari},\ and\ \citenamefont
  {Gibertini}}]{2023PoncePiezo2}%
  \BibitemOpen
  \bibfield  {author} {\bibinfo {author} {\bibfnamefont {S.}~\bibnamefont
  {Ponc{\'e}}}, \bibinfo {author} {\bibfnamefont {M.}~\bibnamefont {Royo}},
  \bibinfo {author} {\bibfnamefont {M.}~\bibnamefont {Stengel}}, \bibinfo
  {author} {\bibfnamefont {N.}~\bibnamefont {Marzari}},\ and\ \bibinfo {author}
  {\bibfnamefont {M.}~\bibnamefont {Gibertini}},\ }\bibfield  {title} {\bibinfo
  {title} {Long-range electrostatic contribution to electron-phonon couplings
  and mobilities of two-dimensional and bulk materials},\ }\href
  {https://doi.org/10.1103/PhysRevB.107.155424} {\bibfield  {journal} {\bibinfo
   {journal} {Phys. Rev. B}\ }\textbf {\bibinfo {volume} {107}},\ \bibinfo
  {pages} {155424} (\bibinfo {year} {2023}{\natexlab{b}})}\BibitemShut
  {NoStop}%
\bibitem [{\citenamefont {Ponc\'e}\ \emph {et~al.}(2021)\citenamefont
  {Ponc\'e}, \citenamefont {Macheda}, \citenamefont {Margine}, \citenamefont
  {Marzari}, \citenamefont {Bonini},\ and\ \citenamefont
  {Giustino}}]{Ponce2021}%
  \BibitemOpen
  \bibfield  {author} {\bibinfo {author} {\bibfnamefont {S.}~\bibnamefont
  {Ponc\'e}}, \bibinfo {author} {\bibfnamefont {F.}~\bibnamefont {Macheda}},
  \bibinfo {author} {\bibfnamefont {E.~R.}\ \bibnamefont {Margine}}, \bibinfo
  {author} {\bibfnamefont {N.}~\bibnamefont {Marzari}}, \bibinfo {author}
  {\bibfnamefont {N.}~\bibnamefont {Bonini}},\ and\ \bibinfo {author}
  {\bibfnamefont {F.}~\bibnamefont {Giustino}},\ }\bibfield  {title} {\bibinfo
  {title} {First-principles predictions of {{Hall}} and drift mobilities in
  semiconductors},\ }\href {https://doi.org/10.1103/PhysRevResearch.3.043022}
  {\bibfield  {journal} {\bibinfo  {journal} {Phys. Rev. Research}\ }\textbf
  {\bibinfo {volume} {3}},\ \bibinfo {pages} {043022} (\bibinfo {year}
  {2021})}\BibitemShut {NoStop}%
\bibitem [{\citenamefont {Giannozzi}\ \emph {et~al.}(2017)\citenamefont
  {Giannozzi} \emph {et~al.}}]{Giannozzi2017}%
  \BibitemOpen
  \bibfield  {author} {\bibinfo {author} {\bibfnamefont {P.}~\bibnamefont
  {Giannozzi}} \emph {et~al.},\ }\bibfield  {title} {\bibinfo {title} {Advanced
  capabilities for materials modelling with {Quantum} {ESPRESSO}},\ }\href
  {https://doi.org/10.1088/1361-648x/aa8f79} {\bibfield  {journal} {\bibinfo
  {journal} {J. Condens. Matter Phys.}\ }\textbf {\bibinfo {volume} {29}},\
  \bibinfo {pages} {465901} (\bibinfo {year} {2017})}\BibitemShut {NoStop}%
\bibitem [{\citenamefont {Gonze}\ \emph {et~al.}(2016)\citenamefont {Gonze}
  \emph {et~al.}}]{Gonze2016}%
  \BibitemOpen
  \bibfield  {author} {\bibinfo {author} {\bibfnamefont {X.}~\bibnamefont
  {Gonze}} \emph {et~al.},\ }\bibfield  {title} {\bibinfo {title} {Recent
  developments in the {ABINIT} software package},\ }\href
  {https://doi.org/http://dx.doi.org/10.1016/j.cpc.2016.04.003} {\bibfield
  {journal} {\bibinfo  {journal} {Comput. Phys. Commun.}\ }\textbf {\bibinfo
  {volume} {205}},\ \bibinfo {pages} {106 } (\bibinfo {year}
  {2016})}\BibitemShut {NoStop}%
\bibitem [{\citenamefont {Gonze}\ \emph {et~al.}(2020)\citenamefont {Gonze}
  \emph {et~al.}}]{Gonze2020}%
  \BibitemOpen
  \bibfield  {author} {\bibinfo {author} {\bibfnamefont {X.}~\bibnamefont
  {Gonze}} \emph {et~al.},\ }\bibfield  {title} {\bibinfo {title} {The
  {{Abinit}} project: Impact, environment and recent developments},\ }\href
  {https://doi.org/https://doi.org/10.1016/j.cpc.2019.107042} {\bibfield
  {journal} {\bibinfo  {journal} {Comput. Phys. Commun.}\ }\textbf {\bibinfo
  {volume} {248}},\ \bibinfo {pages} {107042} (\bibinfo {year}
  {2020})}\BibitemShut {NoStop}%
\bibitem [{\citenamefont {Hamann}(2013)}]{2013HamannONCVPSP}%
  \BibitemOpen
  \bibfield  {author} {\bibinfo {author} {\bibfnamefont {D.~R.}\ \bibnamefont
  {Hamann}},\ }\bibfield  {title} {\bibinfo {title} {Optimized norm-conserving
  {Vanderbilt} pseudopotentials},\ }\href
  {https://doi.org/10.1103/PhysRevB.88.085117} {\bibfield  {journal} {\bibinfo
  {journal} {Phys. Rev. B}\ }\textbf {\bibinfo {volume} {88}},\ \bibinfo
  {pages} {085117} (\bibinfo {year} {2013})}\BibitemShut {NoStop}%
\bibitem [{\citenamefont {van Setten}\ \emph {et~al.}(2018)\citenamefont {van
  Setten}, \citenamefont {Giantomassi}, \citenamefont {Bousquet}, \citenamefont
  {Verstraete}, \citenamefont {Hamann}, \citenamefont {Gonze},\ and\
  \citenamefont {Rignanese}}]{Setten2018}%
  \BibitemOpen
  \bibfield  {author} {\bibinfo {author} {\bibfnamefont {M.}~\bibnamefont {van
  Setten}}, \bibinfo {author} {\bibfnamefont {M.}~\bibnamefont {Giantomassi}},
  \bibinfo {author} {\bibfnamefont {E.}~\bibnamefont {Bousquet}}, \bibinfo
  {author} {\bibfnamefont {M.}~\bibnamefont {Verstraete}}, \bibinfo {author}
  {\bibfnamefont {D.}~\bibnamefont {Hamann}}, \bibinfo {author} {\bibfnamefont
  {X.}~\bibnamefont {Gonze}},\ and\ \bibinfo {author} {\bibfnamefont {G.-M.}\
  \bibnamefont {Rignanese}},\ }\bibfield  {title} {\bibinfo {title} {The
  {PseudoDojo}: Training and grading a 85 element optimized norm-conserving
  pseudopotential table},\ }\href
  {https://doi.org/https://doi.org/10.1016/j.cpc.2018.01.012} {\bibfield
  {journal} {\bibinfo  {journal} {Comput. Phys. Commun.}\ }\textbf {\bibinfo
  {volume} {226}},\ \bibinfo {pages} {39 } (\bibinfo {year}
  {2018})}\BibitemShut {NoStop}%
\bibitem [{\citenamefont {Perdew}\ \emph {et~al.}(1996)\citenamefont {Perdew},
  \citenamefont {Burke},\ and\ \citenamefont {Ernzerhof}}]{Perdew1996}%
  \BibitemOpen
  \bibfield  {author} {\bibinfo {author} {\bibfnamefont {J.~P.}\ \bibnamefont
  {Perdew}}, \bibinfo {author} {\bibfnamefont {K.}~\bibnamefont {Burke}},\ and\
  \bibinfo {author} {\bibfnamefont {M.}~\bibnamefont {Ernzerhof}},\ }\bibfield
  {title} {\bibinfo {title} {Generalized gradient approximation made simple},\
  }\href {https://doi.org/10.1103/PhysRevLett.77.3865} {\bibfield  {journal}
  {\bibinfo  {journal} {Phys. Rev. Lett.}\ }\textbf {\bibinfo {volume} {77}},\
  \bibinfo {pages} {3865} (\bibinfo {year} {1996})}\BibitemShut {NoStop}%
\bibitem [{\citenamefont {Pizzi}\ \emph {et~al.}(2020)\citenamefont {Pizzi}
  \emph {et~al.}}]{2020PizziWannier90}%
  \BibitemOpen
  \bibfield  {author} {\bibinfo {author} {\bibfnamefont {G.}~\bibnamefont
  {Pizzi}} \emph {et~al.},\ }\bibfield  {title} {\bibinfo {title} {Wannier90 as
  a community code: New features and applications},\ }\href
  {https://doi.org/10.1088/1361-648X/ab51ff} {\bibfield  {journal} {\bibinfo
  {journal} {J. Condens. Matter Phys.}\ }\textbf {\bibinfo {volume} {32}},\
  \bibinfo {pages} {165902} (\bibinfo {year} {2020})}\BibitemShut {NoStop}%
\bibitem [{\citenamefont {Giustino}\ \emph {et~al.}(2007)\citenamefont
  {Giustino}, \citenamefont {Cohen},\ and\ \citenamefont
  {Louie}}]{Giustino2007}%
  \BibitemOpen
  \bibfield  {author} {\bibinfo {author} {\bibfnamefont {F.}~\bibnamefont
  {Giustino}}, \bibinfo {author} {\bibfnamefont {M.~L.}\ \bibnamefont
  {Cohen}},\ and\ \bibinfo {author} {\bibfnamefont {S.~G.}\ \bibnamefont
  {Louie}},\ }\bibfield  {title} {\bibinfo {title} {Electron-phonon interaction
  using {Wannier} functions},\ }\href
  {https://doi.org/10.1103/PhysRevB.76.165108} {\bibfield  {journal} {\bibinfo
  {journal} {Phys. Rev. B}\ }\textbf {\bibinfo {volume} {76}},\ \bibinfo
  {pages} {165108} (\bibinfo {year} {2007})}\BibitemShut {NoStop}%
\bibitem [{\citenamefont {Poncé}\ \emph {et~al.}(2016)\citenamefont {Poncé},
  \citenamefont {Margine}, \citenamefont {Verdi},\ and\ \citenamefont
  {Giustino}}]{Ponce2016}%
  \BibitemOpen
  \bibfield  {author} {\bibinfo {author} {\bibfnamefont {S.}~\bibnamefont
  {Poncé}}, \bibinfo {author} {\bibfnamefont {E.}~\bibnamefont {Margine}},
  \bibinfo {author} {\bibfnamefont {C.}~\bibnamefont {Verdi}},\ and\ \bibinfo
  {author} {\bibfnamefont {F.}~\bibnamefont {Giustino}},\ }\bibfield  {title}
  {\bibinfo {title} {{EPW}: Electron–phonon coupling, transport and
  superconducting properties using maximally localized {Wannier} functions},\
  }\href {https://doi.org/http://dx.doi.org/10.1016/j.cpc.2016.07.028}
  {\bibfield  {journal} {\bibinfo  {journal} {Comput. Phys. Commun.}\ }\textbf
  {\bibinfo {volume} {209}},\ \bibinfo {pages} {116 } (\bibinfo {year}
  {2016})}\BibitemShut {NoStop}%
\bibitem [{\citenamefont {Marzari}\ and\ \citenamefont
  {Vanderbilt}(1997)}]{1997Marzari}%
  \BibitemOpen
  \bibfield  {author} {\bibinfo {author} {\bibfnamefont {N.}~\bibnamefont
  {Marzari}}\ and\ \bibinfo {author} {\bibfnamefont {D.}~\bibnamefont
  {Vanderbilt}},\ }\bibfield  {title} {\bibinfo {title} {Maximally-localized
  generalized {{Wannier}} functions for composite energy bands},\ }\href
  {https://doi.org/10.1103/PhysRevB.56.12847} {\bibfield  {journal} {\bibinfo
  {journal} {Phys. Rev. B}\ }\textbf {\bibinfo {volume} {56}},\ \bibinfo
  {pages} {12847} (\bibinfo {year} {1997})}\BibitemShut {NoStop}%
\bibitem [{\citenamefont {Souza}\ \emph {et~al.}(2001)\citenamefont {Souza},
  \citenamefont {Marzari},\ and\ \citenamefont {Vanderbilt}}]{2001Souza}%
  \BibitemOpen
  \bibfield  {author} {\bibinfo {author} {\bibfnamefont {I.}~\bibnamefont
  {Souza}}, \bibinfo {author} {\bibfnamefont {N.}~\bibnamefont {Marzari}},\
  and\ \bibinfo {author} {\bibfnamefont {D.}~\bibnamefont {Vanderbilt}},\
  }\bibfield  {title} {\bibinfo {title} {Maximally localized {{Wannier}}
  functions for entangled energy bands},\ }\href
  {https://doi.org/10.1103/PhysRevB.65.035109} {\bibfield  {journal} {\bibinfo
  {journal} {Phys. Rev. B}\ }\textbf {\bibinfo {volume} {65}},\ \bibinfo
  {pages} {035109} (\bibinfo {year} {2001})}\BibitemShut {NoStop}%
\bibitem [{\citenamefont {Royo}\ and\ \citenamefont
  {Stengel}(2019)}]{2019RoyoPiezo}%
  \BibitemOpen
  \bibfield  {author} {\bibinfo {author} {\bibfnamefont {M.}~\bibnamefont
  {Royo}}\ and\ \bibinfo {author} {\bibfnamefont {M.}~\bibnamefont {Stengel}},\
  }\bibfield  {title} {\bibinfo {title} {First-principles theory of spatial
  dispersion: Dynamical quadrupoles and flexoelectricity},\ }\href
  {https://doi.org/10.1103/PhysRevX.9.021050} {\bibfield  {journal} {\bibinfo
  {journal} {Phys. Rev. X}\ }\textbf {\bibinfo {volume} {9}},\ \bibinfo {pages}
  {021050} (\bibinfo {year} {2019})}\BibitemShut {NoStop}%
\bibitem [{\citenamefont {Bezanson}\ \emph {et~al.}(2017)\citenamefont
  {Bezanson}, \citenamefont {Edelman}, \citenamefont {Karpinski},\ and\
  \citenamefont {Shah}}]{2017BezansonJulia}%
  \BibitemOpen
  \bibfield  {author} {\bibinfo {author} {\bibfnamefont {J.}~\bibnamefont
  {Bezanson}}, \bibinfo {author} {\bibfnamefont {A.}~\bibnamefont {Edelman}},
  \bibinfo {author} {\bibfnamefont {S.}~\bibnamefont {Karpinski}},\ and\
  \bibinfo {author} {\bibfnamefont {V.~B.}\ \bibnamefont {Shah}},\ }\bibfield
  {title} {\bibinfo {title} {Julia: A fresh approach to numerical computing},\
  }\href {https://doi.org/10.1137/141000671} {\bibfield  {journal} {\bibinfo
  {journal} {SIAM Review}\ }\textbf {\bibinfo {volume} {59}},\ \bibinfo {pages}
  {65} (\bibinfo {year} {2017})}\BibitemShut {NoStop}%
\bibitem [{\citenamefont {Verdi}\ \emph {et~al.}(2017)\citenamefont {Verdi},
  \citenamefont {Caruso},\ and\ \citenamefont {Giustino}}]{2017Verdi}%
  \BibitemOpen
  \bibfield  {author} {\bibinfo {author} {\bibfnamefont {C.}~\bibnamefont
  {Verdi}}, \bibinfo {author} {\bibfnamefont {F.}~\bibnamefont {Caruso}},\ and\
  \bibinfo {author} {\bibfnamefont {F.}~\bibnamefont {Giustino}},\ }\bibfield
  {title} {\bibinfo {title} {Origin of the crossover from polarons to {{Fermi}}
  liquids in transition metal oxides},\ }\href
  {https://doi.org/10.1038/ncomms15769} {\bibfield  {journal} {\bibinfo
  {journal} {Nat. Commun.}\ }\textbf {\bibinfo {volume} {8}},\ \bibinfo {pages}
  {15769} (\bibinfo {year} {2017})}\BibitemShut {NoStop}%
\bibitem [{\citenamefont {Mermin}(1970)}]{1970Mermin}%
  \BibitemOpen
  \bibfield  {author} {\bibinfo {author} {\bibfnamefont {N.~D.}\ \bibnamefont
  {Mermin}},\ }\bibfield  {title} {\bibinfo {title} {Lindhard dielectric
  function in the relaxation-time approximation},\ }\href
  {https://doi.org/10.1103/PhysRevB.1.2362} {\bibfield  {journal} {\bibinfo
  {journal} {Phys. Rev. B}\ }\textbf {\bibinfo {volume} {1}},\ \bibinfo {pages}
  {2362} (\bibinfo {year} {1970})}\BibitemShut {NoStop}%
\bibitem [{\citenamefont {Born}\ and\ \citenamefont
  {Huang}(1996)}]{BornHuang1954}%
  \BibitemOpen
  \bibfield  {author} {\bibinfo {author} {\bibfnamefont {M.}~\bibnamefont
  {Born}}\ and\ \bibinfo {author} {\bibfnamefont {K.}~\bibnamefont {Huang}},\
  }\href@noop {} {\emph {\bibinfo {title} {Dynamical theory of crystal
  lattices}}}\ (\bibinfo  {publisher} {Oxford university press},\ \bibinfo
  {address} {Oxford},\ \bibinfo {year} {1996})\BibitemShut {NoStop}%
\bibitem [{\citenamefont {Zhou}\ \emph {et~al.}(2021)\citenamefont {Zhou},
  \citenamefont {Park}, \citenamefont {Lu}, \citenamefont {Maliyov},
  \citenamefont {Tong},\ and\ \citenamefont {Bernardi}}]{2021ZhouPerturbo}%
  \BibitemOpen
  \bibfield  {author} {\bibinfo {author} {\bibfnamefont {J.-J.}\ \bibnamefont
  {Zhou}}, \bibinfo {author} {\bibfnamefont {J.}~\bibnamefont {Park}}, \bibinfo
  {author} {\bibfnamefont {I.-T.}\ \bibnamefont {Lu}}, \bibinfo {author}
  {\bibfnamefont {I.}~\bibnamefont {Maliyov}}, \bibinfo {author} {\bibfnamefont
  {X.}~\bibnamefont {Tong}},\ and\ \bibinfo {author} {\bibfnamefont
  {M.}~\bibnamefont {Bernardi}},\ }\bibfield  {title} {\bibinfo {title}
  {Perturbo: A software package for ab initio electron–phonon interactions,
  charge transport and ultrafast dynamics},\ }\href
  {https://doi.org/https://doi.org/10.1016/j.cpc.2021.107970} {\bibfield
  {journal} {\bibinfo  {journal} {Comput. Phys. Commun.}\ }\textbf {\bibinfo
  {volume} {264}},\ \bibinfo {pages} {107970} (\bibinfo {year}
  {2021})}\BibitemShut {NoStop}%
\bibitem [{\citenamefont {Vanderbilt}(2000)}]{Vanderbilt2000}%
  \BibitemOpen
  \bibfield  {author} {\bibinfo {author} {\bibfnamefont {D.}~\bibnamefont
  {Vanderbilt}},\ }\bibfield  {title} {\bibinfo {title} {Berry-phase theory of
  proper piezoelectric response},\ }\href
  {https://doi.org/https://doi.org/10.1016/S0022-3697(99)00273-5} {\bibfield
  {journal} {\bibinfo  {journal} {J. Phys. Chem. Solids}\ }\textbf {\bibinfo
  {volume} {61}},\ \bibinfo {pages} {147} (\bibinfo {year} {2000})}\BibitemShut
  {NoStop}%
\bibitem [{\citenamefont {Resta}(1992)}]{1992RestaPolarization}%
  \BibitemOpen
  \bibfield  {author} {\bibinfo {author} {\bibfnamefont {R.}~\bibnamefont
  {Resta}},\ }\bibfield  {title} {\bibinfo {title} {Theory of the electric
  polarization in crystals},\ }\href
  {https://doi.org/10.1080/00150199208016065} {\bibfield  {journal} {\bibinfo
  {journal} {Ferroelectrics}\ }\textbf {\bibinfo {volume} {136}},\ \bibinfo
  {pages} {51} (\bibinfo {year} {1992})}\BibitemShut {NoStop}%
\bibitem [{\citenamefont {{King-Smith}}\ and\ \citenamefont
  {Vanderbilt}(1993)}]{1993KingSmithPolarization}%
  \BibitemOpen
  \bibfield  {author} {\bibinfo {author} {\bibfnamefont {R.~D.}\ \bibnamefont
  {{King-Smith}}}\ and\ \bibinfo {author} {\bibfnamefont {D.}~\bibnamefont
  {Vanderbilt}},\ }\bibfield  {title} {\bibinfo {title} {Theory of polarization
  of crystalline solids},\ }\href {https://doi.org/10.1103/PhysRevB.47.1651}
  {\bibfield  {journal} {\bibinfo  {journal} {Phys. Rev. B}\ }\textbf {\bibinfo
  {volume} {47}},\ \bibinfo {pages} {1651} (\bibinfo {year}
  {1993})}\BibitemShut {NoStop}%
\bibitem [{\citenamefont {Vanderbilt}\ and\ \citenamefont
  {{King-Smith}}(1993)}]{1993VanderbiltPolarization}%
  \BibitemOpen
  \bibfield  {author} {\bibinfo {author} {\bibfnamefont {D.}~\bibnamefont
  {Vanderbilt}}\ and\ \bibinfo {author} {\bibfnamefont {R.~D.}\ \bibnamefont
  {{King-Smith}}},\ }\bibfield  {title} {\bibinfo {title} {Electric
  polarization as a bulk quantity and its relation to surface charge},\ }\href
  {https://doi.org/10.1103/PhysRevB.48.4442} {\bibfield  {journal} {\bibinfo
  {journal} {Phys. Rev. B}\ }\textbf {\bibinfo {volume} {48}},\ \bibinfo
  {pages} {4442} (\bibinfo {year} {1993})}\BibitemShut {NoStop}%
\bibitem [{\citenamefont {Resta}(1994)}]{1994RestaPolarization}%
  \BibitemOpen
  \bibfield  {author} {\bibinfo {author} {\bibfnamefont {R.}~\bibnamefont
  {Resta}},\ }\bibfield  {title} {\bibinfo {title} {Macroscopic polarization in
  crystalline dielectrics: The geometric phase approach},\ }\href
  {https://doi.org/10.1103/RevModPhys.66.899} {\bibfield  {journal} {\bibinfo
  {journal} {Rev. Mod. Phys.}\ }\textbf {\bibinfo {volume} {66}},\ \bibinfo
  {pages} {899} (\bibinfo {year} {1994})}\BibitemShut {NoStop}%
\bibitem [{\citenamefont {Marzari}\ \emph {et~al.}(2012)\citenamefont
  {Marzari}, \citenamefont {Mostofi}, \citenamefont {Yates}, \citenamefont
  {Souza},\ and\ \citenamefont {Vanderbilt}}]{2012Marzari}%
  \BibitemOpen
  \bibfield  {author} {\bibinfo {author} {\bibfnamefont {N.}~\bibnamefont
  {Marzari}}, \bibinfo {author} {\bibfnamefont {A.~A.}\ \bibnamefont
  {Mostofi}}, \bibinfo {author} {\bibfnamefont {J.~R.}\ \bibnamefont {Yates}},
  \bibinfo {author} {\bibfnamefont {I.}~\bibnamefont {Souza}},\ and\ \bibinfo
  {author} {\bibfnamefont {D.}~\bibnamefont {Vanderbilt}},\ }\bibfield  {title}
  {\bibinfo {title} {Maximally localized {{Wannier}} functions: {{Theory}} and
  applications},\ }\href {https://doi.org/10.1103/RevModPhys.84.1419}
  {\bibfield  {journal} {\bibinfo  {journal} {Rev. Mod. Phys.}\ }\textbf
  {\bibinfo {volume} {84}},\ \bibinfo {pages} {1419} (\bibinfo {year}
  {2012})}\BibitemShut {NoStop}%
\bibitem [{\citenamefont {Grimsditch}\ \emph {et~al.}(1994)\citenamefont
  {Grimsditch}, \citenamefont {Zouboulis},\ and\ \citenamefont
  {Polian}}]{1994GrimsditchElastic}%
  \BibitemOpen
  \bibfield  {author} {\bibinfo {author} {\bibfnamefont {M.}~\bibnamefont
  {Grimsditch}}, \bibinfo {author} {\bibfnamefont {E.~S.}\ \bibnamefont
  {Zouboulis}},\ and\ \bibinfo {author} {\bibfnamefont {A.}~\bibnamefont
  {Polian}},\ }\bibfield  {title} {\bibinfo {title} {{Elastic constants of
  boron nitride}},\ }\href {https://doi.org/10.1063/1.357757} {\bibfield
  {journal} {\bibinfo  {journal} {Journal of Applied Physics}\ }\textbf
  {\bibinfo {volume} {76}},\ \bibinfo {pages} {832} (\bibinfo {year}
  {1994})}\BibitemShut {NoStop}%
\bibitem [{\citenamefont {Gonze}\ and\ \citenamefont {Lee}(1997)}]{Gonze1997a}%
  \BibitemOpen
  \bibfield  {author} {\bibinfo {author} {\bibfnamefont {X.}~\bibnamefont
  {Gonze}}\ and\ \bibinfo {author} {\bibfnamefont {C.}~\bibnamefont {Lee}},\
  }\bibfield  {title} {\bibinfo {title} {Dynamical matrices, born effective
  charges, dielectric permittivity tensors, and interatomic force constants
  from density-functional perturbation theory},\ }\href
  {https://doi.org/10.1103/PhysRevB.55.10355} {\bibfield  {journal} {\bibinfo
  {journal} {Phys. Rev. B}\ }\textbf {\bibinfo {volume} {55}},\ \bibinfo
  {pages} {10355} (\bibinfo {year} {1997})}\BibitemShut {NoStop}%
\bibitem [{\citenamefont {Hamann}\ \emph {et~al.}(2005)\citenamefont {Hamann},
  \citenamefont {Wu}, \citenamefont {Rabe},\ and\ \citenamefont
  {Vanderbilt}}]{Hamann2005}%
  \BibitemOpen
  \bibfield  {author} {\bibinfo {author} {\bibfnamefont {D.~R.}\ \bibnamefont
  {Hamann}}, \bibinfo {author} {\bibfnamefont {X.}~\bibnamefont {Wu}}, \bibinfo
  {author} {\bibfnamefont {K.~M.}\ \bibnamefont {Rabe}},\ and\ \bibinfo
  {author} {\bibfnamefont {D.}~\bibnamefont {Vanderbilt}},\ }\bibfield  {title}
  {\bibinfo {title} {Metric tensor formulation of strain in density-functional
  perturbation theory},\ }\href {https://doi.org/10.1103/physrevb.71.035117}
  {\bibfield  {journal} {\bibinfo  {journal} {Phys. Rev. B}\ }\textbf {\bibinfo
  {volume} {71}},\ \bibinfo {pages} {035117} (\bibinfo {year}
  {2005})}\BibitemShut {NoStop}%
\bibitem [{\citenamefont {Wu}\ \emph {et~al.}(2005)\citenamefont {Wu},
  \citenamefont {Vanderbilt},\ and\ \citenamefont {Hamann}}]{Wu2005}%
  \BibitemOpen
  \bibfield  {author} {\bibinfo {author} {\bibfnamefont {X.}~\bibnamefont
  {Wu}}, \bibinfo {author} {\bibfnamefont {D.}~\bibnamefont {Vanderbilt}},\
  and\ \bibinfo {author} {\bibfnamefont {D.~R.}\ \bibnamefont {Hamann}},\
  }\bibfield  {title} {\bibinfo {title} {Systematic treatment of displacements,
  strains, and electric fields in density-functional perturbation theory},\
  }\href {https://doi.org/10.1103/PhysRevB.72.035105} {\bibfield  {journal}
  {\bibinfo  {journal} {Phys. Rev. B}\ }\textbf {\bibinfo {volume} {72}},\
  \bibinfo {pages} {035105} (\bibinfo {year} {2005})}\BibitemShut {NoStop}%
\end{thebibliography}%

\end{document}


\title{Supplemental Material:\\
Self-consistent electron lifetimes for electron-phonon scattering}

\author{Jae-Mo Lihm}
\email{jaemo.lihm@gmail.com}
\affiliation{Department of Physics and Astronomy, Seoul National University, Seoul 08826, Korea}
\affiliation{Center for Correlated Electron Systems, Institute for Basic Science, Seoul 08826, Korea}
\affiliation{Center for Theoretical Physics, Seoul National University, Seoul 08826, Korea}
\author{Samuel Ponc\'e}
\email{samuel.ponce@uclouvain.be}
\affiliation{%
European Theoretical Spectroscopy Facility, Institute of Condensed Matter and Nanosciences, Universit\'e catholique de Louvain, Chemin des \'Etoiles 8, B-1348 Louvain-la-Neuve, Belgium. 	
}%
\affiliation{%
WEL Research Institute, avenue Pasteur, 6, 1300 Wavre, Belgique	
}%
\author{Cheol-Hwan Park}
\email{cheolhwan@snu.ac.kr}
\affiliation{Department of Physics and Astronomy, Seoul National University, Seoul 08826, Korea}
\affiliation{Center for Correlated Electron Systems, Institute for Basic Science, Seoul 08826, Korea}
\affiliation{Center for Theoretical Physics, Seoul National University, Seoul 08826, Korea}

\date{\today}

\maketitle


\section{Logarithmic divergence of the linewidth for piezoelectric EPC}
\subsection{Model}
In this section, we derive the electron linewidth due to piezoelectric EPC.
For the analytical derivation, we use an isotropic long-wavelength model with electron and phonon dispersions
\begin{equation}
    \veps_k = \frac{\hbar^2\mb{k}^2}{2m},\quad
    \omega_q = v q,
\end{equation}
where $m$ is the electron effective mass and $v$ the phonon velocity.
We let the piezoelectric tensor $e_{abc}$ have the symmetry of zincblende crystals, which gives~\cite{1968ArltPiezo}
\begin{equation}
    e_{abc} = \begin{cases}
        e^* & \text{if $a \neq b \neq c \neq a$} \\
        0 & \text{otherwise}.
    \end{cases}
\end{equation}
In the long wavelength limit, the piezoelectric electron-phonon matrix elements read~\cite{Ridley1982}
\begin{equation} \label{eq:deriv_ep}
    g(\mb{k}, \mb{q})
    = \sqrt{\frac{\hbar}{2M\omega_q}} e^*f(\theta, \phi) + \mcO(\sqrt{q}),
\end{equation}
where $M$ is the total mass of the unit cell and $\theta$ and $\phi$ are the spherical coordinate variables with $\mb{q} = q(\sin\theta \cos\phi, \sin\theta \sin\phi, \cos\theta)$.
Finally, we assume that the temperature $T$ is high enough so that $k_{\rm B} T \gg \hbar \omega_q$ holds, where $k_{\rm B}$ is the Boltzmann constant.
Then, the phonon occupation $n_q$ can be approximated as
\begin{equation} \label{eq:deriv_n}
    n_q
    = \frac{1}{e^{\hbar\omega_q / k_{\rm B} T} - 1}
    \approx \frac{ k_{\rm B} T}{\hbar\omega_q} \gg 1.
\end{equation}

\subsection{Wavevector-dependent divergence of the linewidths}
Now, we calculate the linewidth of an electron state at $\mb{k}$.
Without loss of generality, we choose $\mb{k}$ to be parallel to the $z$ axis: $\mb{k} = k \hat{\mb{z}}$.
Substituting the model parameters into \Eq{eq:gamma} of the main manuscript, we find
\begin{align} \label{eq:inv_tau_1}
    \gamma_k
    \approx & \frac{2\pi}{\hbar} \sum_{\pm} \int \frac{d\mb{q}}{\Omega^{\rm BZ}} \abs{g(\mb{k}, \mb{q})}^2 n_q \delta(\veps_k - \veps_{\abs{\mb{k+q}}} \pm \hbar\omega_q) \nnnl
    \approx & \frac{2\pi}{\hbar} \sum_{\pm} \int \frac{d\mb{q}}{\Omega^{\rm BZ}} \frac{\hbar e^{*2}}{2M \omega_q} \abs{f(\theta, \phi)}^2 \frac{ k_{\rm B} T}{\hbar\omega_q} \nonumber \\
    & \quad \quad \quad\quad \quad\quad\quad\quad\quad \times  \delta(\veps_k - \veps_{\abs{\mb{k+q}}} \pm \hbar\omega_q).
\end{align}
In the first line, we neglect the $f^\pm_\mkq$ factors because the phonon occupation is much larger than 1 [\Eq{eq:deriv_n}].
The upper and lower signs correspond to phonon absorption and emission, respectively.
Since we are considering a continuum long-wavelength model, the integral domain spans the entire three-dimensional reciprocal space, but the integrand natually becomes zero at large $\mb{q}$ due to the delta function.

We separate the three-dimensional integral into radial and angular parts.
Performing the azimuthal integration in \Eq{eq:inv_tau_1} gives
\begin{equation}
    \int_{0}^{2\pi} d\phi \abs{f(\theta, \phi)}^2 = g(\theta).
\end{equation}
This angle-dependent factor only contributes as a constant multiplicative factor to the linewidth and does not alter its divergence.
Hence, when discussing the divergence of the linewidths, we set $g = 1$.
We explicitly denote ``$\stackrel{g = 1}{\approx}$'' when we make use of this approximation.
Defining $z = \cos\theta$ and $\widetilde{g}(z) = g(\cos^{-1} z)$, we find
\begin{align} \label{eq:inv_tau_2}
    \gamma_k
    \approx &  \frac{\pi e^{*2} k_{\rm B} T}{\hbar \Omega^{\rm BZ} M} \sum_{\pm} \int_0^\infty dq q^2 \int_0^\pi d\theta \sin\theta g(\theta) \frac{1}{\omega_q^2} \nnnl
      & \times \delta \Big( \frac{\hbar^2 k^2}{2m} - \frac{\hbar^2 (k^2 + q^2 + 2kq \cos \theta)}{2m} \pm \hbar vq \Big)
    \nnnl
    = & \frac{\pi e^{*2} k_{\rm B} T}{\hbar \Omega^{\rm BZ} M v^2} \sum_{\pm} \int_{0}^{\infty} dq q^2 \int_{-1}^1 dz \frac{1}{q^2} \widetilde{g}(z) \nnnl
      & \times \delta \Big( \frac{\hbar^2 (-q^2 - 2kq z)}{2m} \pm \hbar v q \Big)
    \nnnl
    = & C \! \sum_{\pm} \! \int_{0}^{\infty} \frac{dq}{q} \int_{-1}^1 dz \widetilde{g}(z) \delta\Big( q \!+\! 2k z \mp \frac{2m v}{\hbar} \Big),
\end{align}
where
\begin{equation}
    C = \frac{2\pi e^{*2} k_{\rm B} T m}{\hbar^3 \Omega^{\rm BZ} M v^2}.
\end{equation}

The integrand of \Eq{eq:inv_tau_2} is zero except when
\begin{equation}
    z_\pm(q) = -\frac{q}{2k} \pm \frac{mv}{\hbar k}.
\end{equation}
So, the integral over $z$ gives
\begin{multline} \label{eq:deriv_int_z}
    \int_{-1}^1 dz \widetilde{g}(z) \delta \Big( q + 2k z \mp \frac{2m v}{\hbar} \Big)
     \\
    =  \begin{cases}
        \frac{1}{2k} \widetilde{g}(z_\pm(q)) & \text{if $-1 \leq z_\pm(q) \leq 1$} \\
        0 & \text{otherwise}.
    \end{cases}
\end{multline}
The condition $-1 \leq z_\pm(q) \leq 1$ is equivalent to
\begin{equation}
    -2k \pm \frac{2mv}{\hbar} \leq q \leq 2k \pm \frac{2mv}{\hbar}.
\end{equation}

Now, let us consider two different cases depending on whether $k$ is greater or lesser than $m v / \hbar$.
First, suppose $k < m v / \hbar$.
Then, \Eq{eq:deriv_int_z} is nonzero only for the upper sign (phonon absorption) with
\begin{equation}
    0 < -2k + \frac{2mv}{\hbar} \leq q \leq 2k + \frac{2mv}{\hbar}.
\end{equation}
The electron linewidth is finite with value
\begin{align} \label{eq:inv_tau_k_small}
    \gamma_k
    &= C \int_{-2k + 2mv / \hbar}^{2k + 2mv / \hbar} \frac{dq}{q} \frac{1}{2k} \widetilde{g}(z_+(q))
    \stackrel{g = 1}{\approx} \frac{C}{2k} \ln \frac{mv + \hbar k}{mv - \hbar k}.
\end{align}

Next, suppose $k \geq m v / \hbar$.
Then, \Eq{eq:deriv_int_z} can be nonzero for both the upper and lower signs (phonon absorption and emission, respectively), and the lower bound of $q$ extends to $q=0$.
The resulting electron linewidth is logarithmically divergent:
\begin{align} \label{eq:inv_tau_k_large}
    \gamma_k
    &= C \sum_{\pm} \int_{0}^{2k \pm 2mv / \hbar} \frac{dq}{q} \frac{1}{2k} \widetilde{g}(z_\pm(q)) \nnnl
    &\stackrel{g = 1}{\approx} \frac{C}{2k} \sum_\pm \ln \frac{2k \pm 2mv / \hbar}{0}.
\end{align}

This electron-wavevector-dependent dichotomy of divergence and non-divergence of the linewidth is reported here for the first time.
The second, divergent case was considered in Ref.~\cite{Ridley1982} (see Sec.~3.6); however, the necessary condition for this divergence ($k>mv/\hbar$) was not reported therein.

\subsection{Linewidths with a finite broadening}
When a finite, fixed broadening is used, we replace the delta function in \Eq{eq:inv_tau_2} with a regularized one, finding
\begin{align} \label{eq:inv_tau_3}
    \gamma_k
    \approx & \frac{\hbar^2 C}{2m} \sum_{\pm} \int_{0}^{\infty} dq q^2 \int_{-1}^1 dz \frac{1}{q^2} \widetilde{g}(z) \nonumber \\
    & \quad \quad \times \delta_\eta \left( \frac{\hbar^2 (-q^2 - 2kq z)}{2m} \pm \hbar v q \right) \nnnl
    =& C \sum_{\pm} \int_{0}^{\infty} \frac{dq}{q} \int_{-1}^1 dz \widetilde{g}(z) \nonumber \\
    & \quad \quad \times \delta_{2m\eta / \hbar^2 q} \Big( q + 2k z \mp \frac{2m v}{\hbar} \Big),
\end{align}
where we used $\delta_{\eta}(\alpha x) = \frac{1}{\alpha} \delta_{\eta / \alpha}(x)$.
For concreteness, we consider a Lorentzian function:
\begin{equation}
    \delta_\eta(x) = \frac{1}{\pi} \frac{\eta/2}{x^2 + \eta^2/4}.
\end{equation}
Other choices such as Gaussians should give the same behavior in the $\eta \to 0$ limit.

\begin{figure}[b]
  \centering
  \includegraphics[width=0.99\linewidth]{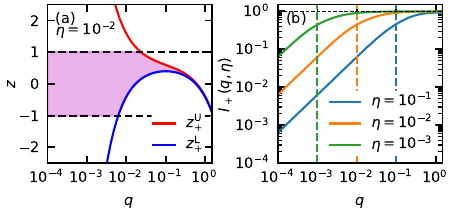}
  \caption{
  (a) Lower and upper values of $z$ [\Eq{eq:supp_z_LR}] at which the Lorentzian delta function of \Eq{eq:inv_tau_3} is half of its maximum value.
  The overall domain to be integrated, taking also the integration domain $[-1, 1]$ into account, is indicated with the colored background.
  We set $m=2k=\hbar=1$ and $v=0.3$.
  (b) $I_+(k, q, \eta)$ [\Eq{eq:supp_I_q}] for $g=1$. The vertical dashed lines indicate $q^{\rm IR}(k,\eta)$ [\Eq{eq:supp_q_IR}].
  The horizontal dashed line marks $I_+ = 1 / 2k = 1$.
  }
  \label{fig:supp_div_schematic}
\end{figure}

We focus on the $k > mv/\hbar$ case and see how the divergence in the $\eta = 0$ case is regularized when a finite $\eta$ is used.
The lower (L) and upper (U) values of $z$ at which the Lorentzian delta function in \Eq{eq:inv_tau_3} decays to half are
\begin{align} \label{eq:supp_z_LR}
\begin{split}
    z^{\rm L}_{\pm}(q, \eta) &= -\frac{m\eta}{2\hbar^2 kq} - \frac{q}{2k} \pm \frac{m v}{\hbar k}, \\
    z^{\rm U}_{\pm}(q, \eta) &= \hphantom{-} \frac{m\eta}{2\hbar^2 kq} - \frac{q}{2k} \pm \frac{m v}{\hbar k}.
\end{split}
\end{align}
As shown in Fig.~\ref{fig:supp_div_schematic}(a), when $q$ is small, $\abs{z^{\rm L,U}_{\pm}} \gg 1$ and only a small portion of the broadened delta function contributes to the integral.
In this case, the Lorentzian can be approximated by a constant: $\delta_\eta(x) \approx 2 / \pi\eta$ for $\abs{x} \ll \eta$.
On the other hand, if $q$ is large so that the Lorentzian delta function in \Eq{eq:inv_tau_3} has a narrow width, but not much larger than $mv/\hbar$ so that $z^{\rm L}_{\pm}(q, \eta)$ and $z^{\rm U}_{\pm}(q, \eta)$ are still in $[-1, 1]$, the full delta function lies within the integration domain.
These two limiting cases for the integral of \Eq{eq:inv_tau_3} read
\begin{multline} \label{eq:supp_I_q}
    I_\pm(k, q, \eta) = \int_{-1}^1 dz \widetilde{g}(z)
    \delta_{2m\eta / \hbar^2 q} \left( q + 2k z \mp \frac{2m v}{\hbar} \right) \\
    \approx \begin{cases}
        \frac{1}{2k} \widetilde{g}(z_\pm(q))
        & \mathrm{if} \ q^{\rm IR}(k, \eta) \ll q < 2k \pm 2mv/\hbar \\
        \frac{\hbar^2 q}{2\pi m\eta} \int_{-1}^1 dz \widetilde{g}(z)
        & \mathrm{if} \ q \ll q^{\rm IR}(k, \eta)
    \end{cases},
\end{multline}
where we introduced an $\eta$-dependent IR cutoff
\begin{equation} \label{eq:supp_q_IR}
    q^{\rm IR}(k, \eta) = \frac{m\eta}{\hbar^2 k}
\end{equation}
which distinguishes the two limits, $\abs{z^{\rm L,U}_\pm} \ll 1$ and $\abs{z^{\rm L,U}_\pm} \gg 1$.
Figure~\ref{fig:supp_div_schematic}(b) shows that the integral is nearly constant for $q \gg q^{\rm IR}(k,\eta)$ and decays linearly for $q \ll q^{\rm IR}(k,\eta)$, demonstrating numerically that $q^{\rm IR}(k, \eta)$ is the effective IR cutoff.

We now insert \Eq{eq:supp_I_q} into \Eq{eq:inv_tau_3}.
Approximately, the $q < q^{\rm IR}(k,\eta)$ part gives an $\eta$-independent contribution:
\begin{equation}
    \int_0^{q^{\rm IR}(k, \eta)} \frac{dq}{q} \frac{\hbar^2 q}{2\pi m\eta} \int_{-1}^1 dz \widetilde{g}(z)
    \stackrel{g=1}{\approx} \frac{1}{\pi k}.
\end{equation}
In contrast, the $q > q^{\rm IR}(k,\eta)$ part gives a contribution logarithmic in $\eta$:
\begin{align}
    \int_{q^{\rm IR}(k, \eta)}^{2k \pm 2mv/\hbar} \frac{dq}{q} \frac{1}{2k} \widetilde{g}(z_\pm(q))
    \stackrel{g = 1}{\approx} & \frac{1}{2k} \ln \frac{2k \pm 2mv/\hbar}{q^{\rm IR}(k, \eta)}  \nonumber \\
        \approx & \frac{1}{2k} \ln \frac{\veps_k}{\eta}.
\end{align}
In the small-broadening limit, the latter contribution dominates, yielding
\begin{align} \label{eq:inv_tau_4}
    \gamma_k
    \stackrel{g = 1}{\approx} \frac{C}{k} \ln \frac{\veps_k}{\eta}.
\end{align}

For a sufficiently small $\eta$, the linewidth is peaked at $k = mv/\hbar$.
For smaller $k$, the logarithmic divergence does not occur, and the linewidths are given by the $\eta$-independent value \Eq{eq:inv_tau_k_small}.
For larger $k$, the linewidths are determined by the logarithmic divergence \Eq{eq:inv_tau_4}, but the prefactor becomes smaller as it is inversely proportional to $k$.
To accurately obtain the linewidth at $k = mv/\hbar$, one would need to resolve phonons with wavevector as small as
\begin{equation} \label{eq:supp_q_IR_global}
    q^{\rm IR}\left( k=\frac{mv}{\hbar}, \eta \right) = \frac{\eta}{\hbar v}.
\end{equation}
This quantity provides an estimate of the minimum $\mb{q}$-point spacing near $q=0$ to capture the correct momentum dependence of the linewidth.

\section{Self-consistent linewidths} \label{sec:supp_sc}

\begin{figure}[tb]
  \centering
  \includegraphics[width=0.99\linewidth]{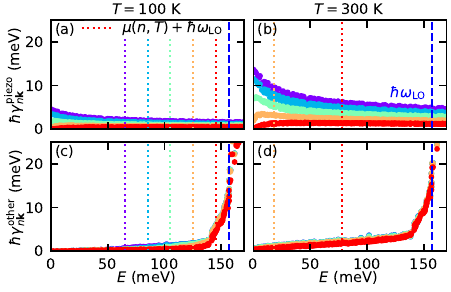}
  \caption{
  Doping-dependent linewidths of c-BN, analogous to Fig.~\ref{fig:lifetime_temperature} of the main text, but calculated with a fixed broadening of 2~meV [\Eq{eq:sc_lifetime_approx1}].
  }
  \label{fig:lifetime_temperature_fixed}
\end{figure}

\begin{figure}[tb]
  \centering
  \includegraphics[width=0.99\linewidth]{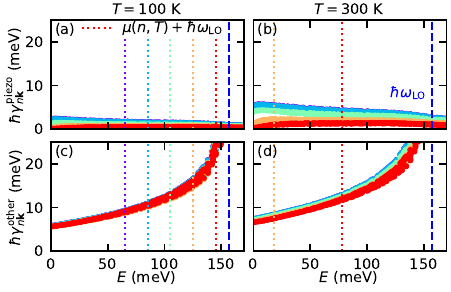}
  \caption{
  Doping-dependent linewidths of c-BN, analogous to Fig.~\ref{fig:lifetime_temperature} of the main text, but 
  calculated using the self-consistent formula with an additional approximation 
  $f(\veps_\nk \pm \hbar\omega_\qnu) \approx f(\veps_\mkq)$ [\Eq{eq:sc_lifetime_approx2}].
  }
  \label{fig:lifetime_temperature_occ_kq}
\end{figure}

In this section, we derive \Eq{eq:gamma_sc} of the main text, the self-consistent equation for the linewidths.
The main idea is to evaluate the Fan--Migdal self-energy in the quasiparticle approximation and extract the linewidth from the on-shell self-energy.
The electron self-energy in the Fan--Migdal approximation reads~\cite{2017Giustino,Ponce2020}
\begin{multline} \label{eq:self_energy}
    \Sigma_\nk^{\gtrless}(\omega) \approx i\hbar \sum_{m,\nu} \iint \frac{d\mb{q}}{\Omega^{\rm BZ}} \frac{d\omega'}{2\pi} \abs{g^{\rm R}_{mn\nu}(\mb{k},\mb{q};\omega')}^2  \\
    \times G^{\gtrless}_\mkq(\omega+\omega') D^{\lessgtr}_\qnu(\omega').
\end{multline}
Here, the superscripts $>$ and $<$ denote the greater and lesser components of the self-energy or Green functions, $g^{\rm R}_{mn\nu}(\mb{k},\mb{q};\omega')$ is the retarded electron-phonon vertex, $G$ the electron Green function, and $D$ the phonon Green function.
We approximate that the electron-phonon vertex is frequency-independent and equal to the static electron-phonon matrix element, as done in most \textit{ab initio} treatments of EPC~\cite{2017Giustino,Ponce2020}:
\begin{equation} \label{eq:e_ph_vertex}
    g^{\rm R}_{mn\nu}(\mb{k},\mb{q};\omega') \approx g_{mn\nu}(\mb{k},\mb{q}).
\end{equation}
In addition, we use the Born--Oppenheimer approximation in which the noninteracting phonon Green function is used for $D$:
\begin{equation} \label{eq:phonon_Green}
    D^{\gtrless}_\qnu(\omega) \approx \frac{-2\pi i}{\hbar} \left[ (n_\qnu + 1) \delta(\omega \mp \omega_\qnu) + n_\qnu \delta(\omega \pm \omega_\qnu) \right].
\end{equation}
Substituting \Eqs{eq:e_ph_vertex} and \eqref{eq:phonon_Green} into \Eq{eq:self_energy} yields
\begin{multline} \label{eq:self_energy_2}
    \Sigma_\nk^{\gtrless}(\omega) \approx \sum_{m,\nu} \intdq  \abs{g_{mn\nu}(\mb{k},\mb{q})}^2  \Big[ (n_\qnu + 1) \\
    \times  G^{\gtrless}_\mkq(\omega \mp \omega_\qnu) + n_\qnu G^{\gtrless}_\mkq(\omega \pm \omega_\qnu) \Big].
\end{multline}
The self-energies are related to the Green function by the Dyson equation~\cite{2017Giustino}.
A complete self-consistent solution of \Eq{eq:self_energy_2} and the Dyson equation requires a self-consistent update of the frequency-dependent electron Green function.

We simplify the problem further by adopting the quasiparticle approximation.
Specifically, we assume that the electron Green function takes the quasiparticle form, where the only variable parameter is the quasiparticle linewidth $\gamma_\nk$~\cite{StefanucciBook}:
\begin{equation}\label{eq:qp_green}
    G^{\gtrless}_\nk(\omega) = \frac{\mp i\hbar\gamma_\nk}{(\hbar\omega - \veps_\nk)^2 + \hbar^2\gamma_\nk^2/4} f^\mp(\hbar\omega),
\end{equation}
where $f^+(\hbar\omega) = 1 / [\exp(\hbar\omega / k_{\rm B} T) -1]$ is the Fermi-Dirac function and $f^- = 1 - f^+$.
The retarded (R) and advanced (A) components in the quasiparticle approximation are given by~\cite{StefanucciBook}
\begin{align}
    G^{\rm R/A}_\nk(\omega) &= \frac{1}{\hbar\omega - \veps_\nk \pm i\hbar\gamma_\nk/2}.
\end{align}
The corresponding self-energies are frequency-independent and proportional to the linewidth~\cite{2017Giustino}:
\begin{equation} \label{eq:qp_self_energy}
    \Sigma^{\rm R,A}_\nk(\omega) \approx \mp i\hbar \gamma_\nk/2.
\end{equation}
We have neglected the shift of the quasiparticle energy since, in the materials investigated here, the dominant impact of phonons on the electron quasiparticle energy is a rigid shift of the band structure with weak mass enhancement.
Such a shift does not affect the quasiparticle lifetime that is the focus of this work.

By substituting \Eq{eq:qp_green} into \Eq{eq:self_energy_2}, we find
\begin{widetext}
\begin{equation}\label{eq:self_energy_3}
\begin{split}
    \Sigma_\nk^{>}(\omega) &\approx -i\hbar \sum_{m,\nu} \intdq  \abs{g_{mn\nu}(\mb{k},\mb{q})}^2
    \!\bigg[ \frac{(n_\qnu + 1)f^-(\hbar\omega-\hbar\omega_\qnu) \gamma_\mkq}{(\hbar\omega \!-\! \veps_\mkq - \hbar\omega_\qnu)^2 + \hbar^2\gamma_\mkq^2/4}
    \!+ \! \frac{n_\qnu f^-(\hbar\omega+\hbar\omega_\qnu) \gamma_\mkq}{(\hbar\omega - \veps_\mkq \!+\! \hbar\omega_\qnu)^2 + \hbar^2\gamma_\mkq^2/4} \bigg], \\
    \Sigma_\nk^{<}(\omega) &\approx i\hbar \sum_{m,\nu} \intdq  \abs{g_{mn\nu}(\mb{k},\mb{q})}^2
    \!\left[ \frac{(n_\qnu + 1) f^+(\hbar\omega+\hbar\omega_\qnu) \gamma_\mkq}{(\hbar\omega - \veps_\mkq + \hbar\omega_\qnu)^2 + \hbar^2\gamma_\mkq^2/4}
    \!+\! \frac{n_\qnu f^+(\hbar\omega - \hbar\omega_\qnu) \gamma_\mkq}{(\hbar \omega \! -\! \veps_\mkq - \hbar\omega_\qnu)^2 + \hbar^2\gamma_\mkq^2/4} \right].
\end{split}
\end{equation}
\end{widetext}
To recover the quasiparticle approximation, we need to extract a scalar, the linewidth $\gamma_\nk$, from these frequency-dependent self-energies.
To do so, we evaluate the difference between the lesser and greater self-energy at the quasiparticle energy $\hbar\omega = \veps_\nk$, and calculate the linewidth as
\begin{align} \label{eq:supp_gamma_from_sigma}
    \gamma_\nk
    &\approx \Im \left[ \Sigma^{\rm A}(\veps_\nk / \hbar) - \Sigma^{\rm R}(\veps_\nk / \hbar) \right] / \hbar \nonumber \\
    &= \Im \left[ \Sigma^{\rm <}(\veps_\nk / \hbar) - \Sigma^{\rm >}(\veps_\nk / \hbar) \right] / \hbar.
\end{align}
Evaluating \Eq{eq:self_energy_3} at $\omega = \veps_\nk / \hbar$ and substituting it into \Eq{eq:supp_gamma_from_sigma} gives
\begin{align} \label{eq:self_consistent_lifetime}
    &\gamma_\nk \approx \sum_{m,\nu} \intdq \abs{g_{mn\nu}(\mb{k},\mb{q})}^2 \nonumber \\
    &\quad \times \bigg[ \frac{[n_\qnu + f^+(\veps_\nk+\hbar\omega_\qnu)] \gamma_\mkq}{(\veps_\nk - \veps_\mkq + \hbar\omega_\qnu)^2 + \hbar^2\gamma_\mkq^2/4} \nonumber \\
    &\qquad + \frac{[n_\qnu + f^-(\veps_\nk - \hbar\omega_\qnu)] \gamma_\mkq}{(\veps_\nk - \veps_\mkq - \hbar\omega_\qnu)^2 + \hbar^2 \gamma_\mkq^2/4} \bigg]
    \nnnl
    &= \frac{2\pi}{\hbar} \sum_{m,\nu} \intdq \abs{g_{mn\nu}(\mb{k},\mb{q})}^2 \sum_\pm [n_\qnu + f^\pm(\veps_\nk \pm \hbar\omega_\qnu)]  \nonumber \\
    &\quad \times  \delta_{\hbar\gamma_\mkq}(\veps_\nk - \veps_\mkq \pm \hbar\omega_\qnu),
\end{align}
which is \Eq{eq:gamma_sc} of the main text.

\Equ{eq:self_consistent_lifetime} gives a practical formula for the self-consistent evaluation of the electron lifetime.
This equation reduces to the linewidth formula for transport calculations (Eqs.~(40, 41) of \citet{Ponce2020}) in the limit $\gamma_\mkq \to 0^+$.

The usual formula for the linewidths with a finite, fixed broadening $\eta$ reads
\begin{align} \label{eq:sc_lifetime_approx1}
    \gamma_\nk 
    &= 2\pi \sum_{m,\nu} \intdq \abs{g_{mn\nu}(\mb{k},\mb{q})}^2 \sum_\pm (n_\qnu + f^\pm_\mkq)  \nonumber \\
    &\quad \times  \delta_{\eta}(\veps_\nk - \veps_\mkq \pm \hbar\omega_\qnu).
\end{align}
A simple generalization of this formula to use a self-consistent broadening gives
\begin{align} \label{eq:sc_lifetime_approx2}
    \gamma_\nk 
    &\stackrel{?}{=} 2\pi \sum_{m,\nu} \intdq \abs{g_{mn\nu}(\mb{k},\mb{q})}^2 \sum_\pm (n_\qnu + f^\pm_\mkq)  \nonumber \\
    &\quad \times  \delta_{\hbar\gamma_\mkq}(\veps_\nk - \veps_\mkq \pm \hbar\omega_\qnu).
\end{align}
However, as we indicate with the `?' sign above the equality, \Eq{eq:sc_lifetime_approx2} is not the correct formula for the self-consistent broadening.
Compared to \Eq{eq:self_consistent_lifetime}, \Eq{eq:sc_lifetime_approx2} contains an additional approximation
$f^\pm(\veps_\nk + \hbar\omega_\qnu) \approx f^\pm(\veps_\mkq)$,
which is valid only if $\gamma_\mkq$ is much smaller than the temperature.
While this approximation is exact in the zero-broadening limit, it leads to a sizable change in the linewidths in the self-consistent calculations.
The effect of this approximation can be seen by comparing Fig.~\ref{fig:lifetime_temperature} and Figs.~\ref{fig:lifetime_temperature_fixed}, \ref{fig:lifetime_temperature_occ_kq}, which are obtained without and with the approximation, respectively.
In particular, after the approximation, the doping dependence of the LO phonon contribution to the linewidths disappears for both fixed-broadening calculations (Figs.~\ref{fig:lifetime_temperature_fixed}(c,d))
and self-consistent calculations (Figs.~\ref{fig:lifetime_temperature_occ_kq}(c,d)).

In addition, at $T=0$, \Eq{eq:self_consistent_lifetime} gives zero linewidths for states at the Fermi surface, $\veps_\nk = \mu$, because $n_\qnu = 0$ and $f^\pm(\veps_\nk \pm \hbar\omega_\qnu) = \Theta(\pm(\mu - (\veps_\nk \pm \hbar\omega_\qnu))) = \Theta(- \hbar\omega_\qnu) = 0$ where $\Theta$ is a step function.
In contrast, the approximate equation \eqref{eq:sc_lifetime_approx2} with $f^\pm(\veps_\mkq)$ instead of $f^\pm(\veps_\nk + \hbar\omega_\qnu)$ breaks this property.

\section{Spectral function} \label{sec:sc_spectral}
To assess the validity of the quasiparticle approximation, we calculate the spectral functions with a full frequency resolution in a partially self-consistent way.
We perform a one-shot calculation of the frequency-dependent self-energy, using the quasiparticle Green functions with the self-consistent linewidths.

To obtain the spectral function, we first compute the retarded self-energy from the lesser self-energy of \Eq{eq:self_consistent_lifetime} using the fluctuation-dissipation theorem~\cite{StefanucciBook}
\begin{equation} \label{eq:sigma_der_Sigma_fdt}
    \mathrm{Im} \Sigma^{\rm R}_\nk(\omega) = -\frac{1}{2f^+(\hbar\omega)} \mathrm{Im} \Sigma^{<}_\nk(\omega).
\end{equation}
With the help of the identities
\begin{gather}
    f^+(\hbar\omega + \hbar\omega_\qnu) \bigl[ 1 + n_\qnu \bigr]
    = f^+(\hbar\omega) \bigl[f^+(\hbar\omega + \hbar\omega_\qnu) + n_\qnu \bigr],
    \nnnl
    f^+(\hbar\omega - \hbar\omega_\qnu) n_\qnu
    = f^+(\hbar\omega) \bigl[f^-(\hbar\omega - \hbar\omega_\qnu) + n_\qnu \bigr],
\end{gather}
\begin{widetext}
\noindent we find
\begin{align} \label{eq:self_energy_R}
    \mathrm{Im} \Sigma_\nk^{\rm R}(\omega)
    &= -\frac{\hbar}{2} \sum_{m,\nu} \intdq  \abs{g_{mn\nu}(\mb{k},\mb{q})}^2
    \Bigl[ \frac{\bigl[f^+(\hbar\omega + \hbar\omega_\qnu) + n_\qnu \bigr] \gamma_\mkq}{(\hbar\omega - \veps_\mkq + \hbar \omega_\qnu)^2 + \hbar^2\gamma_\mkq^2/4} 
    + \frac{\bigl[f^-(\hbar\omega - \hbar\omega_\qnu) + n_\qnu \bigr] \gamma_\mkq}{(\hbar\omega - \veps_\mkq - \hbar\omega_\qnu)^2 + \hbar^2\gamma_\mkq^2/4}  \Bigr]
    \nnnl
    &= -\pi \sum_{m,\nu} \intdq  \abs{g_{mn\nu}(\mb{k},\mb{q})}^2
    \sum_\pm \bigl[f^\pm(\hbar\omega \pm \hbar\omega_\qnu) + n_\qnu \bigr] \delta_{\hbar\gamma_\mkq}(\hbar\omega - \veps_\mkq  \pm \hbar \omega_\qnu).
\end{align}
We note that this equation is consistent with Eq.~(A2) of \citet{Abramovitch2023}.

The real part of the self-energy can be obtained by analytically evaluating the Kramers--Kronig relation
\begin{equation} \label{eq:spectral_selfen_real}
    \Re \Sigma^{\rm R}_\nk(\omega)
    = \mathcal{P} \frac{1}{\pi} \!\int_{-\infty}^{\infty}\! d\omega' \frac{\Im \Sigma^{\rm R}_\nk(\omega')}{\omega' - \omega}.
\end{equation}
We use the Hilbert transform of a Lorentzian
\begin{equation} \label{eq:spectral_delta_hilbert}
    -\mathcal{P} \!\int_{-\infty}^{\infty}\! dz' \, \frac{\delta_{\hbar\gamma}(z' - z_0)}{z' - z}
    = \frac{z - z_0}{(z - z_0)^2 + (\hbar\gamma / 2)^2},
\end{equation}
and the Hilbert transform of a Lorentzian multiplied by a Fermi--Dirac function
\begin{align} \label{eq:spectral_delta_f_hilbert}
    & \hphantom{=} - \mathcal{P} \! \!\int_{-\infty}^{\infty}\! dz' \frac{ f^\pm(z') \delta_{\hbar\gamma}(z' - z_0)}{z' - z}
    \nnnl
    &= \pm \frac{1}{2\pi i \beta} \mathcal{P} \! \!\int_{-\infty}^{\infty}\! dz' \!\! \sum_{n=-\infty}^{\infty} \frac{1}{z' - \mu - i\nu_n} \frac{1}{z' - z} 
    \Bigl( \frac{1}{z' - z_0 - i\hbar\gamma/2} - \frac{1}{z' - z_0 + i\hbar\gamma/2} \Bigr)
    + \frac{1}{2} \frac{z - z_0}{(z - z_0)^2 + (\hbar\gamma / 2)^2}
    \nnnl
    &= \frac{ f^\pm(z) (z - z_0)}{(z - z_0)^2 + (\hbar\gamma/2)^2}
    \mp \frac{1}{\beta} \sum_{n=-\infty}^{-1} \frac{1}{i\nu_n + \mu - z} 
    \frac{1}{i\nu_n +\mu - z_0 - i\hbar\gamma/2}
    \mp \frac{1}{\beta} \sum_{n=0}^{\infty} \frac{1}{i\nu_n + \mu - z} \frac{1}{i\nu_n +\mu - z_0 + i\hbar\gamma/2}
    \nnnl
    &= \frac{ f^\pm(z) (z - z_0)}{(z - z_0)^2 + (\hbar\gamma/2)^2}
    \mp \frac{1}{\beta} \sum_{n=0}^{\infty} \Bigl( \frac{1}{i\nu_n + \mu - z} 
    \frac{1}{i\nu_n + \mu - z_0 + i\hbar\gamma/2}
    + \cc \Bigr)
    \nnnl
    &= \frac{ f^\pm(z) (z - z_0)}{(z - z_0)^2 + (\hbar\gamma/2)^2}
    \mp \Biggl[ \frac{1}{2\pi i} \frac{\psi(\tfrac{1}{2} + \tfrac{\hbar\gamma \beta}{4\pi} - \tfrac{(z_0 - \mu) \beta}{2\pi i}) - \psi(\tfrac{1}{2} - \tfrac{(z - \mu) \beta}{2\pi i})}{z - z_0 + i\hbar\gamma/2}
    + \cc \Biggr].
    \nnnl
    &= \frac{ f^\pm(z) (z - z_0)}{(z - z_0)^2 + (\hbar\gamma/2)^2}
    \mp \frac{1}{\pi} \mathrm{Im}\, \frac{\psi(\tfrac{1}{2} + \tfrac{\hbar\gamma \beta}{4\pi} - \tfrac{(z_0 - \mu) \beta}{2\pi i}) - \psi(\tfrac{1}{2} - \tfrac{(z - \mu) \beta}{2\pi i})}{z - z_0 + i\hbar\gamma/2}
    .
\end{align}
Here, $\beta = 1/(k_{\rm B} T)$, $\cc$ denotes complex conjugation, and $\psi$ is the digamma function
\begin{equation} \label{eq:sc_digamma_def}
    \psi(z) = -\gamma + \sum_{n=1}^{\infty} \Bigl( \frac{1}{n} - \frac{1}{n + z - 1} \Bigr),
\end{equation}
where $\gamma$ is the Euler--Mascheroni constant.
To derive \Eq{eq:spectral_delta_f_hilbert}, in the first equality, we used
\begin{equation}
    f^\pm(z) = \frac{1}{2} \mp \frac{1}{\beta} \sum_{n=-\infty}^{\infty} \frac{1}{z - \mu - i\nu_n}
\end{equation}
and used \Eq{eq:spectral_delta_hilbert}.
In the second equality of \Eq{eq:spectral_delta_f_hilbert}, we performed contour integration along the lower (upper) semicircle for the first (second) term in the parenthesis, so that the poles $z' = i\nu_n$ with $n \leq -1$ ($n \geq 0$) are enclosed by the contour.
Finally, in the last equality, we used the identity
\begin{equation}
    \frac{1}{\beta} \sum_{n=0}^{\infty} \frac{1}{(i\nu_n + a)(i\nu_n + b)}
    = \frac{\beta}{(2\pi i)^2} \sum_{n=0}^{\infty} \frac{1}{(n + \tfrac{1}{2} + a \tfrac{\beta}{2\pi i})(n + \tfrac{1}{2} + b \tfrac{\beta}{2\pi i})}
    = \frac{1}{2\pi i} \frac{\psi(\tfrac{1}{2} + a \tfrac{\beta}{2\pi i}) - \psi(\tfrac{1}{2} + b \tfrac{\beta}{2\pi i})}{a-b}\,,
\end{equation}
which has been derived from
\begin{equation}
    \sum_{n=0}^{\infty} \frac{1}{(n+a)(n+b)} = \frac{\psi(a) - \psi(b)}{a-b}\,,
\end{equation}
which follows directly from \Eq{eq:sc_digamma_def}.

By substituting \Eq{eq:self_energy_R} into \Eq{eq:spectral_selfen_real}
and using \Eqs{eq:spectral_delta_hilbert} and \eqref{eq:spectral_delta_f_hilbert}, we find
\begin{align} \label{eq:self_energy_R_re}
    \Re \Sigma^{\rm R}_\nk(\omega)
    &= - \sum_{m,\nu} \intdq  \abs{g_{mn\nu}(\mb{k},\mb{q})}^2
    \sum_\pm \mathcal{P} \int_{-\infty}^{\infty} d\omega'
    \frac{\bigl[f^\pm(\hbar\omega' \pm \hbar\omega_\qnu) + n_\qnu \bigr] \delta_{\hbar\gamma_\mkq}(\hbar\omega' - \veps_\mkq  \pm \hbar \omega_\qnu)}{\omega' - \omega}
    \nnnl
    &= - \sum_{m,\nu} \intdq  \abs{g_{mn\nu}(\mb{k},\mb{q})}^2
    \sum_\pm \mathcal{P} \int_{-\infty}^{\infty} d\hbar\omega'
    \frac{\bigl[f^\pm(\hbar\omega') + n_\qnu \bigr] \delta_{\hbar\gamma_\mkq}(\hbar\omega' - \veps_\mkq)}{\hbar\omega' - (\hbar\omega \pm \hbar\omega_\qnu)}
    \nnnl
    &= \sum_{m,\nu} \intdq  \abs{g_{mn\nu}(\mb{k},\mb{q})}^2 \sum_\pm
    \Biggl\{
    \bigl[ f^\pm(\hbar\omega \pm \hbar\omega_\qnu) + n_\qnu \bigr]
    \frac{\hbar\omega - \veps_\mkq  \pm \hbar \omega_\qnu}{(\hbar\omega - \veps_\mkq  \pm \hbar \omega_\qnu)^2 + (\hbar \gamma_\mkq / 2)^2}
    \nnnl
    &\hspace{13em} \mp \mathrm{Im}\,
    \frac{\psi\bigl( \tfrac{1}{2} + \tfrac{\hbar\gamma_\mkq\beta}{4\pi} - \frac{(\veps_\mkq - \mu) \beta}{2\pi i} \bigr) - \psi\bigl(\tfrac{1}{2} - \tfrac{(\hbar\omega - \mu \pm \hbar\omega_\qnu) \beta}{2\pi i} \bigr) }
    {\hbar\omega - \veps_\mkq \pm \hbar\omega_\qnu + i \hbar \gamma_\mkq / 2} \Biggr\}
\end{align}
To take into account the effect of a rigid shift of the bands, which does not affect the self-energy and the spectral function when considered self-consistently, we rigidly shift the real part of the self-energy at all $\mb{k}$ points by the same amount $-\Re \Sigma^{\rm R}_{\rm CBM}(0)$.
\end{widetext}

Using the frequency-dependent self-energy, we calculate the spectral function, which we call the ``full-frequency'' (FF) spectral function:
\begin{align} \label{eq:spectral_specfun}
    &A_\nk^{\rm FF}(\omega) = -\frac{1}{\pi} \Im \frac{1}{\hbar \omega - \veps_\nk - \Sigma^{\rm R}_\nk(\omega) + \Re \Sigma^{\rm R}_{\rm CBM}(0)} \nonumber \\
    &= \frac{-\Im \Sigma^{\rm R}_\nk(\omega) / \pi}{[\hbar \omega \!-\! \veps_\nk \!-\! \Re \Sigma^{\rm R}_\nk(\omega) \!+\! \Re \Sigma^{\rm R}_{\rm CBM}(0)]^2 \!+\! [\Im \Sigma^{\rm R}_\nk(\omega)]^2}.
\end{align}
The spectral function in the quasiparticle (QP) approxmation reads
\begin{align}
    A_\nk^{\rm QP}(\omega)
    = \frac{\hbar\gamma_\nk / 2\pi}{(\hbar \omega - \veps_\nk)^2 + \hbar^2\gamma_\nk^2/4}
    = \delta_{\hbar\gamma_\nk}(\hbar\omega - \veps_\nk).
\end{align}
The quasiparticle spectral function can be derived from the full-frequency spectral function by applying two approximations, (i) neglecting the real part of the self-energy, $\Sigma^{\rm R}_\nk(\omega) - \Re \Sigma^{\rm R}_{\rm CBM}(0) = 0$, and (ii) evaluating the self-energy at the bare energy, $\Im \Sigma(\omega) = \Im \Sigma(\veps_\nk)$.
We assess the validity of these approximations by comparing the quasiparticle and full-frequency spectral functions.

\begin{figure}[tb]
    \centering
    \includegraphics[width=0.99\linewidth]{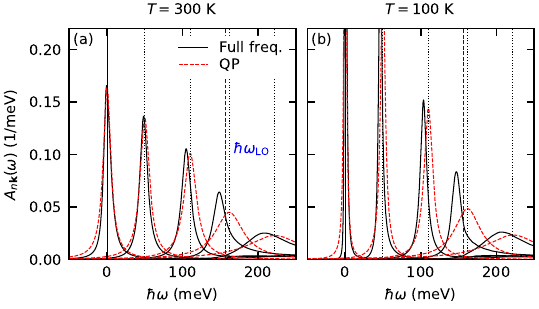}
    \caption{
    Spectral function of c-BN at $n=10^{13}~\mathrm{cm}^{-3}$ and (a) $T=300~\mathrm{K}$ and (b) $T=100~\mathrm{K}$
    for a few $\mb{k}$ points along $\Gamma$X ($|\mb{k}| / |\mb{k}_\mathrm{X}|$ = 1, 0.94, 0.91, 0.89, and 0.87, from left to right), computed using the quasiparticle approximation (QP) or the full frequency-dependent self-energy.
    The vertical black dotted lines denote the bare electron energies.
    The blue vertical dashed line denotes the LO phonon energy.
    $\omega = 0$ is set at the CBM energy.
    }
    \label{fig:spectral_function_cBN}
\end{figure}

Figure~\ref{fig:spectral_function_cBN} shows the calculated spectral functions for cBN.
At both $T=300~$K and $T=100~$K, the quasiparticle spectral function remains a good approximation of the full-frequency spectral function.
The strongest deviation is found for states at energy $\veps_\nk \approx \hbar\omega_{\rm LO}$, for which a resonant LO phonon emission to the conduction band minimum~\cite{2022Kandolf}.
Similar results are found for Si, PbTe, and NaCl as shown in Fig.~\ref{fig:lifetime_Si_PbTe}.

\section{Hybrid integration method for calculating the linewidths} \label{sec:supp_hybrid}

\begin{figure}[t]
  \centering
  \includegraphics[width=0.99\linewidth]{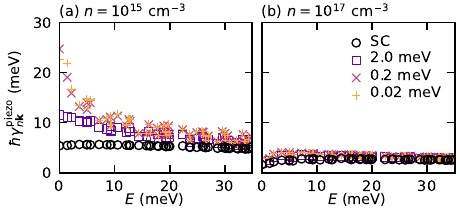}
  \caption{Piezoelectric contribution to the linewidth of c-BN at $T=300$~K at different doping levels.
  }
  \label{fig:lifetime_screen_piezo_only}
\end{figure}

To fully converge the $\mb{q}$-point integration in the calculation of the linewidths for a tiny broadening as small as 0.002~meV, we use a hybrid adaptive-uniform integration method.
We treat the near-divergent piezoelectric contribution using a long-range model with parameters derived from \textit{ab initio} calculations,
and compute the remainder using standard \textit{ab initio} Wannier interpolation on a regular $\mb{q}$-point grid.
Below, we discuss the details of the piezoelectric model and the hybrid integration method.

The parameters of the long-wavelength model are the electron effective mass $m^{\rm el}_{\alpha\beta}$, the short-circuit elastic tensor $C_{\alpha\beta\gamma\delta}$, the piezoelectric tensor $e_{\alpha\beta\gamma}$,
electronic dielectric matrix $\epsilon^{\rm \infty}_{\alpha\beta}$, the $\mb{q}=\mb{0}$ dynamical matrix $\Phi^{{\rm AN},(0)}_{\kappa\alpha,\kappa'\beta}$,
Born effective charge $Z_{\kappa\alpha\beta}$, and the total atomic mass per unit cell $M$.
The superscript `AN' in $\Phi^{{\rm AN},(0)}_{\kappa\alpha,\kappa'\beta}$ indicates this term corresponds to the analytical part of the dynamical matrix for $\mb{q} \neq \mb{0}$.
The dynamical matrix $\Phi^{{\rm AN},(0)}$ has three zero eigenvalues.
We write its Moore--Penrose pseudoinverse with a tilde as $\widetilde{\Phi}^{{\rm AN},(0)}$.
These parameters are computed using \textit{ab initio} density functional perturbation theory or finite-difference calculations, as detailed in Sec.~\ref{sec:supp_comp_piezo}.

\begin{figure*}[htb]
  \centering
  \includegraphics[width=0.99\linewidth]{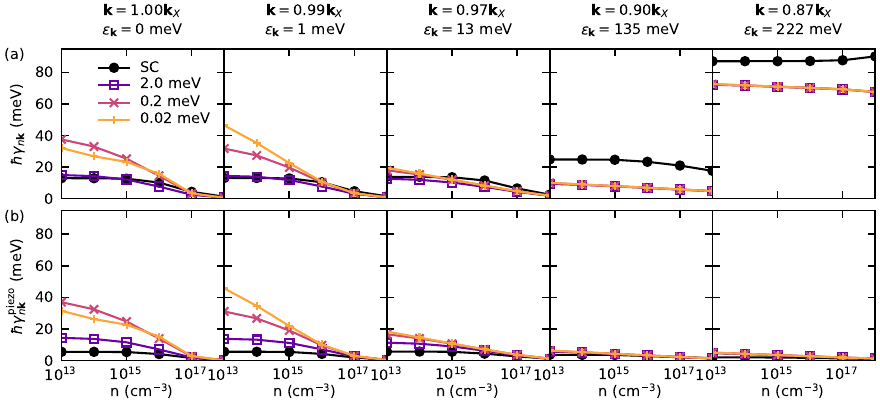}
  \caption{
  (a) Doping dependence of the linewidth of c-BN for $\mb{k}$ points along $\Gamma \mathrm{X}$.
  (b) Same as (a), but for the piezoelectric contribution only.
  }
  \label{fig:lifetime_screen_various_k}
\end{figure*}

The electrons are assumed to have an anisotropic parabolic dispersion in the long-range limit.
Intervalley electron-phonon scatterings are not included in the long-wavelength model.
For the phonons, we only consider the acoustic phonon branch.
The phonon frequencies and eigenmodes are calculated by solving the eigenvalue problem~\cite{Royo2020}
\begin{equation}\label{eq:secularEq}
    \sum_\beta (q^2 K_{\alpha\beta}(\hatq) - M \omega_{\qnu}^2 \delta_{\alpha\beta}) u_{\beta, \qnu} = 0,
\end{equation}
where
\begin{subequations}
\begin{align}
    K_{\alpha\beta}(\hatq) &= V \sum_{\gamma\delta} \hat{q}_\gamma \hat{q}_\delta C_{\alpha\gamma\beta\delta} + 4\pi V \frac{e_\alpha(\hatq) e_\beta(\hatq)}{\epsilon(\hatq)},
    \\
    e_\alpha(\hatq) &= \sum_{\gamma\delta} e_{\gamma\delta\alpha} \hatq_\gamma \hatq_\delta,
    \\
    \epsilon(\hatq) &= \sum_{\alpha\beta} \hatq_\alpha \hatq_\beta \epsilon^{\rm \infty}_{\alpha\beta}  + \frac{4\pi}{V} \sum_{\alpha\beta\kappa\kappa'} Z_{\kappa\alpha}(\hatq) \widetilde{\Phi}^{{\rm AN},(0)}_{\kappa\alpha,\kappa'\beta} Z_{\kappa'\beta}(\hatq),
    \\
    Z_{\kappa\alpha}(\hatq) &= \frac{1}{e} \sum_\gamma Z_{\kappa\alpha\gamma} \hatq_\gamma.
\end{align}
\end{subequations}
Here, $\epsilon(\hatq)$ is the static dielectric matrix which includes both the electronic contribution (first term) and the lattice contribution (second term).
We normalize the eigenmodes to satisfy
\begin{equation}
    \sum_\alpha u_{\alpha,\qnu} u_{\alpha,\mb{q}\nu'} = \frac{1}{M} \delta_{\nu,\nu'}.
\end{equation}
Finally, the electron-phonon matrix element of this long-range model reads~\cite{Ridley1982}:
\begin{equation}
    g_\nu(\mb{k},\mb{q})  = - \sum_\alpha \sqrt{\frac{\hbar}{2\omega_\qnu}} \frac{ 4\pi e(\hatq)_\alpha u_{\alpha,\qnu}}{\epsilon(\hatq)}
\end{equation}

Now, let us discuss how the linewidths are calculated.
In the usual regular grid method, one replaces the integral in \Eq{eq:gamma}, $\int \frac{d\mb{q}}{\Omega^{\rm BZ}}$, with a discrete summation $\frac{1}{N_q} \sum_\mb{q}$ to find
\begin{align} \label{eq:supp_gamma_grid}
    \gamma_\nk &=  \frac{1}{N_q} \sum_{\mathbf{q},m,\nu} \gamma_{\nk;m\qnu},\\
     \gamma_{\nk;m\qnu} &= \frac{2\pi}{\hbar} |g_{mn\nu}(\mb{k,q})|^2 \sum_\pm (n_{\qnu}  +   f^\pm_{\mkq}) \nonumber \\ 
     &\quad \times \delta(\veps_{n\mathbf{k}} - \veps_{\mkq}  \pm  \hbar \omega_{\qnu} ). \label{eq:supp_gamma_qnu}
\end{align}

In our hybrid method, we start by writing the linewidth as the sum of the piezoelectric contribution and the remainder:
\begin{equation}
    \gamma_\nk = \gamma_\nk^{\rm piezo} + \gamma_\nk^{\rm other}.
\end{equation}
where $\gamma_\nk^{\rm piezo}$ is the linewidth for an anisotropic long-wavelength model with piezoelectric EPC.
We evaluate the piezoelectric contribution in spherical coordinates:
\begin{equation} \label{eq:supp_hybrid_piezo_spherical}
    \gamma^{\rm piezo}_\nk = \int_0^{q_{\rm cut}} dq q^2 \int_0^\pi d\theta \sin\theta \int_0^{2\pi} d\phi\, \gamma^{\rm piezo}_{\nk;m\qnu}\,
\end{equation}
where $\mb{q} = (q\sin\theta\cos\phi, q\sin\theta\sin\phi, q\cos\theta)$.
$\gamma^{\rm piezo}_{\nk;m\qnu}$ is the partial scattering rate computed using the \textit{ab initio} long-wavelength model for piezoelectric EPC  and hence can be evaluated very efficiently.
We use an upper bound $q_{\rm cut} = 0.3~\mathrm{bohr}$ for the piezoelectric contribution, as the long-wavelength model is valid only for small $q$.
For the angular integration, we use the h-adaptive integration~\cite{1980GenzAdaptive} implemented in the \texttt{Hcubature.jl} package~\cite{JuliaHCubature}.
For the radial integration, we use the one-dimensional adaptive Gauss--Kronrod quadrature method from the \texttt{QuadGK.jl} package~\cite{JuliaQuadGK}.
To deal efficiently with the $1/q$ behavior of the integrand, we change the variable as $q = e^x$:
\begin{equation}
    \int_{0}^{q_{\rm cut}} dq f(q)
    = \int_{-\infty}^{\ln q_{\rm cut}} dx e^x f(e^x).
\end{equation}
Then, one can sample $\mb{q}$ points near $q=0$ very densely and thus better resolve the logarithmic divergence.
The remainder is calculated on a regular grid:
\begin{multline}
    \gamma^{\rm other}_\nk = \\
    \frac{1}{N_q} \sum_{\mathbf{q},m,\nu} \Big[ \gamma_{\nk;m\qnu}
    - \gamma^{\rm piezo}_{\nk;m\qnu} \Theta(q^{\rm cut} - q) \Big].
\end{multline}
The partial scattering rate $\gamma_{\nk;m\qnu}$ is calculated using \textit{ab initio} Wannier interpolation.
The piezoelectric contribution that is already taken into account in \Eq{eq:supp_hybrid_piezo_spherical} is subtracted using the same regular grid to avoid double counting.

\begin{figure}[tb]
  \centering
  \includegraphics[width=0.99\linewidth]{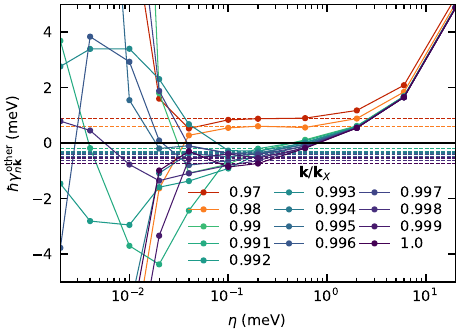}
  \caption{
  Broadening dependence of $\gamma^{\rm other}_\nk$ computed using a $800^3$ $\mb{q}$-point grid.
  The horizontal lines indicate the linewidth at $\eta = 0.1~$meV, which we take to approximate those at $\eta < 0.1$~meV (see Sec.~\ref{sec:supp_hybrid} for details).
  We note that negative contributions may arise due to the subtraction of the piezoelectric contribution, which is calculated using a long-range model.
  The total linewidth, $\gamma^{\rm piezo}_\nk + \gamma^{\rm other}_\nk$, is always positive for any $\eta$ when converged with respect to the Brillouin-zone sampling.
  }
  \label{fig:conv_other_broadening}
\end{figure}

For fixed-broadening calculations, we exploit the fact that $\gamma^{\rm other}_\nk$ converges smoothly in $\eta$ and use the value obtained at $\eta = 0.1~\mathrm{meV}$ to approximate that at a smaller $\eta$.
We find that the error resulting from this approximation is less than 2~meV (see Fig.~\ref{fig:conv_other_broadening}).
This approximation is not used for $\gamma^{\rm piezo}_\nk$ because it shows the logarithmic divergence.

Figure~\ref{fig:lifetime_screen_piezo_only} shows the linewidths calculated using only the piezoelectric model, without the non-piezoelectric contributions.
Figure~\ref{fig:lifetime_screen_various_k} shows the density dependence of the total and piezoelectric linewidths at various $\mb{k}$ points.
The piezoelectric contribution dominates the linewidths near the CBM and becomes negligible away from the CBM.

\begin{figure}[tb]
  \centering
  \includegraphics[width=0.99\linewidth]{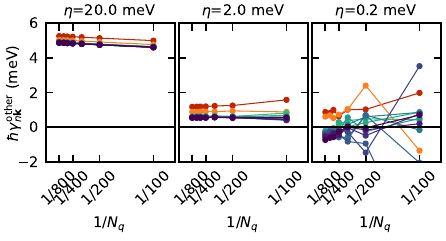}
  \caption{
    Convergence of the non-piezoelectric linewidth $\gamma^{\rm other}_\nk$ at fixed broadening.
    Colors indicate different $\mb{k}$ points as in Fig.~\ref{fig:conv_other_broadening}.
  }
  \label{fig:conv_other_nk}
\end{figure}

Since we converged the adaptive integration to a relative error tolerance of 1~\% by construction, the convergence of the hybrid adaptive-uniform method is controlled by the broadening parameter and the $\mb{q}$-point grid used to compute the remainder, $\gamma^{\rm other}_\nk$.
Figure~\ref{fig:conv_other_nk} shows that for $\eta \geq 0.1~$meV, a $800^3$ grid gives $\gamma^{\rm other}_\nk$ converged within 1~meV.
We approximate $\gamma^{\rm other}_\nk$ for $\eta < 0.1~$meV with those computed at $\eta = 0.1$~meV, because smaller $\eta$ requires a much denser $\mb{q}$-point grid to converge, while the $\eta$-dependence is weak once converged, as shown in Fig.~\ref{fig:conv_other_broadening}.

\begin{figure}[tb]
  \centering
  \includegraphics[width=0.99\linewidth]{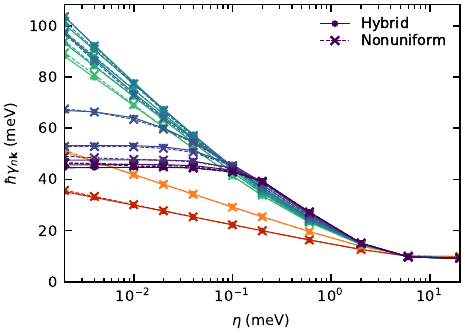}
  \caption{
  Comparison of the electron linewidths of c-BN at $T=300~$K computed using the hybrid adaptive-uniform integration method and the brute-force \textit{ab initio} calculation with cubic nonuniform grids.
  Colors indicate different $\mb{k}$ points as in Fig.~\ref{fig:conv_other_broadening}.
  }
  \label{fig:conv_nonuniform_compare}
\end{figure}

\begin{figure*}[tb]
  \centering
  \includegraphics[width=0.99\linewidth]{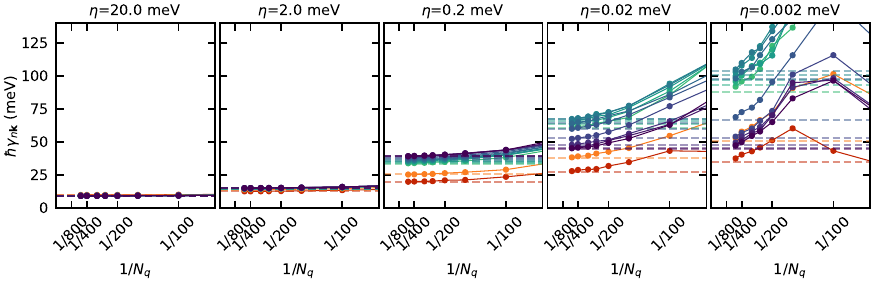}
  \caption{
  Convergence of the linewidth at fixed broadening calculated using cubic nonuniform grids.
  The horizontal dashed lines indicate the linewidths calculated using the hybrid adaptive-uniform method.
  Colors indicate different $\mb{k}$ points as in Fig.~\ref{fig:conv_other_broadening}.
  }
  \label{fig:conv_nonuniform_nk}
\end{figure*}

Finally, to validate our long-wavelength model for the piezoelectric EPC and the hybrid integration method, we compare it with a brute-force calculation of the linewidths using standard \textit{ab initio} Wannier interpolation.
Since using a uniform $\mb{q}$-point grid would require an impractically high grid density, we use a cubically-scaled nonuniform grid.
The aim is to more densely sample the $\mb{q}$ points near $\Gamma$ than those away from $\Gamma$ so that the large contribution from the piezoelectric EPC at small $q$ can be captured with less total number of $\mb{q}$ points.
For a given grid size $N=2N'$, we define a one-dimensional, cubically-scaled nonuniform grid $x_1 < \cdots < x_N$ in $[-0.5, 0.5]$:
\begin{equation}
	x_{N'+i} = \tfrac{1}{2} (i / N')^3
	\quad (i = -N'+1, \cdots, N')
\end{equation}
Then, we assign a weight to each point as
\begin{equation}
	w_i = \frac{x_{i+1} - x_{i-1}}{2},
\end{equation}
where we define $x_0 = x_N - 1$ and $x_{N+1} = x_1 + 1$ according to the periodic boundary condition.
The weights satisfy
\begin{equation}
	\sum_{i=1}^{N} w_i = \frac{x_{N+1} + x_N - x_1 - x_0}{2} = 1.
\end{equation}
Finally, we form a three-dimensional product of the nonuniform grid and use it to sample the Brillouin zone in the crystal (reduced) coordinates:
\begin{multline}
	\int \frac{d\mb{q}}{\Omega^{\rm BZ}} 	f(\mb{q}) \\
	\approx \sum_{i=1}^N \sum_{j=1}^N \sum_{k=1}^N w_{i} w_{j} w_{k} f(\mb{b}_1 x_{i} + \mb{b}_2 x_{j} + \mb{b}_3 x_{k}),
\end{multline}
where $\mb{b}_i$ is the $i$-th reciprocal lattice vector.

Figure~\ref{fig:conv_nonuniform_compare} shows that the brute-force calculation using the nonuniform grid agree well with the results of the hybrid method.
The small disagreement at $\eta = 2\times10^{-3}~$meV is attributed to the underconvergence of the brute-force calculation (Fig.~\ref{fig:conv_nonuniform_nk})
and to the approximation of $\gamma^{\rm other}_\nk$ with that calculated at $\eta = 0.2~\mathrm{meV}$.
Yet, the disagreement does not alter the qualitative features of the linewidths, in particular the presence or absence of the logarithic divergence.

\section{Piezoelectric EPC from dipoles and quadrupoles} \label{sec:supp_piezo_epc}
In this section, we derive the electron-phonon coupling constant for acoustic phonons in the $\mb{q}\to\mb{0}$ limit in terms of the dipoles and quadrupoles.
We can understand the varying impact of piezoelectricity in different materials by separating the piezoelectric term into two contributions: the clamped-ion term, $e^{\rm CI}$, and the internal-strain term, $e^{\rm IS}$~\cite{1994DalCorsoPiezo,1997BernardiniPiezo,1998SaghiSzaboPiezo}.
The former comes from a purely electronic response to strain, and the latter comes from lattice relaxation.
In most materials, the two contributions have opposite signs and similar magnitudes, so that they partially cancel each other out.
In \textit{ab initio} Wannier interpolation, these contributions are included by the long-range dipolar and quadrupolar potentials of the phonons.
Using a long-wavelength analysis~\cite{Stengel2013}, we find here that the calculations including only the dipole contributions~\cite{2015VerdiFrohlich} give a piezoelectric EPC proportional to the internal-strain term $e^{\rm IS}$, and that one recovers the EPC proportional to the total piezoelectric tensor by including the quadrupole contributions~\cite{2020BruninPiezo1,2020BruninPiezo2,2020JhalaniPiezo,2020ParkPiezo,2021GanoseAMSET,2023PoncePiezo1,2023PoncePiezo2}.

In most materials, the internal-strain contribution is larger than the clamped-ion one.
Therefore, including quadrupoles decreases the piezoelectric EPC $\abs{e^{\rm IS} + e^{\rm CI}} < \abs{e^{\rm IS}}$.
In contrast, the piezoelectricity of c-BN is dominated by the clamped-ion term, and the inclusion of the quadrupole reverses the sign and increases the EPC $\abs{e^{\rm IS} + e^{\rm CI}} > \abs{e^{\rm IS}}$.
This difference explains why neglecting the quadrupole term leads to an overestimation of the EPC and an underestimation of the mobility in many materials such as GaN~\cite{2020JhalaniPiezo}, GaP~\cite{2020BruninPiezo2,Ponce2021}, 3C-SiC, AlP, AlSb~\cite{Ponce2021}, PbTiO3~\cite{2020ParkPiezo}, and the electron mobility of GaAs~\cite{2020BruninPiezo2}, while leading to an overestimation of the mobility in c-BN~\cite{Ponce2021}.

\begin{figure}[b]
  \centering
  \includegraphics[width=0.99\linewidth]{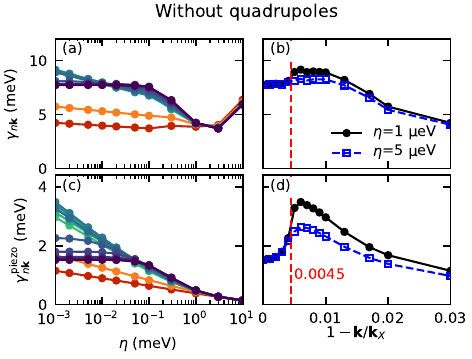}
  \caption{
      (a) Scattering rate in the conduction band of intrinsic c-BN computed using Wannier interpolation without the quadrupolar contribution to the EPC and dynamical matrix, at $T=300$~K, and as a function of broadening for $k$ points along the $\Gamma \mathrm{X}$ line.
      Different colors indicate different $k$ points in the same scheme as Fig.~\ref{fig:lifetime_fixed}(c).
      Compared to Fig.~\ref{fig:lifetime_fixed}(c), the lifetimes are significantly underestimated due to the neglect of the quadrupole contribution to EPC.
      (b) Scattering rate as a function of the wavevector for two broadening values.
      The red dashed line indicates the onset of the divergence.
      (c, d) Same as (a, b), but for the piezoelectric contribution to the scattering rate.
  }
  \label{fig:lifetime_fixed_noquadrupole}
\end{figure}

Figure~\ref{fig:lifetime_fixed_noquadrupole} shows the scattering rates of c-BN calculated without the quadrupolar contribution to the EPC and the dynamical matrix.
The scattering rates are much smaller than those shown in Fig.~\ref{fig:lifetime_fixed} which are computed with the quadrupoles, approximately by a factor of $\abs{e^{\rm IS} / (e^{\rm IS} + e^{\rm CI})}^2 \approx 0.03$ (Sec.~\ref{sec:supp_comp_piezo}).
This result demonstrates the importance of including the quadrupolar contribution in describing the EPC of c-BN.

The scattering rates calculated without quadrupoles still display the momentum-dependent divergence in the $\eta \to 0$ limit, which becomes more evident when plotting only the contribution of the piezoelectric EPC (Fig.~\ref{fig:lifetime_fixed_noquadrupole}(c,d)).
The onset of the divergence does not change as it is determined only by the electron and phonon band structure, which are hardly affected by the inclusion of the quadrupoles.

\subsection{Phonon eigenmode for analytic force constants}
Let us first study the phonon eigenmodes including only the analytic long-range part of the force-constant matrix.
We follow Sec.~IV of Ref.~\cite{Stengel2013} where they use the short-circuit electrical boundary conditions, which is equivalent to taking the analytic part of the dynamical matrix.

The phonon eigenvalue equation is
\begin{equation}
    M_\kappa \omega_\mb{q}^2 U_{\ka}(\mb{q}) = \sum_{\kappa'\beta} \Phi_{\ka,\kpb}^{{\rm AN},\mb{q}} U_{\kpb}(\mb{q}),
\end{equation}
where we consider here only acoustic modes. 
We define the Taylor expansion of the force-constant matrix as
\begin{equation}
    \Phi_{\ka,\kpb}^{{\rm AN},\mb{q}} = \Phi_{\ka,\kpb}^{{\rm AN},(0)} - iq_\gamma \Phi_{\ka,\kpb}^{{\rm AN},(1,\gamma)} + \mcO(q^2),
\end{equation}
where we note that $\Phi^{\rm AN}$ is analytic in $\mb{q}$ and can be Taylor expanded.
For a given direction $\hatq$ we write the phonon eigenmode at $\lambda\hatq$ as a Taylor expansion of $\lambda$:
\begin{equation}
    U_{\ka}^{\rm AN}(\lambda\hatq) = U_{\ka}^{{\rm AN},(0)}(\hatq) + i\lambda U_{\ka}^{{\rm AN},(1)}(\hatq) + \mcO(\lambda^2).
\end{equation}

At $\mcO(\lambda^0)$ the phonon eigenvector should be a uniform translation:
\begin{equation} \label{eq:ph_mode_AN_0}
    U_{\ka}^{{\rm AN},(0,\hatq)} = U_\alpha.
\end{equation}
For $\mcO(\lambda^1)$, using first-order perturbation theory, we find
\begin{equation}
    U_{\ka}^{{\rm AN},(1)}(\hatq) = \sum_{\beta\gamma} \Gamma^{{\rm AN},\kappa}_{\alpha\beta\gamma} U_\beta \hat{q}_\gamma,
\end{equation}
where
\begin{align}
    \Gamma^{{\rm AN},\kappa}_{\alpha\beta\gamma} \equiv & \sum_{\kappa'\delta} \widetilde{\Phi}^{{\rm AN},(0)}_{\ka,\kappa'\delta} \Lambda^{{\rm AN},\kappa'}_{\delta\beta\gamma} \\
     \Lambda^{{\rm AN},\kappa}_{\alpha\beta\gamma} \equiv & \sum_{\kappa'} \Phi^{{\rm AN},(1,\gamma)}_{\ka,\kpb},
\end{align}
and $\widetilde{\Phi}^{{\rm AN},(0)}$ is the pseudoinverse of $\Phi^{{\rm AN},(0)}$.

\subsection{Phonon eigenmode for the full force-constant matrix}
Now, we include the nonanalytic part following the supplemental material of Ref.~\cite{Royo2020}.
For the phonon momentum $\lambda\hatq$, the Taylor expansion of the force-constant matrix is
\begin{equation} \label{eq:ph_full_dyn0}
    \Phi^{(0)}_{\ka,\kpb}(\hatq) = \Phi^{{\rm AN},(0)}_{\ka,\kpb} + \frac{4\pi}{\Omega} \frac{Z_\ka(\hatq) Z_\kpb(\hatq)}{\bar{\epsilon}(\hatq)}
\end{equation}
where we defined
\begin{align}
    \bar{\epsilon}(\hatq) \equiv & \sum_{\alpha\beta} \hat{q}_\alpha \bar{\epsilon}_{\alpha\beta} \hat{q}_\beta \\
    Z_\ka(\hatq) \equiv & \sum_{\beta}  Z^*_{\kappa\alpha\beta} \hat{q}_\beta,
\end{align}
where $\bar{\epsilon}$ is the electronic (clamped-ion) dielectric tensor and $Z^*$ the Born effective charge tensor.
The eigenvectors for the acoustic phonons are still uniform translations:
\begin{equation}
    U_{\ka}^{(0)}(\hatq) = U_\alpha.
\end{equation}
This result can be easily derived as follows: 
\begin{multline}
    \sum_{\kappa'\beta} \Phi^{(0)}_{\ka,\kpb}(\hatq) U_\beta
    = \sum_{\kappa'\beta} \Phi^{{\rm AN},(0)}_{\ka,\kpb}(\hatq) U_\beta \\
    + \sum_{\kappa'\beta} \frac{4\pi}{\Omega} \frac{Z_\ka(\hatq) Z_\kpb(\hatq)}{\bar{\epsilon}(\hatq)} U_\beta
    = 0,
\end{multline}
where we have used \Eq{eq:ph_mode_AN_0} and the charge neutrality condition
\begin{equation}
    \sum_{\kappa'\beta} Z_\kpb (\hatq)
    = \sum_{\kappa'\alpha\beta} Z^*_{\kappa'\alpha\beta} \hat{q}_\beta
    = 0.
\end{equation}

For $\mcO(\lambda^1)$, first-order perturbation theory gives
\begin{align}
    U_{\ka}^{(1)}(\hatq) =& \sum_\beta \Gamma_{\kappa\alpha\beta}(\hatq) U_\beta, \\
    \Gamma^{\kappa}_{\alpha\beta}(\hatq) \equiv & \sum_{\kappa'\gamma} \widetilde{\Phi}^{(0)}_{\ka,\kappa'\gamma}(\hatq)  \Lambda^{\kappa'}_{\gamma\beta}(\hatq), \\
    \Lambda^{\kappa}_{\alpha\beta}(\hatq) \equiv & \sum_{\kappa'} \Phi^{(1)}_{\ka,\kpb}(\hatq) \nonumber \\
    =& \sum_\gamma \Lambda^{{\rm AN},\kappa}_{\alpha\beta\gamma} \hat{q}_\gamma 
    - 4\pi \frac{Z_{\kappa\alpha}(\hatq) \bar{e}_\beta(\hatq)}{\bar{\epsilon}(\hatq)},    \label{eq:ph_full_Lambda}   \\ 
\bar{e}_\beta(\hatq) \equiv & \sum_{\alpha\gamma} \bar{e}_{\alpha\beta\gamma} \hatq_\alpha \hatq_\gamma.       
\end{align}
We note that the piezoelectric tensors $e$ and $\bar{e}$ are symmetric in the exchange of the second and third indices.
From the Sherman--Morrison formula, $\widetilde{\Phi}^{(0,\hatq)}$, the pseudoinverse of $\widetilde{\Phi}^{(0),\hatq}$,  \Eq{eq:ph_full_dyn0}, becomes
\begin{multline}
    \widetilde{\Phi}^{(0)}_{\ka,\kpb}(\hatq) = \widetilde{\Phi}^{{\rm AN},(0)}_{\ka,\kpb} \\
    - \frac{4\pi}{\Omega}  \sum_{\lambda\gamma\lambda'\gamma'} \frac{\widetilde{\Phi}^{{\rm AN},(0)}_{\ka,\lambda\gamma} Z_{\lambda\gamma}(\hatq) Z_{\lambda'\gamma'}(\hatq) \widetilde{\Phi}^{{\rm AN},(0)}_{\lambda'\gamma',\kpb}}{\epsilon^{\rm s}(\hatq)},
\end{multline}
where $\epsilon^{\rm s}(\hatq)$ is the static dielectric tensor
\begin{equation}
    \epsilon^{\rm s}(\hatq) = \bar{\epsilon}(\hatq) + \frac{4\pi}{\Omega} \sum_{\kappa\kappa'\alpha\beta} Z_{\ka}(\hatq) \widetilde{\Phi}^{{\rm AN},(0)}_{\ka,\kpb} Z_{\kappa'\beta}(\hat{q}).
\end{equation}

\subsection{Electron-phonon coupling without quadrupoles}
We now calculate the electron-phonon coupling constants.
First, we neglect the quadrupole contributions to the EPC and to the dynamical matrix.
Then, the $\Lambda$ tensor [\Eq{eq:ph_full_Lambda}], which appears in the eigenmode, is rewritten as
\begin{equation} \label{eq:ph_noQ_Lambda}
    \Lambda^{{\rm noQ},\kappa}_{\alpha\beta}(\hatq)
    = \sum_\gamma \Lambda^{{\rm AN},\kappa}_{\alpha\beta\gamma} \hat{q}_\gamma.
\end{equation}

The $\mcO(\lambda^0)$ term of the long-wavelength dipole potential is (Eq.~(24) of Ref.~\cite{2020BruninPiezo2})
\begin{align}
    V^{{\rm D},(0)}(\mb{r}; \hatq) =& \frac{4\pi}{\Omega} \sum_{\kappa\alpha} \frac{i\lambda Z_{\ka}(\hatq)}{\lambda^2 \bar{\epsilon}(\hatq)} i\lambda U^{{\rm noQ},(1)}_{\ka}(\hatq) e^{i\mb{q}\cdot\mb{r}} \nonumber \\
    =& -\frac{4\pi}{\Omega} \sum_{\kappa\alpha} \frac{Z_{\ka}(\hatq) U^{{\rm noQ},(1)}_{\ka}(\hatq)}{\bar{\epsilon}(\hatq)} e^{i\mb{q}\cdot\mb{r}}.
\end{align}
The numerator becomes
\begin{align}
    \sum_{\kappa\alpha} & Z_{\ka}(\hatq) U^{{\rm noQ},(1)}_{\ka}(\hatq) \nonumber \\
    =& \sum_{\kappa\kappa'\alpha\beta\gamma} Z_{\ka}(\hatq) \widetilde{\Phi}^{(0)}_{\ka,\kappa'\gamma}(\hatq) \Lambda^{{\rm noQ},\kappa'}_{\gamma\beta}(\hatq) U_\beta \nnnl
    =& \sum_\beta \bra{Z(\hatq)} \bigg[ \widetilde{\Phi}_{\kappa\alpha,\kappa'\beta}^{{\rm AN},(0)} \nonumber \\
    &- \frac{4\pi}{\Omega} \frac{\widetilde{\Phi}_{\kappa\alpha,\kappa'\beta}^{{\rm AN},(0)} \ket{Z(\hatq)} \bra{Z(\hatq)} \widetilde{\Phi}_{\kappa\alpha,\kappa'\beta}^{{\rm AN},(0)}}{\epsilon_{\rm s}(\hatq)} \bigg]
    \ket{\Lambda^{{\rm noQ}}_\beta(\hatq)} U_\beta \nnnl
    =& \sum_\beta \frac{\bar{\epsilon}(\hatq)}{\epsilon^{\rm s}(\hatq)} \bra{Z(\hatq)} \widetilde{\Phi}_{\kappa\alpha,\kappa'\beta}^{{\rm AN},(0)}
    \ket{\Lambda^{{\rm noQ}}_\beta(\hatq)} U_\beta \nnnl
    =& \Omega \frac{\bar{\epsilon}(\hatq)}{\epsilon^{\rm s}(\hatq)} \sum_{\alpha\beta\gamma} e_{\alpha\beta\gamma} U_\beta \hatq_\alpha \hatq_\gamma.
\end{align}
Here, we used
\begin{align}
    \Big\langle Z(\hatq) \Big|  \widetilde{\Phi}_{\kappa\alpha,\kappa'\beta}^{{\rm AN},(0)}  \Big| & \Lambda^{{\rm noQ}}_\beta(\hatq) \Big\rangle \nonumber \\
      &= \sum_{\kappa\kappa'\alpha\gamma\delta} \hatq_\gamma Z_{\kappa\alpha\gamma} \widetilde{\Phi}^{{\rm AN},(0)}_{\ka,\kappa'\delta'}
    \Lambda^{{\rm AN},\kappa}_{\delta'\beta\delta} \hatq_\delta \nonumber \\
    &= \sum_{\kappa\alpha\gamma\delta} \hatq_\gamma Z_{\kappa\alpha\gamma} \Gamma^{{\rm AN},\kappa}_{\alpha\beta\delta} \hatq_\delta \\
    &= \Omega \sum_{\gamma\delta} (e - \bar{e})_{\gamma\beta\delta} \hatq_\gamma \hatq_\delta.
\end{align}
In the last equality, we used Eq.~(48) of \cite{Stengel2013}.

Therefore, the EPC is
\begin{multline} \label{eq:eph_noQ_D}
    V^{{\rm noQ,D},(0)}(\mb{r}; \hatq)\\
      =- 4\pi \sum_{\alpha\beta\gamma} \frac{(e - \bar{e})_{\alpha\beta\gamma} U_\beta \hatq_\alpha \hatq_\gamma}{\epsilon_{\rm s}(\hatq)} e^{i\mb{q}\cdot\mb{r}}.
\end{multline}
The piezoelectric EPC is proportional to $e - \bar{e}$, which is the internal-strain contribution to the piezoelectric tensor.

\subsection{Electron-phonon coupling with quadrupoles}
Now, we include the quadrupole contribution to the force constants and to the EPC.
The $\mcO(\lambda^0)$ term of the EPC is now the sum of the dipole term
\begin{align}
    V^{{\rm D},(0)}(\mb{r}; \hatq)
    = \frac{4\pi i}{\Omega} \sum_{\kappa\alpha} \frac{Z_{\ka}(\hatq) iU^{(1)}_{\ka}(\hatq)}{\bar{\epsilon}(\hatq)} e^{i\mb{q}\cdot\mb{r}}.
\end{align}
and the quadrupole term
\begin{align}
    V^{{\rm Q},(0)}(\mb{r}; \hatq)
    = \frac{4\pi}{\Omega} \sum_{\kappa\alpha\beta\gamma} \frac{\frac{1}{2} \hatq_\beta \hatq_\gamma Q_{\kappa\alpha\beta\gamma} U^{(0)}_{\ka}(\hatq)}{\bar{\epsilon}(\hatq)} e^{i\mb{q}\cdot\mb{r}}.
\end{align}
Here, we neglect the electric-field term ($\mathcal{Q}v^{{\rm Hxc},\mathcal{E}_\gamma}$) in Eq.~(24) of Ref.~\cite{2020BruninPiezo2}.

The dipole term now has an additional contribution from the second term of \Eq{eq:ph_full_Lambda}.
This contribution is
\begin{align} \label{eq:eph_quad_D}
    -\frac{4\pi}{\Omega} & \sum_{\kappa\alpha}  \frac{Z_{\ka}(\hatq) (U^{(1)}_{\ka}(\hatq) - U^{{\rm noQ},(1)}_{\ka}(\hatq))}{\bar{\epsilon}(\hatq)} e^{i\mb{q}\cdot\mb{r}} \nonumber \\
    =& -\frac{4\pi}{\Omega} \sum_\beta \frac{- 4\pi \frac{\bar{\epsilon}(\hatq)}{\epsilon^{\rm s}(\hatq)} \bra{Z(\hatq)} \widetilde{\Phi}^{(0)}(\hatq) \ket{Z(\hatq)}}{{\bar{\epsilon}(\hatq)}} \frac{\bar{e}_\beta(\hatq) U_\beta}{\bar{\epsilon}(\hatq)} e^{i\mb{q}\cdot\mb{r}} \nnnl
    =& -4\pi \frac{\bar{\epsilon}(\hatq) - \epsilon_{\rm s}(\hatq)}{\bar{\epsilon}(\hatq) \epsilon_{\rm s}(\hatq)} \sum_\beta \bar{e}_\beta(\hatq) U_\beta e^{i\mb{q}\cdot\mb{r}}
\end{align}

Using Eqs.~(91, 95) of Ref.~\cite{Stengel2013}, we find
\begin{align}
    \sum_{\alpha\beta\gamma}
    \bar{e}_{\alpha\beta\gamma} \hatq_\alpha U_\beta \hatq_\gamma 
    =& \frac{-1}{2\Omega} \sum_{\kappa\alpha\beta\gamma} (Q_{\kappa\beta\alpha \gamma} \!+\! Q_{\kappa\alpha\gamma\beta} \!-\! Q_{\kappa\gamma\alpha\beta}) \hatq_\alpha U_\beta \hatq_\gamma \nonumber \\
    =& \frac{-1}{2\Omega} \sum_{\kappa\alpha\beta\gamma} Q_{\kappa\beta\alpha\gamma} \hatq_\alpha U_\beta \hatq_\gamma.
\end{align}
The second equality holds because the second and third terms inside the parenthesis cancel out when we swap $\alpha$ and $\gamma$.
\begin{align} \label{eq:eph_quad_Q}
    V^{{\rm Q},(0)}(\mb{r}; \hatq)
    =& \sum_{\kappa\alpha\beta\gamma} \frac{4\pi}{\Omega} \frac{\frac{1}{2} \hatq_\beta \hatq_\gamma
    Q_{\kappa\alpha\beta\gamma} U_\alpha}{\bar{\epsilon}(\hatq)} e^{i\mb{q}\cdot\mb{r}} \nonumber \\
    =& - 4\pi \sum_{\alpha\beta\gamma} \frac{\bar{e}_{\alpha\beta\gamma} \hatq_\alpha U_\beta \hatq_\gamma}{\bar{\epsilon}(\hatq)} e^{i\mb{q}\cdot\mb{r}}.
\end{align}

Adding Eqs.~(\ref{eq:eph_noQ_D}, \ref{eq:eph_quad_D}, \ref{eq:eph_quad_Q}), we find
\begin{align}
    V^{{\rm D+Q},(0)}(\mb{r}; \hatq) =& - 4\pi e^{i\mb{q}\cdot\mb{r}} \sum_{\alpha\beta\gamma} U_\beta \hatq_\alpha \hatq_\gamma \Big[ \frac{(e - \bar{e})_{\alpha\beta\gamma} }{\epsilon_{\rm s}(\hatq)} \nonumber \\
   & + \frac{\bar{e}_{\alpha\beta\gamma} (\bar{\epsilon}(\hatq) - \epsilon_{\rm s}(\hatq))}{\bar{\epsilon}(\hatq) \epsilon_{\rm s}(\hatq)} 
    + \frac{\bar{e}_{\alpha\beta\gamma}}{\bar{\epsilon}(\hatq)} \Big] \nonumber \\
    =& - 4\pi e^{i\mb{q}\cdot\mb{r}} \sum_{\alpha\beta\gamma} \frac{ e_{\alpha\beta\gamma} U_\beta \hatq_\alpha \hatq_\gamma}{\epsilon_{\rm s}(\hatq)}.
\end{align}
Contrary to \Eq{eq:eph_noQ_D}, the piezoelectric EPC is proportional to $e$, the total piezoelectric tensor.

\section{Computational details} \label{sec:supp_comp_details}

\begin{table*}[ht]
\begin{tabular}{c|cccccccccccc}
\hline
Material & $a$ (bohr) & $E_{\rm cut}$ (Ry) & $N_k^{\rm DFT}$ & $N_{k,q}^{\rm Wannier}$ & Initial guess & $E_{\rm froz}$ (eV) & $E_{\rm dis}$ (eV) & $N_{\rm band}$ & $N_{\rm k}$ & $N_{\rm q}$ & $\veps^{\rm max}_\mb{k}$ (eV)& $\veps^{\rm max}_\mb{k+q}$  (eV)
\\ \hline
c-BN & 6.847589 & 100 & $20^3$ & $8^3$ & B: sp3  & -   & -    & 6  & $200^3$ & $200^3$ & 0.3 & 0.4   \\
Si   & 10.336   & 40  & $20^3$ & $8^3$ & Si: sp3 & 3.9 & -    & 14 & $100^3$ & $200^3$ & 0.13 & 0.23  \\
PbTe & 12.2076  & 80  & $16^3$ & $8^3$ & Pb, Te: s,p   & 2.2 & 6.7  & 15  & $200^3$ & $400^3$$^*$ & 0.05 & 0.2$^*$ \\
NaCl & 10.5636  & 80  & $20^3$ & $8^3$ & Na: s   & 1.5 & 6.5  & 5  & $100^3$ & $400^3$$^*$ & 0.1 & 0.4$^*$ \\
\hline
\end{tabular}
\caption{
    Computational parameters used in this work: cubic lattice parameter $a$, kinetic energy cutoff $E_{\rm cut}$
    size of the $\mb{k}$-point grid for DFT and DFPT calculations $N_k^{\rm DFT}$, 
    size of the $\mb{k}$- and $\mb{q}$-point grid for constructing the Wannier functions $N_{k,q}^{\rm Wannier}$,
    initial guess for constructing the Wannier functions,
    upper bound of the frozen (inner) window $E_{\rm froz}$,
    upper bound of the disentanglement (outer) window $E_{\rm dis}$,
    number of bands included for Wannierization $N_{\rm band}$,
    size of the $\mb{k}$-point ($\mb{q}$-point) grid for calculating self-consistent linewidths $N_k$ ($N_q$),
    maximum energy where the linewidths are calculated $\veps^{\rm max}_\mb{k+q}$,
    and the maximum energy of the intermediate states included in the calculation of the linewidths $\veps^{\rm max}_\mb{k+q}$.
    All energies, except for the kinetic energy cutoff, are evaluated with respect to the CBM energy.
    A hyphen denotes that the corresponding window was not set.
    ($^*$ For fixed-smearing calculations, we used $N_q=1600^3$ and $\veps^{\rm max}_\mb{k+q} = 0.08~\mathrm{eV}$ for PbTe and
    $N_q=600^3$ and $\veps^{\rm max}_\mb{k+q} = 0.2~\mathrm{eV}$ for NaCl.)
}
\label{tab:comp_params}
\end{table*}

\subsection{\textit{Ab initio} Wannier interpolation of electrons and phonons}

The density functional theory (DFT) and density functional perturbtion theory (DFPT) calculations were performed using the \textsc{Quantum ESPRESSO} package~\cite{Giannozzi2017}, except for the DFPT calculation of the piezoelectric tensor for which we used \textsc{Abinit}~\cite{Gonze2016,Gonze2020}.
We used a 20$\times$20$\times$20 unshifted $\mb{k}$-point grid, a kinetic energy cutoff of 100~Ry,
and fully-relativistic norm-conserving pseudopotentials~\cite{2013HamannONCVPSP} from \textsc{PseudoDojo} (v0.4)~\cite{Setten2018} in the Perdew-Burke-Ernzerhof (PBE) parametrization of the exchange-correlation functional in the generalized-gradient approximation~\cite{Perdew1996}.
The cubic lattice parameter of 6.8476~Bohr was obtained by a full relaxation of the cell, in consistence with 6.848~Bohr reported in Ref.~\onlinecite{Ponce2021}.

To construct the Wannier functions and the real-space matrix elements, we used \textsc{Wannier90}~\cite{2020PizziWannier90} and EPW~\cite{Giustino2007,Ponce2016}.
The Brillouin zone was sampled with a 8$\times$8$\times$8 grid for both electrons and phonons.
We generated 4 Wannier functions using the atomic $sp^3$ orbitals at the $B$ atom as initial guesses and performing disentanglement and maximal localization~\cite{1997Marzari,2001Souza}.
We set the upper bound of the frozen (inner) and disentanglement (outer) windows at 7.4~eV and 17.4~eV above the CBM energy, respectively.
The dipolar and quadrupolar contributions to the phonon dynamical matrix~\cite{2019RoyoPiezo} and the electron-phonon coupling~\cite{2020BruninPiezo1,2020JhalaniPiezo} were explicitly included during Wannier interpolation.
The quadrupoles of $Q_{\rm B} = 3.46$ and $Q_{\rm N} = -0.63$ were obtained by fitting the DFPT calculations at small $\mb{q}$ to the long-range model~\cite{Ponce2021}.

We performed Wannier interpolation and computed the linewidths using an in-house developed code \texttt{ElectronPhonon.jl}, which is written in the Julia programming language~\cite{2017BezansonJulia}.
Details of the implementation will be discussed elsewhere.
We calculated the linewidths only for states with energy in the range $[E_{\rm CBM},\, E_{\rm CBM} + 0.3~\mathrm{eV}]$, and summed over only the states with energy in the range $[E_{\rm CBM},\, E_{\rm CBM} + 0.4~\mathrm{eV}]$ for the evaluation of the sum-over-states in \Eq{eq:gamma}.
For the self-consistent calculation, the linewidths were linearly interpolated.
For the intermediate states outside the window $[E_{\rm CBM},\, E_{\rm CBM} + 0.3~\mathrm{eV}]$, we used a constant extrapolation from a nearest $\mb{k}$-point in the Brillouin zone.
Doping and free-carrier screening were included using the Lindhard model and the rigid-band approximation, following Ref.~\cite{2017Verdi}.
The renormalized vertex $\widetilde{g}_\qnu(\omega)$ is computed by screening the long-range part of the vertex using the free-carrier dielectric function evaluated at the phonon energy:
\begin{equation} \label{eq:supp_g_screen}
    \widetilde{g}_\qnu(\omega_\qnu) = g^{\rm sr}_\qnu + \frac{1}{\epsilon(\omega_\qnu)} g^{\rm lr}_\qnu \,.
\end{equation}
We use the Lindhard dielectric function in the relaxation-time approximation~\cite{1970Mermin} with a fixed linewidth of 14~meV, which is the smallest self-consistent linewidth of cubic BN at $300~\mathrm{K}$.

Calculations for silicon, PbTe, and NaCl were performed similarly.
The computational parameters are summarized in Tab.~\ref{tab:comp_params}.

\begin{figure}[b]
    \centering
    \includegraphics[width=0.99\linewidth]{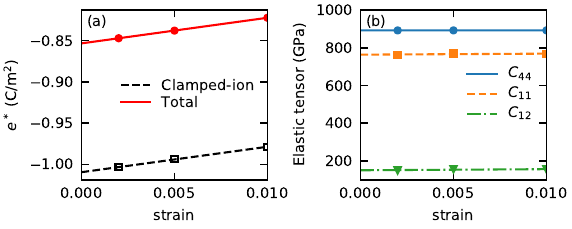}
    \caption{
        (a) Piezoelectric and (b) elastic coefficients obtained from finite-difference calculations.
        Symbols denote values obtained from DFT calculations under a finite strain.
        Lines denote the linear least-squares fit of the finite-strain data for zero-strain extrapolation.
    }
    \label{fig:piezo_and_elastic}
\end{figure}

In electron-phonon calculations, one sets a cutoff for acoustic phonon frequency and ignore phonon modes with frequency below that cutoff to avoid numerical problems with vanishing or negative phonon frequencies that may occur due to numerical inaccuracy or a small violation of the acoustic sum rule~\cite{BornHuang1954}.
The cutoff is typically set to a value in the scale of 1~meV (EPW~\cite{Ponce2016} and Perturbo~\cite{2021ZhouPerturbo} codes use $5~\mathrm{cm^{-1}} \approx 0.6~\mathrm{meV}$ and 1~meV as default, respectively).
In piezoelectric materials, the low-energy acoustic phonon modes play a crucial role in the electron linewidths, so it is important to accurately describe the phonons with small wavevectors.
To this aim, we set a very small cutoff of $10^{-5}$~meV.

We find that the use of the minimal-distance replica in the Wigner--Seitz supercell for Wannier interpolation of the force constants~\cite{Ponce2021} is crucial to avoid negative acoustic phonons at very small wavevectors ($q \sim 10^{-4}~\mathrm{\AA}^{-1}$).
In the Wannier interpolation of force constants, the dynamical matrix calculated on a coarse $\mb{q}$-point grid, say $N_q \times N_q \times N_q$, is Fourier transformed to real space force constants.
Due to limited resolution in momentum space, the real-space force constant between an atom $a$ at position $\mb{\tau}_a$ and an atom $b$ at $\mb{\tau}_b$ separated by a lattice vector $\mb{R}_p$ can be assigned to be any periodic image of the lattice vector in the Wigner--Seitz supercell, $\mb{R}'_p = \mb{R}_p + N_q \sum_{i=1}^3 n_i \mb{a}_i$, where $\mb{a}_i$ is a lattice vector and $n_i$ an integer.
The minimal distance replica method is a method to assign the real-space force constants to a lattice vector $\mb{R}'_p$ that minimizes the distance between the atom pairs $\left| \mb{\tau}_a - \mb{\tau}_b - \mb{R}'_p \right|$.
Note that all DFT and DFPT calculations are done in the primitive unit cell.
Details of this method are described in App.~A of Ref.~\cite{Ponce2021}.

In the self-consistent calculations, we fix the chemical potential to the value for which the corresponding carrier density of the electrons matches the desired carrier density.
The convergence criteria is that, for every $\mb{k}$ point the absolute change in the linewidth is less than 0.1~meV, or the relative change is less than 1~\%.
The typical number of self-consistent iterations was between 5 to 10; the maximum number of iterations required in this work was 18.

Compared to conventional, state-of-the-art fixed- or adaptive-smearing methods, the cost of the whole calculation is \textit{even lower} when using our self-consistent method because a coarser $\mb{q}$-point grid can be used thanks to the sizable magnitude of self-consistent linewidth.
The typical $\mb{q}$-grid size needed for convergence was $200^3$ for the self-consistent calculation and $800^3$ for a fixed broadening of 0.02~meV.
Thus, our scheme significantly reduces the cost of computing the electron-phonon matrix elements, which takes more than 99~\% of the total computational cost.
These matrix elements are reused during the self-consistent iterations, and hence the computational overhead of self-consistency is almost negligible.

\subsection{Calculation of piezoelectric and elastic coefficients} \label{sec:supp_comp_piezo}
\subsubsection{Using \textsc{Quantum ESPRESSO}: Finite difference}

The piezoelectric and elastic tensor can be obtained as the change in polarization and stress, respectively, due to a change in strain.
The piezoelectric tensor is \textit{improper} when obtained from such a numerical derivative:
\begin{equation} \label{eq:piezo_improper}
    e^{\rm improper}_{\gamma\delta\alpha} = \frac{\partial P_\alpha}{\partial \eta_{\gamma\delta}}\bigg|_{\mathcal{E}} = - \frac{\partial \sigma_{\alpha\gamma}}{\partial \mathcal{E}_{\delta}}\bigg|_{\eta}.
\end{equation}
The improper piezoelectric tensor leads to a dependence on the choice of branch of the polarization which can be eliminated with the standard relationship~\cite{Vanderbilt2000}
\begin{equation}
    e_{\gamma\delta\alpha} = e^{\rm improper}_{\gamma\delta\alpha} + \delta_{\gamma\delta}P_\alpha - \delta_{\alpha\gamma}P_{\delta}.
\end{equation}

For a cubic crystal with a zincblende structure and space group F$\bar{4}$3m (\#216) such as c-BN,
the elastic tensor has three independent parameters, $C_{11}$, $C_{12}$, and $C_{44}$.
The tensor elements read
\begin{equation}
    C_{\alpha\gamma\beta\delta} = \begin{cases}
        C_{11} & \text{ if $\alpha=\beta = \gamma=\delta$} \\
        C_{12} & \text{ if $\alpha=\beta \neq\gamma=\delta$} \\
        C_{44} & \text{ if $\alpha \neq \beta, \gamma \neq\delta, (\alpha, \beta) = (\gamma, \delta)$ or $(\delta, \gamma)$} \\
        0 & \text{ otherwise}.
    \end{cases}
\end{equation}
The piezoelectric tensor has one free parameter $e^*$ and reads
\begin{equation}
    e_{\alpha\beta\gamma} = \begin{cases}
        e^* & \text{ if $\alpha \neq \beta \neq \gamma \neq \alpha$} \\
        0 & \text{ otherwise}.
    \end{cases}
\end{equation}

These parameters can be obtained from calculations with two different strain. 
For fully relaxed cubic lattice vectors $\mb{a}^0_i$ ($i = 1, 2, 3$), the strained lattice vectors under strain $\mb{\eta}$ read
\begin{equation}
    \mb{a}_i(\lambda \boldsymbol{\eta}) = (1 + \lambda \boldsymbol{\eta}) \mb{a}^0_i.
\end{equation}
By calculating the electric polarization $P$ and stress $S$ from the strained structures, we get the piezoelectric and elastic tensors:
\begin{subequations}
\begin{align}
    e^{\rm improper}_{yzx} &= \frac{1}{2} \lim_{\lambda \to 0} \frac{P_x(\lambda \mb{\eta}) - P_x(0)}{\lambda},
    \\
    \label{eq:supp_elastic_C44}
    C_{44} &= \frac{1}{2} \lim_{\lambda \to 0} \frac{S_{yz}(\lambda \mb{\eta}) - S_{yz}(0)}{\lambda}, \\
    \mb{\eta} &= \begin{pmatrix} 0 & 0 & 0 \\ 0 & 0 & 1 \\ 0 & 1 & 0\end{pmatrix},
\end{align}
\end{subequations}
as well as
\begin{subequations}
\begin{align}
    C_{11} =& \lim_{\lambda \to 0} \frac{S_{zz}(\lambda \mb{\eta}) - S_{zz}(0)}{\lambda}, \\
    C_{12} =& \lim_{\lambda \to 0} \frac{S_{xx}(\lambda \mb{\eta}) - S_{xx}(0)}{\lambda}, \\
    \mb{\eta} =& \begin{pmatrix} 0 & 0 & 0 \\ 0 & 0 & 0 \\ 0 & 0 & 1\end{pmatrix}.
\end{align}
\end{subequations}
Using the modern theory of polarization~\cite{1992RestaPolarization,1993KingSmithPolarization,1993VanderbiltPolarization,1994RestaPolarization}, we calculate the polarization of each structure as the sum of the Wannier function centers plus the contribution of the atomic nuclei~\cite{2012Marzari}.
To reduce the finite-difference error in the derivative with respect to strain, we repeat the calculation at three different values of $\lambda$, 0.002, 0.005, 0.01, and linearly extrapolate the piezoelectric and elastic tensors to $\lambda = 0$.

When we take the strained structures without further relaxation of atomic coordinates, we get the clamped-ion (CI) contribution.
By performing the calculation with a structural relaxation at each strain, we obtain the total piezoelectric and elastic tensors.
The difference between the two is called the internal-strain (IS) contribution.

Figure~\ref{fig:piezo_and_elastic} shows the result of the finite-difference calculations.
For the piezoelectric tensor, we obtain $e^{*,{\rm CI}} = 1.0095~\mathrm{C/m^2}$ without structural relaxation and $e^{*} = 0.8534~\mathrm{C/m^2}$ with relaxation.
The difference between the two is the internal-strain contribution $e^{*,{\rm IS}} = -0.1561~\mathrm{C/m^2}$.
For the elastic tensor, we obtain $C_{11} = 764$, $C_{12}=151$, and $C_{44} = 446~$GPa from relaxed calculations, in good agreement with the experimental values $C_{11} = 820$, $C_{12}=190$, and $C_{44} = 480~$GPa~\cite{1994GrimsditchElastic}.

\subsubsection{Using \textsc{Abinit}: Density-functional perturbation theory}
To validate our finite-difference calculation, we also calculated the piezoelectric tensor using the DFPT~\cite{Gonze1997a}.
In the context of first-principles simulation, an efficient way to compute the piezoelectric tensor is by using DFPT, treating a homogeneous strain as perturbation through a metric tensor formulation~\cite{Hamann2005}.
Such DFPT-based methods, as implemented in the \textsc{Abinit}~\cite{Gonze2016,Gonze2020} code, are a superior way of computing piezoelectric tensor as they do not rely on finite-difference derivatives. 
Also, while finite difference yields the \textit{improper} piezoelectric tensor \Eq{eq:piezo_improper}~\cite{Vanderbilt2000}, DFPT directly yields the \textit{proper} piezoelectric tensor $e_{\gamma\delta\alpha}$.

The mixed derivative of the total electronic energy with respect to electric field and strain gives the clamped-ion piezoelectric tensor~\cite{Hamann2005}:
\begin{align}
    e^{\rm CI}_{\gamma\delta\alpha} = \frac{\partial^2 E}{\partial \mathcal{E}_\alpha \partial \eta_{\gamma\delta}}
    = \frac{2\Omega}{(2\pi)^3} \int \sum_n \langle i \Psi_{n\mb{k}}^{(k_\alpha)} | \Psi_{n\mb{k}}^{(\eta_{\gamma\delta})} \rangle d\mb{k},
\end{align}
where we have the first-order derivative of the wavefunction with respect to $\mb{k}$ and to strain. 
We obtain in Cartesian coordinate the clamped-ion proper piezoelectric tensor which is zero except for shear strain $e^{*,\mathrm{CI}}=1.0094$~C/m$^2$.
The addition of atom-relaxation effects is done following Ref.~\cite{Wu2005} and gives the total proper piezoelectric constant of $e^*= 0.8541~\mathrm{C/m^2}$.
These results are in good agreement with those obtained by finite difference.

\FloatBarrier  

\makeatletter\@input{xx.tex}\makeatother
\bibliography{Bibliography}